\documentclass[acmtog]{acmart}

\usepackage{graphics}
\usepackage{algorithm, algorithmic, makecell, multirow, amssymb}
\usepackage{url}

\begin{document}
\title{Real-time Image Smoothing via Iterative Least Squares}

\author{Wei Liu}
\affiliation{
  \institution{School of Computer Science, The University of Adelaide}
  \city{Adelaide}
  \postcode{5005}
  \country{Australia}
  }
\email{wei.liu02@adelaide.edu.au}

\author{Pingping Zhang}
\affiliation{
  \institution{School of Information and Communication Engineering, Dalian University of Technology}
  \city{Dalian}
  \postcode{116024}
  \country{China}}
\email{jssxzhpp@mail.dlut.edu.cn}

\author{Xiaolin Huang}
\affiliation{
  \institution{Institute of Image Processing and Pattern Recognition \&  Institute of Medical Robotics, Shanghai Jiao Tong University}
  \city{Shanghai}
  \postcode{200240}
  \country{China}}
\email{xiaolinhuang@sjtu.edu.cn}

\author{Jie Yang}
\affiliation{
  \institution{Institute of Image Processing and Pattern Recognition \&  Institute of Medical Robotics, Shanghai Jiao Tong University}
  \city{Shanghai}
  \postcode{200240}
  \country{China}}
\email{jieyang@sjtu.edu.cn}

\author{Chunhua Shen}
\affiliation{
  \institution{School of Computer Science, The University of Adelaide}
  \city{Adelaide}
  \postcode{5005}
  \country{Australia}}
\email{chunhua.shen@adelaide.edu.au}

\author{Ian Reid}
\affiliation{
  \institution{School of Computer Science, The University of Adelaide}
  \city{Adelaide}
  \postcode{5005}
  \country{Australia}}
\email{ian.reid@adelaide.edu.au}

\renewcommand{\shortauthors}{Wei Liu, Pingping Zhang, Xiaolin Huang, Jie Yang, Chunhua Shen, and Ian Reid}

\begin{abstract}
Edge-preserving image smoothing is a fundamental procedure for many computer vision and graphic applications. There is a tradeoff between the smoothing quality and the processing speed: the high smoothing quality usually requires a high computational cost which leads to the low processing speed. In this paper, we propose a new global optimization based method, named iterative least squares (ILS), for efficient edge-preserving image smoothing. Our approach can produce high-quality results but at a much lower computational cost. Comprehensive experiments demonstrate that the propose method can produce results with little visible artifacts. Moreover, the computation of ILS can be highly parallel, which can be easily accelerated through either multi-thread computing or the GPU hardware. With the acceleration of a GTX 1080 GPU, it is able to process images of 1080p resolution ($1920\times1080$) at the rate of 20fps for color images and 47fps for gray images. In addition, the ILS is flexible and can be modified to handle more applications that require different smoothing properties. Experimental results of several applications show the effectiveness and efficiency of the proposed method. The code is available at \url{https://github.com/wliusjtu/Real-time-Image-Smoothing-via-Iterative-Least-Squares}
\end{abstract}



\maketitle

\section{Introduction}

Edge-preserving smoothing (EPS) has attracted increasing research interests in the fields of both computer vision and graphics for decades. The main aim of EPS is to smooth out small details in images and preserve the major edges and structures. Due to the wide applications, many EPS approaches have been proposed in the literature. Among these approaches, two kinds of methods have been widely developed: weighted average based methods and global optimization based methods.

Weighted average based methods are also known as filter based methods or local methods \cite{kim2017fast, min2014fast}. These approaches include the well-known bilateral filter \cite{tomasi1998bilateral} and joint bilateral filter \cite{eisemann2004flash, petschnigg2004digital}. A number of approaches have also been proposed to either accelerate bilateral filter \cite{adams2010fast, durand2002fast, paris2006fast, porikli2008constant, yang2009real} or introduce fast alternatives \cite{gastal2011domain, gastal2012adaptive}. The weighted median filter \cite{ma2013constant, zhang2014100+}, tree filter \cite{bao2014tree} and guided filter \cite{he2013guided} are also widely used in various applications. Generally, weighted average based methods are computationally efficient. Some of them can even achieve real-time or near real-time image smoothing \cite{gastal2011domain, yang2009real}. However, the main drawback of these methods is that there is a tradeoff between their edge-preserving abilities and the smoothing abilities: large smoothing strength can blur edges, and this can lead to halo artifacts. Some methods such as bilateral filter \cite{tomasi1998bilateral} and its alternatives \cite{gastal2011domain, gastal2012adaptive} can also produce results with gradient reversal artifacts.

Global optimization based methods usually formulate the image smoothing with global optimization frameworks. Methods in this category include total variation smoothing \cite{rudin1992nonlinear}, weighted least squares smoothing \cite{farbman2008edge}, gradient $L_0$ norm smoothing \cite{xu2011image} and relative total variation smoothing \cite{xu2012structure}, to name a few. Generally, global optimization based methods can achieve superior performance over the weighted average based ones in avoiding artifacts such as gradient reversals and halos. However, the superior performance is achieved at the expense of high computational costs arising from solving the global optimization objective function.

In this paper, we propose a new global optimization based method that is able to achieve high smoothing quality and high processing speed at the same time. The effectiveness, efficiency and flexibility of the proposed approach are validated through applications in various tasks. The main contributions of this paper are as follows:

\begin{itemize}
  \item[1.] We propose a new global optimization based method, named iterative least squares (ILS), for efficient edge-preserving image smoothing. It can produce high-quality results that are on par with that of the state-of-the-art approaches, but at a much lower computational cost which is comparable with that of some state-of-the-art weighted average based methods. Comprehensive experiments demonstrate that the proposed ILS can produce results with little visible artifacts.

  \item[2.]  The computation of ILS is simple and highly parallel. It only needs to iteratively perform two steps a few times: a point-wise manipulation step followed by one fast Fourier transform (FFT) and one inverse fast Fourier transform (IFFT). Both these two steps are suitable for multi-thread computing and GPU acceleration. When running on a GTX 1080 GPU, the ILS can process images of 1080p resolution ($1920\times1080$) at the rate of 20fps for color images and 47fps for gray images.

  \item[3.] The ILS is also flexible. With slight modifications, it is capable of more applications that require different smoothing properties.
\end{itemize}

The rest of this paper is organized as follows: Sec.~\ref{SecRelatedWork} describes the related approaches. Sec.~\ref{SecOurILS} is devoted to the proposed method, including how it is derived, parameter analysis and a deep comparison with highly related approaches. In Sec.~\ref{SecComp}, we show comparisons between our approach and state-of-the-art approaches in terms of running speed and smoothing quality. Our approach is further extended in Sec.~\ref{SecExtension} with the application to more tasks. We draw the conclusion and limitation of our approach in Sec.~\ref{SecConclusion}.

\section{Related Work}
\label{SecRelatedWork}


\subsection{Weighted Average Based Methods}

Weighted average based methods have been widely developed in past decades. These approaches usually calculate the output pixel value as a weighted average of input pixel values inside a local window or a non-local window. The very early work is the bilateral filter \cite{tomasi1998bilateral} which has been used in a range of applications such as HDR tone mapping \cite{durand2002fast}, image detail enhancement \cite{fattal2007multiscale}, and etc. As its variant, joint bilateral filter has also been used in flash/no flash filtering \cite{petschnigg2004digital, eisemann2004flash} and depth map upsampling \cite{Kopf2007Joint}. Since the brute-force implementation of bilateral filter is computationally expensive, a variety of approaches have been proposed to accelerate bilateral filter \cite{adams2010fast, durand2002fast, paris2006fast, porikli2008constant, yang2009real}. The domain transform filter \cite{gastal2011domain} and adaptive manifold filter \cite{gastal2012adaptive} are also alternatives of bilateral filter that are fast and able to handle high-dimension data such as color images. Weighted median filter \cite{ma2013constant, zhang2014100+} is another widely used edge-preserving filter due to its robustness against outliers and the ability to preserve sharp edges. The above filters calculate output pixel values in local windows of input images. The tree filter \cite{bao2014tree} and segment graph based filter \cite{zhang2015segment} consider images as minimum spanning trees. They calculate output pixel values in non-local widows of input images. Different from the above approaches which are based on piecewise constant models, the guided filter \cite{he2010guided, he2013guided} and its variants \cite{dai2015fully, lu2012cross, tan2014multipoint} are based on piecewise linear models. Although weighted average based methods remain popular, they retain their own inherent drawbacks. There is usually a tradeoff between their edge-preserving abilities and the smoothing abilities: large smoothing strength can blur edges, and this can lead to halo artifacts. Some methods such as bilateral filter \cite{tomasi1998bilateral} and its alternatives \cite{gastal2011domain, gastal2012adaptive} can also produce results with gradient reversal artifacts.

\subsection{Global Optimization Based Methods}

Global optimization based methods usually formulate the image smoothing with a global optimization framework consisting of a data term and a regularization term. The regularization term embeds the prior of the output images, and it usually imposes a certain penalty on image gradients. The output image is the optimum of the object function. The very early work is the total variation smoothing \cite{rudin1992nonlinear} which regularizes image gradients with the $L_1$ norm penalty. Farbman et~al. \shortcite{farbman2008edge} proposed a weighted least squares (WLS) framework which imposes a weighted $L_2$ norm penalty on image gradients. Their method shows superior performance over bilateral filter \cite{tomasi1998bilateral} and the methods based on bilateral filter \cite{chen2007real, fattal2007multiscale}. Xu et~al. \shortcite{xu2011image} adopted gradient $L_0$ norm smoothing to sharpen salient edges while smoothing out weak edges. They also proposed relative total variation smoothing \cite{xu2012structure} for efficient image texture smoothing. Ham et~al. \shortcite{ham2015robust} proposed a static/dynamic filter which combined both static guidance weights and non-convex potentials as the priors for output images. Generally, global optimization based methods can achieve superior performance over the weighted average based ones in several tasks, however, this is achieved at the expense of much higher computational costs.

The method proposed by Badri et~al. \shortcite{badri2013fast, badri2015fast} is an exception which is very fast. It shares a similar two-step smoothing procedure with our approach. However, as we will show in Sec.~\ref{SecComp}, the mathematical mechanisms behind these two methods are different. Besides, their method performs poorly in avoiding gradient reversals and halos while our method shows better performance, shown in Fig.~\ref{FigILSVSL0L1} and Fig.~\ref{FigILSvsFOPA_HDR}.

Preconditioned techniques \cite{krishnan2013efficient, afonso2010fast} have also been developed to accelerate the conjugate gradient (CG) optimization method \cite{boyd2004convex} which can be used to solve the general WLS smoothing problem. Although they greatly reduce the iteration number, the cost of constructing preconditioners is also considerable. Recently, the fast global smoother \cite{min2014fast} and semi-global weighted least squares \cite{liu2017semi} were proposed to accelerate the WLS smoothing by decomposing the large linear system into a series of small linear subsystems. However, as we will demonstrate in Sec.~\ref{SecComp} (Fig.~\ref{FigILSvsFGSandSGWLS}), the decomposition can lead to visible blocky artifacts in HDR tone mapping.  Barron et~al. \shortcite{barron2016fast} proposed the bilateral solver to accelerate the WLS smoothing. However, it is only capable of Gaussian guidance weights. As we will show in Sec.~\ref{SecComp} (Fig.~\ref{FigILSvsFBS}), it is also prone to produce artifacts in the results. Besides, its computational cost is also sensitive to the ``spatial bandwidth'' parameter, and a small value of this parameter can greatly increase the memory and computational cost \cite{mazumdar2017hardware}.

\subsection{Pyramid Based Methods}

In contrast to the approaches in the above two categories, the local Laplacian filter \cite{paris2011local} is a pyramid based one. It performs edge-preserving smoothing with a Laplacian pyramid and shows state-of-the-art performance. However, the promising performance is also achieved at the expense of high computational costs. The fast local Laplacian filter \cite{aubry2014fast} achieves a great acceleration. However, as shown in Sec.~\ref{SecComp}, it is still much slower than the state-of-the-art weighted average based methods \cite{gastal2011domain, gastal2012adaptive, he2013guided}. Based on the second-generation wavelets \cite{sweldens1998lifting}, Fattal \shortcite{fattal2009edge} proposed edge-avoiding wavelets for efficient edge-preserving image smoothing. As shown in Sec.~\ref{SecComp}, its main drawback is that it can produce noticeable artifacts in tone and detail manipulation tasks.

\subsection{Deep Learning Based Methods}

In recent years, a large majority of deep learning based approaches have also been proposed \cite{chen2017fast,gharbi2017deep, gharbi2015transform, xu2015deep, isola2017image, liu2016learning}. Generally, these methods adopt different deep neural network architectures to imitate the smoothing effects of existing filters. Liu et~al. \shortcite{liu2016learning} combined recurrent neural networks (RNN) and convolutional neural networks (CNN) for recursive filters learning. Conditional adversarial networks were investigated in \cite{isola2017image} for general image-to-image translation problems. Fully convolutional networks (FCN) were used by Chen et~al. \shortcite{chen2017fast} to accelerate a wide variety of image processing operators. These approaches can yield significant speedups over the original filters with the help of GPU acceleration. However, their main drawback is that different models usually need to be trained separately for different parameter settings. Chen et~al. \shortcite{chen2017fast} proposed a network that could handle various parameter settings, however, it only works on the filter with one parameter, e.g., gradient $L_0$ norm smoothing \cite{xu2011image}. For the filter with more than one parameter, e.g., WLS smoothing \cite{farbman2008edge}, there is seldom work that can handle the different parameter settings.

\section{Iterative Least Squares for Efficient Edge-preserving Smoothing}
\label{SecOurILS}

Our approach starts from the minimization procedure of the following objective function:
\begin{equation}\label{EqLpSmoothObj}
    E(u, f) = \sum\limits_{s}\left((u_s - f_s)^2 +\lambda\sum\limits_{\ast\in\{x,y\}}\phi_{p} \left(\nabla u_{\ast,s}\right) \right),
\end{equation}
where $f$ is the input image, $u$ is the smoothed output image, $s$ denotes the pixel position and $\nabla u_\ast (\ast\in\{x,y\})$ represents the gradient of $u$ along $x$-axis/$y$-axis. In this paper, we adopt the discrete differential operators $[1,-1]$ and $[1, -1]^T$ for the computation of gradients along $x$-axis and $y$-axis, respectively. These discrete differential operators are the standard way to compute image gradients, and they are also widely used in the literature \cite{farbman2008edge, xu2011image, xu2012structure, xu2013unnatural, wang2008new}. The penalty function $\phi_{p}(\cdot)$ is defined as:
\begin{equation}\label{EqSmoothedLpNorm}
    \phi_{p}(x)=(x^2 + \epsilon)^{\frac{p}{2}},
\end{equation}
where $\epsilon$ is a small constant. Eq.~(\ref{EqSmoothedLpNorm}), which is differentiable at the origin, is called generalized Charbonnier penalty in the literature \cite{krahenbuhl2012efficient, sun2010secrets}. In this paper, if not specified, we fix $\epsilon=0.0001$ in all the experiments. The norm power $p$ is usually set as $0<p\leq1$ for edge-preserving smoothing.

When solving Eq.~(\ref{EqLpSmoothObj}), iterative re-weighted least squares (IRLS) is the most widely adopted technique in the literature \cite{dong2017blind, pan2016robust, pan2016soft}. In contrast, Eq.~(\ref{EqLpSmoothObj}) can also be solved through the additive half-quadratic minimization \cite{geman1995nonlinear, nikolova2005analysis, he2013half}. The additive half-quadratic minimization has not received much attention in the past decades \cite{nikolova2005analysis}. However, in this paper, we show that based on this technique and our experimental observations (detailed in Sec.~\ref{SecILS}), a new global optimization based approach can be obtained for edge-preserving image smoothing. It can achieve high smoothing quality while maintaining the high processing speed at the same time.

\subsection{Additive Half-quadratic Minimization}
\label{SecHalfQuadratic}

The additive half-quadratic minimization was first introduced by Geman and Yang \shortcite{geman1995nonlinear} for solving the objective functions regularized with non-convex penalties in image restoration. It was further analyzed by Nikolova and Ng \shortcite{nikolova2005analysis} for solving the objective functions regularized with convex penalties. He et~al. \shortcite{he2013half} further shows its applications in face recognition.

In this subsection, we briefly present the optimization procedure of solving Eq.~(\ref{EqLpSmoothObj}) with the additive half-quadratic minimization technique. For the generalized Charbonnier penalty in Eq.~(\ref{EqSmoothedLpNorm}), a constant $c=p\epsilon^{\frac{p}{2}-1}>0$ (a detailed proof is given in Sec.~\ref{SecCAnalysis} and Appendix B) exists such that $g(x)=\frac{c}{2}x^2-\phi_{p}(x)$ is strictly convex. Clearly, $g(x)$ is even and the property of $g(x)$ for $x\leq0$ is the same as that for $x\geq0$. We thus focus on the property of $g(x)$ for $x\geq0$. According to Eq.~(\ref{EqAddFormInequality}) in Appendix A, a $\psi(\mu)$ exists such that:
\begin{equation}\label{EqLpUpperBound}
\phi_{p}(x)=\underset{\mu}{\min} \left\{\frac{1}{2}(\sqrt{c}x - \frac{1}{\sqrt{c}}\mu)^2 + \psi(\mu)\right\},
 \end{equation}
with $\mu=cx - \phi_{p}'(x)$ as the optimum condition, and $\phi_{p}'(x)$ is the derivative of $\phi_{p}(x)$ with respect to $x$. Thus, we can define the following energy function:
\begin{equation}\label{EqAugmentedAdd}
\small
{
\begin{split}
    &\widetilde{E}_A(u,f,\mu_x, \mu_y) = \\
    &\sum\limits_{s}\left((u_s - f_s)^2  + \lambda \sum\limits_{\ast\in\{x,y\}} \left(\frac{1}{2}\left(\sqrt{c}\nabla u_{\ast,s} - \frac{1}{\sqrt{c}}\mu_{\ast,s} \right)^2+ \psi(\mu_{\ast,s})\right)\right).
\end{split}
}
\end{equation}
Then we have:
\begin{equation}\label{EqEnergyInequalityAdd}
{
    E(u, f)=\underset{\mu_x,\mu_y}{\min} \widetilde{E}_A(u, f, \mu_x, \mu_y),
}
\end{equation}
with $\mu_{x,s} = c\nabla u_{x,s} - \phi_{p}'(\nabla u_{x,s})$ and $\mu_{y,s} = c\nabla u_{y,s}- \phi_{p}'(\nabla u_{y,s})$ as the optimum condition. Based on Eq.~(\ref{EqAugmentedAdd}) and Eq.~(\ref{EqEnergyInequalityAdd}), the output $u$ can be obtained by alternatively updating $u$ and $\mu_x,\mu_y$ in Eq.~(\ref{EqAugmentedAdd}). Since $\mu_{x,s}$ and $\mu_{y,s}$ have closed-form solutions as described above, the value of $u$ in each iteration can be obtained as:
\begin{equation}\label{EqIterSolutionEnergyAdd}
{
    u^{n + 1} = \underset{u}{\mathop{\arg\min }} \widetilde{E}_A(u,f,\mu^n_x, \mu^n_y).
}
\end{equation}
The value of $\mu^n_x$ and $\mu^n_y$ in each iteration is computed as:
\begin{equation}\label{EqIntermidiateVar}
{
\begin{split}
    & \mu^n_{\ast,s}=c\nabla u^n_{\ast,s} - \phi_{p}'\left(\nabla u^n_{\ast,s}\right)\\
    & \ \ \ \ \ \ =c\nabla u^n_{\ast,s} - p\nabla u_{\ast,s}\left(\left(\nabla u^n_{\ast,s}\right)^2 +\epsilon \right)^{\frac{p}{2}-1}, \ast\in\{x,y\}.
\end{split}
}
\end{equation}

\begin{figure*}
\centering
\setlength{\tabcolsep}{0.25mm}
\begin{tabular}{ccc}
\includegraphics[width=0.31\linewidth]{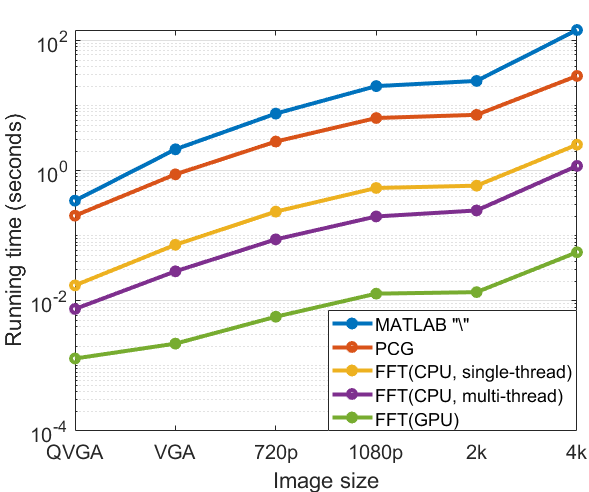} &
\includegraphics[width=0.31\linewidth]{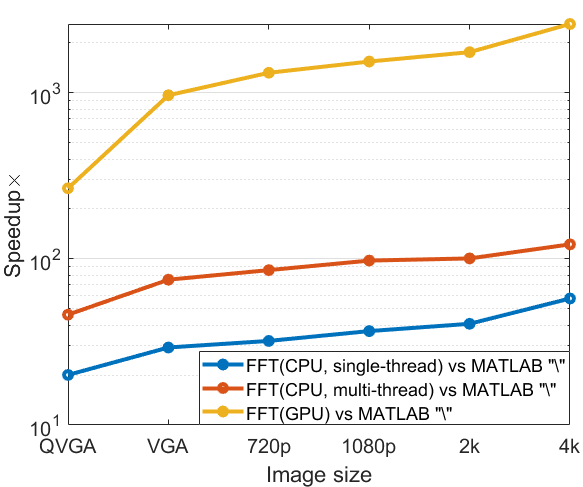} &
\includegraphics[width=0.31\linewidth]{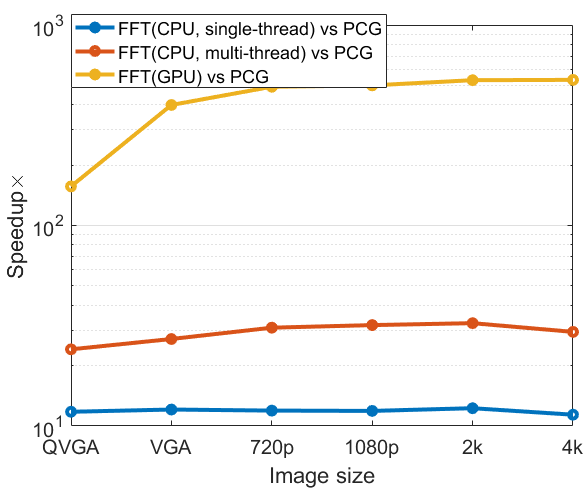}\\
(a) & (b) & (c)
\end{tabular}
\caption{(a) Running time of different solvers for different image sizes. ``FFT(CPU, single-thread)/FFT(CPU, multi-thread)'' refers to the single-thread/multi-thread CPU implementation of Eq.~(\ref{EqILSSolution}), and ``FFT(GPU)'' refers to the GPU counterpart. Time is evaluated for one iteration of Eq.~(\ref{EqILSObj}). Input images are RGB color images. The speedup of Eq.~(\ref{EqILSSolution}) over the intensity domain solvers of (b) the MATLAB ``$\setminus$" operator and (c) the PCG solver.}\label{FigTimeComp}
\end{figure*}

\begin{figure*}
\centering
 \setlength{\tabcolsep}{0.25mm}
\begin{tabular}{cccc}
\includegraphics[width=0.253\linewidth]{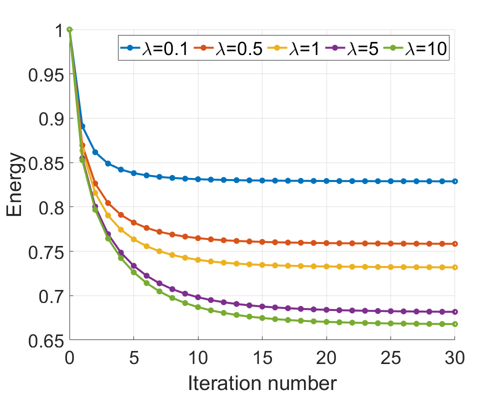}&
\includegraphics[width=0.253\linewidth]{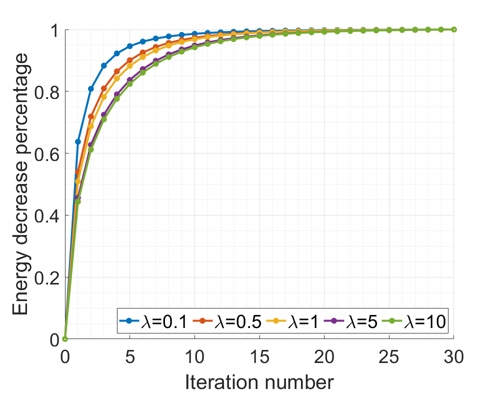}&
\includegraphics[width=0.253\linewidth]{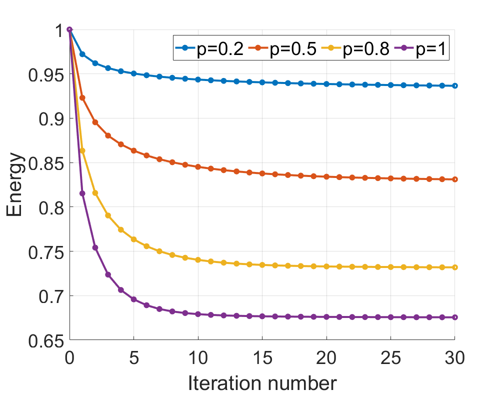}&
\includegraphics[width=0.253\linewidth]{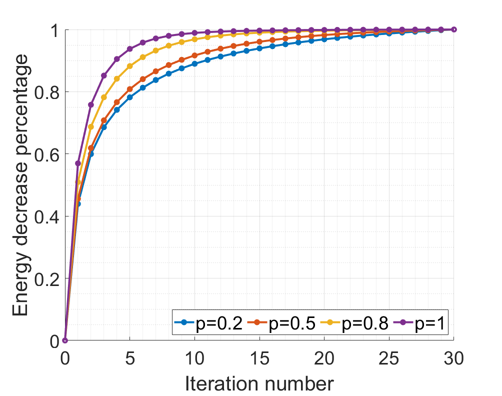}\\
(a) $p=0.8$ & (b) $p=0.8$ & (c) $\lambda=1$ & (d) $\lambda=1$
\end{tabular}
\caption{Energy plots of the objective function in Eq.~(\ref{EqLpSmoothObj}) with respect to the iteration number of Eq.~(\ref{EqILSObj}) and the corresponding plots of the relative energy decrease percentage. (a) and (b) $p=0.8$, $\lambda$ varies in $\{0.1, 0.5, 1, 5, 10\}$. (c) and (d) $\lambda=1$, $p$ varies in $\{0.2, 0.5, 0.8, 1\}$. Energy values of each plot in (a) and (c) are normalized to make the energy value of the input image equal 1. Values of the relative energy decrease percentage in each plot in (b) and (d) are computed as the ratio between the current energy decrease with respect to the final energy decrease. }\label{FigEnergyVSIter}
\end{figure*}

In this way, the objective function in Eq.~(\ref{EqLpSmoothObj}) can be minimized by computing Eq.~(\ref{EqIntermidiateVar}) and solving Eq.~(\ref{EqIterSolutionEnergyAdd}) in an iterative manner. The main advantage of this minimization procedure is that Eq.~(\ref{EqIterSolutionEnergyAdd}) in each iteration is a least squares (LS) problem. As we will demonstrate in Sec.~\ref{SecILS}, the LS problem can be efficiently solved and be easily accelerated through either multi-thread computing or the GPU hardware. In contrast, if Eq.~(\ref{EqLpSmoothObj}) is minimized through the IRLS technique, then a WLS problem needs to be solved in each iteration, which does not enjoy the efficient solver and easy acceleration described in Sec.~\ref{SecILS}.

\subsection{Iterative Least Squares}
\label{SecILS}

When solving $u^{n + 1}$, the values of $\mu^n_{x,s}$ and $\mu^n_{y,s}$ have been fixed as described in Eq.~(\ref{EqIntermidiateVar}). Accordingly, $\psi(\mu^n_{x,s})$ and $\psi(\mu^n_{y,s})$ also become constants. Thus, both $\psi(\mu^n_{x,s})$ and $\psi(\mu^n_{y,s})$ can be omitted in Eq.~(\ref{EqIterSolutionEnergyAdd}) which can be re-written explicitly as:
\begin{equation}\label{EqILSObj}
{
\begin{split}
    &u^{n+1}=\\
    &\underset{u}{\mathop{\arg }\min}\sum\limits_{s}\left((u_s - f_s)^2  + \lambda \sum\limits_{\ast\in\{x,y\}} \frac{1}{2}\left(\sqrt{c}\nabla u_{\ast, s} - \frac{1}{\sqrt{c}}\mu^n_{\ast,s} \right)^2\right).
\end{split}
}
\end{equation}

Because each iteration in Eq.~(\ref{EqILSObj}) is a LS problem, and Eq.~(\ref{EqILSObj}) computes $u$ in an iterative manner, we denote Eq.~(\ref{EqILSObj}) as \emph{iterative least squares (ILS)} in this paper. The iteration number is denoted as $N$, i.e. $n=0, \cdots, N-1$. $u^N$ is the final output of the ILS.

In this subsection, we show that the ILS can be adopted as a fundamental tool for efficient edge-preserving smoothing, which is based on the following two observations: First, the LS problem in each iteration of Eq.~(\ref{EqILSObj}) has an efficient solver. Second, only a few iterations of Eq.~(\ref{EqILSObj}) are able to achieve most of the energy decrease, which is sufficient for edge-preserving smoothing.

We first detail the efficient solver for the LS problem in each iteration of Eq.~(\ref{EqILSObj}). One straightforward way to solve Eq.~(\ref{EqILSObj}) is to directly solve it as a linear system in the intensity domain. However, if we assume the periodic boundary condition for $f$ and $u$, which has been widely used in the literature \cite{wang2008new, xu2010two, xu2011image, xu2013unnatural}, then the solution can also be obtained with the help of FFT and IFFT as follows:
\begin{equation}\label{EqILSSolution}
\small
{
    u^{n + 1}=\mathcal{F}^{-1}\left(\frac{\mathcal{F}(f) + \frac{\lambda}{2}(\overline{\mathcal{F}(\nabla_x})\cdot\mathcal{F}(\mu^n_x) + \overline{\mathcal{F}(\nabla_y)}\cdot\mathcal{F}(\mu^n_y))}{\mathcal{F}(1) + \frac{c}{2}\cdot\lambda(\overline{\mathcal{F}(\nabla_x)}\cdot\mathcal{F}(\nabla_x) + \overline{\mathcal{F}(\nabla_y)}\cdot\mathcal{F}(\nabla_y))}\right),
}
\end{equation}
where $\mathcal{F}(\cdot)$ and $\mathcal{F}^{-1}(\cdot)$ are the FFT and IFFT operators, $\overline{\mathcal{F}(\cdot)}$ denotes the complex conjugate of $\mathcal{F}(\cdot)$, $\mathcal{F}(1)$ is the FFT of the delta function. The plus, multiplication and division are all point-wise operations.

There are three advantages of solving Eq.~(\ref{EqILSObj}) with Eq.~(\ref{EqILSSolution}). First, Eq.~(\ref{EqILSSolution}) is more efficient than directly solving Eq.~(\ref{EqILSObj}) in the intensity domain. This is because when minimizing Eq.~(\ref{EqILSObj}) as a linear system in the intensity domain, an inverse of a very large matrix is required. In contrast, Eq.~(\ref{EqILSSolution}) transforms this matrix inverse into point-wise divisions in the Fourier domain, which is much faster. Second, only $\overline{\mathcal{F}(\nabla_x})\cdot\mathcal{F}(\mu^n_x)$ and $\overline{\mathcal{F}(\nabla_y)}\cdot\mathcal{F}(\mu^n_y)$ need to be computed in each iteration because the other components are constants and can be pre-computed. Third, computing FFT and IFFT in Eq.~(\ref{EqILSSolution}) can be highly parallel. Thus, Eq.~(\ref{EqILSSolution}) can be accelerated through multi-thread computing. In addition, the GPU acceleration of FFT has been a mature technique, e.g., the \texttt{cuFFT} \cite{cuFFT} library which is widely used on the modern GPU hardware, Thus, it is quite easy to accelerate Eq.~(\ref{EqILSSolution}) with the GPU hardware.

It is worthwhile to point out that $\overline{\mathcal{F}(\nabla_x})\cdot\mathcal{F}(\mu^n_x)$ can be computed as $\mathcal{F}\left(\nabla_-\mu^n_x\right)$, where $\nabla_-\mu^n_x$ denotes the inverse first order derivative of $\mu^n_x$ along $x$-axis, i.e., $\nabla_-\mu^n_x$  is computed in an inverse direction of the computation of $\nabla\mu^n_x$. For example, if $[1, -1]$ represents the differential operator of $\nabla\mu^n_x$, then the differential operator of $\nabla_-\mu^n_x$ will be $[-1, 1]$. Similarly, we have $\overline{\mathcal{F}(\nabla_y)}\cdot\mathcal{F}(\mu^n_y)=\mathcal{F}\left(\nabla_-\mu^n_y\right)$. In this way, we further have $\overline{\mathcal{F}(\nabla_x})\cdot\mathcal{F}(\mu^n_x) + \overline{\mathcal{F}(\nabla_y)}\cdot\mathcal{F}(\mu^n_y)=\mathcal{F}\left(\nabla_-\mu^n_x\right) + \mathcal{F}\left(\nabla_-\mu^n_y\right)=\mathcal{F}\left(\nabla_-\mu^n_x + \nabla_-\mu^n_y\right)$, where only one FFT actually needs to be performed. Thus, Eq.~(\ref{EqILSSolution}) will only need one FFT and one IFFT in each iteration once the constant terms have been pre-computed.

To demonstrate the efficiency of Eq.~(\ref{EqILSSolution}), we compare its running time with that of directly solving Eq.~(\ref{EqILSObj}) in the intensity domain. Two intensity domain solvers are compared: the backslash ``$\setminus$'' in MATLAB and the pre-conditioned conjugate gradient (PCG) solver.  All the solvers are implemented in MATLAB. For the PCG solver, we adopt the MATLAB build-in incomplete Cholesky factorization preconditioner which is faster than the one presented in \cite{krishnan2013efficient} as reported by Badri et~al. \cite{badri2013fast, badri2015fast}. For both of the intensity domain solvers, the time of constructing the Laplacian matrix is also included in the measured time. Note that constructing the Laplacian matrix is similar to computing $\overline{\mathcal{F}(\nabla_x)}\cdot\mathcal{F}(\nabla_x) + \overline{\mathcal{F}(\nabla_y)}\cdot\mathcal{F}(\nabla_y)$ in Eq.~(\ref{EqILSSolution}). We test each solver with $\lambda$ varying in $\{0.1, 0.5, 1, 5, 10\}$, the average of all the measured time is plotted in Fig.~\ref{FigTimeComp}(a). As shown in the figure, Eq.~(\ref{EqILSSolution}) is consistently faster than the compared intensity domain solvers by a large margin. To be more explicit, the speedups of Eq.~(\ref{EqILSSolution}) over the intensity domain solvers are illustrated in Fig.~\ref{FigTimeComp}(b) and (c). As shown in the figures, the single-thread CPU implementation\footnote{The implementation of FFT in MATLAB is in parallel and multi-thread computing is adopted under the default configuration, we enable the single-thread computing by using the \texttt{maxNumCompThreads(1)} function (or the \texttt{-singleCompThread} function in the latest version of MATLAB).} of Eq.~(\ref{EqILSSolution}) is generally more than $10\times$ times faster than the PCG solver, it is over $20\times$ faster if computed in multi-thread. When compared with the MATLAB ``$\setminus$'' operator, in most cases, Eq.~(\ref{EqILSSolution}) is more than $30\times$ faster and $60\times$ faster when computed in single-thread and multi-thread, respectively. With the GPU acceleration, Eq.~(\ref{EqILSSolution}) is more than $400\times$ faster than the PCG solver and over $1000\times$ times faster than the MATLAB ``$\setminus$'' operator in most cases.

Our second observation is that only a few iterations of Eq.~(\ref{EqILSObj}) is able to achieve most of the energy decrease. Fig.~\ref{FigEnergyVSIter} shows the energy decrease of Eq.~(\ref{EqLpSmoothObj}) with respect to the iteration number of Eq.~(\ref{EqILSObj}). As shown in the figure, for a large range of $p$ and $\lambda$, Eq.~(\ref{EqILSObj}) can almost converge within 30 iterations ($N=30$). However, 4 iterations ($N=4$) of Eq.~(\ref{EqILSObj}) are able to achieve $74\%\sim90\%$ of the total energy decrease, and 6 iterations ($N=6$) can reach the percentage of $81\%\sim96\%$. In fact, for tasks of tone and detail manipulation, we do not need to iterate Eq.~(\ref{EqILSObj}) for many times, and a few iterations of Eq.~(\ref{EqILSObj}) are able to produce promising results. As we will show in the next subsection, with the iteration number $N$ fixed as $N=4$, the smoothing strength of ILS can be easily controlled by simply varying the value of $\lambda$ in Eq.~(\ref{EqILSObj}).

\begin{figure*}
\centering
\setlength{\tabcolsep}{0.25mm}
\begin{tabular}{cccc}
\includegraphics[width=0.215\linewidth]{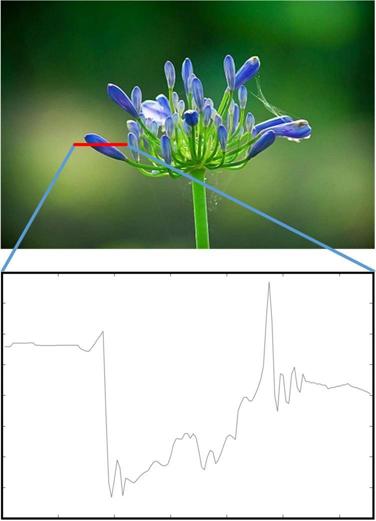} &
\includegraphics[width=0.215\linewidth]{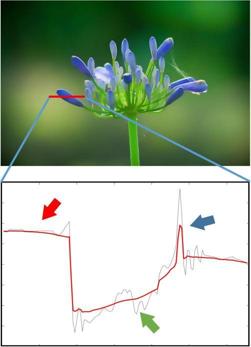} &
\includegraphics[width=0.215\linewidth]{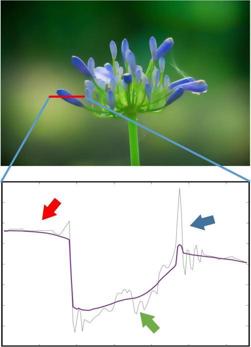} &
\includegraphics[width=0.215\linewidth]{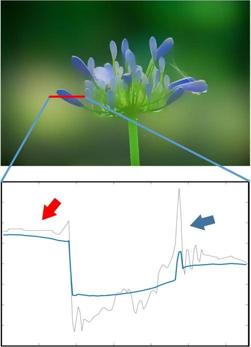}\\
(a) Input & (b) $p=0.8, \lambda=0.5, N=4$ & (c) $p=0.8, \lambda=0.5, N=30$ & (d) $p=0.8, \lambda=3.5, N=4$
\end{tabular}
\caption{Smoothing results of the ILS with different $\lambda$ and $N$. The first row shows the input image and smoothed images. The second row shows 1D plots of the regions labeled with the red lines in the first row. For details of small amplitudes, both larger $N$ and smaller $N$ lead to similar smoothing effects. A larger $N$ can result in larger smoothing on the large-amplitude details than a smaller $N$ does. A larger $\lambda$ can yield larger smoothing on the large-amplitude details and enhance the intensity shift at the same time. Input image courtesy of the flickr user Amanda Slater.}\label{FigIterNumVSLambda}
\end{figure*}

\begin{figure*}
  \centering
  \setlength{\tabcolsep}{0.7mm}
  \begin{tabular}{cccc}
  \includegraphics[width=0.22\linewidth]{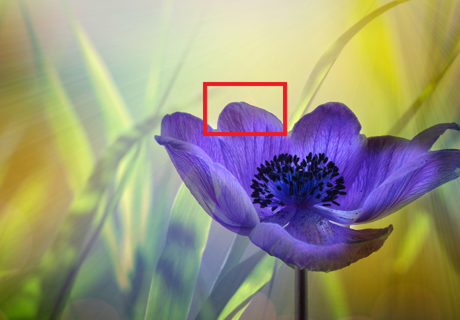}&
  \includegraphics[width=0.22\linewidth]{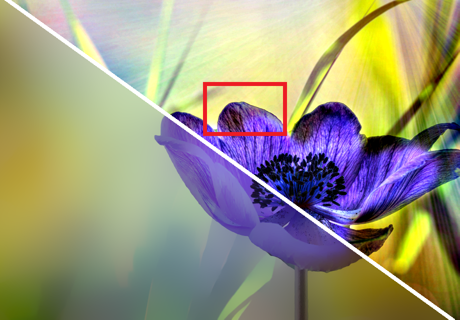}&
  \includegraphics[width=0.204\linewidth]{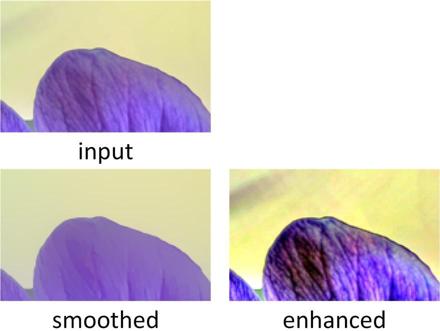}&
  \includegraphics[width=0.2\linewidth]{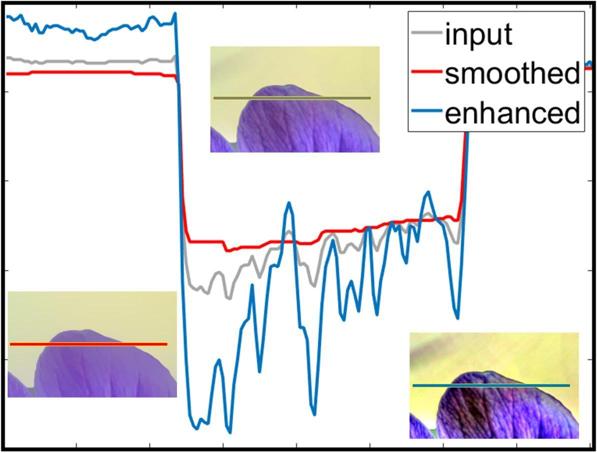}\\

   \includegraphics[width=0.22\linewidth]{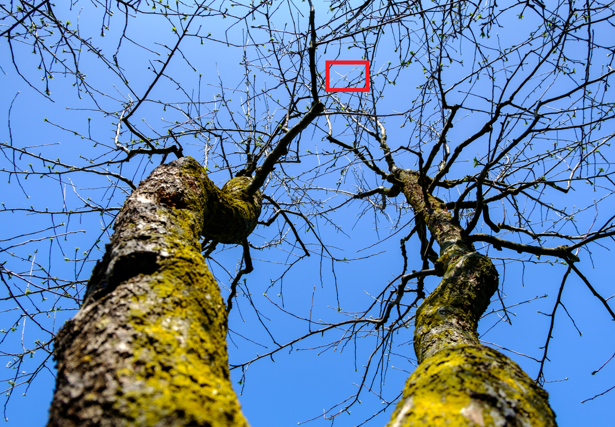}&
  \includegraphics[width=0.22\linewidth]{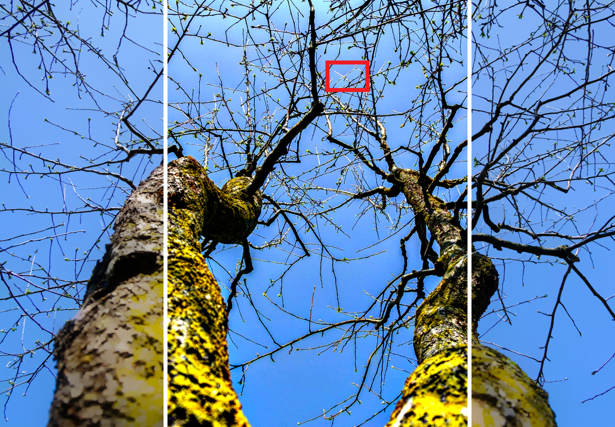}&
  \includegraphics[width=0.2035\linewidth]{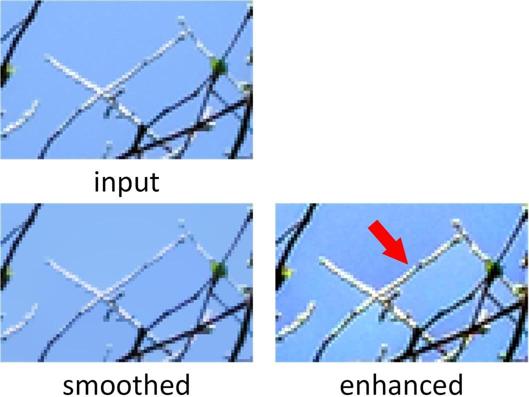}&
  \includegraphics[width=0.2\linewidth]{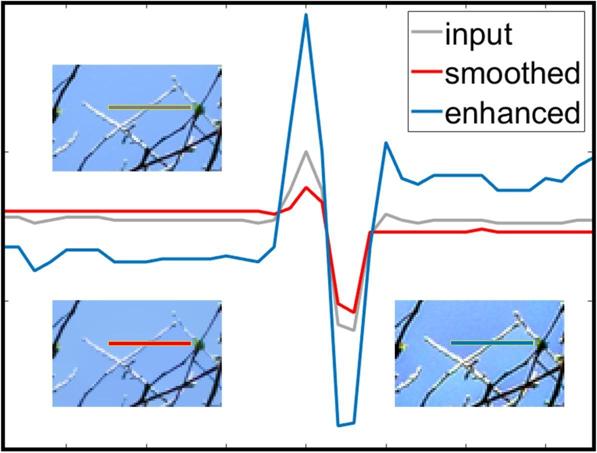}\\
  (a) Input & (b) Smoothed/detail  & (c) Close-up & (d) Intensity 1D plot\\
  & \ \ \ \ \ enhanced image & &
  \end{tabular}
  \caption{The intensity shift caused by the ILS can be used to enhance image contrast as shown in the first row. It can also incorrectly result in compartmentalization artifacts as illustrated in the second row.  The smoothed images are obtained with $p=0.8,\lambda=1$, details are $3\times$ boosted in the detail enhanced images. Input image (top to bottom) courtesy of the flickr user Christophe Brutel and Ivan Vrani{\'c} hvranic.}\label{FigIntensityShift}
\end{figure*}
\begin{figure*}
\centering
\setlength{\tabcolsep}{0.5mm}
\begin{tabular}{ccccc}
\includegraphics[width=0.15\linewidth]{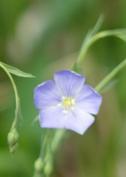}&
\includegraphics[width=0.15\linewidth]{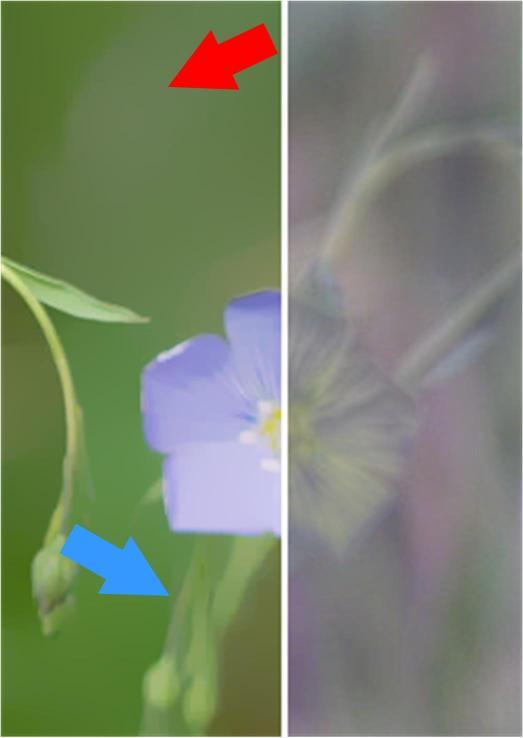}&
\includegraphics[width=0.15\linewidth]{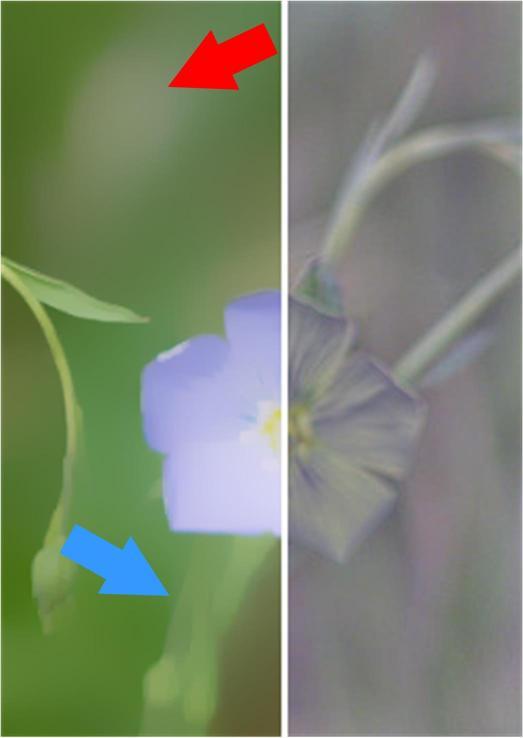}&
\includegraphics[width=0.195\linewidth]{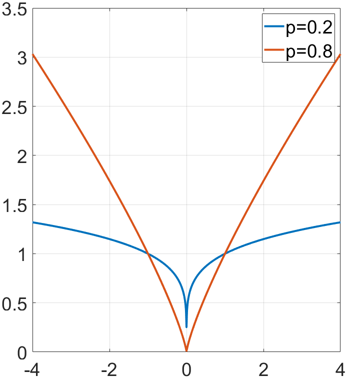}&
\includegraphics[width=0.1915\linewidth]{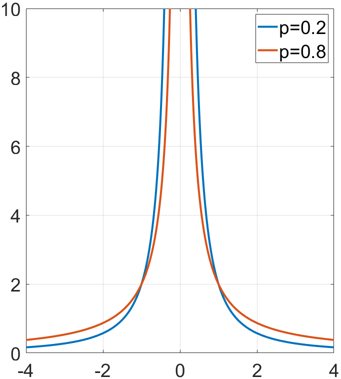}\\
(a) Input & (b) $p=0.2$  & (c) $p=0.8$ & (d) Penalty function & (d) Edge stopping function
\end{tabular}
\caption{Smoothing results of the ILS with different norm power $p$. A smaller $p$ tends to produce results with more piecewise constant regions separated by strong edges. The right parts in (b) and (c) show the residual detail layers correspond to the smoothed images on the left side. All the results are obtained with $\lambda=1$. (d) Plots of the generalized Charbonnier penalty with $p=0.8$ and $p=0.2$, and (e) their corresponding edge stopping functions.}\label{FigPValue}
\end{figure*}
\begin{figure}
\centering
\setlength{\tabcolsep}{0.25mm}
\begin{tabular}{cc}
\includegraphics[width=0.5\linewidth]{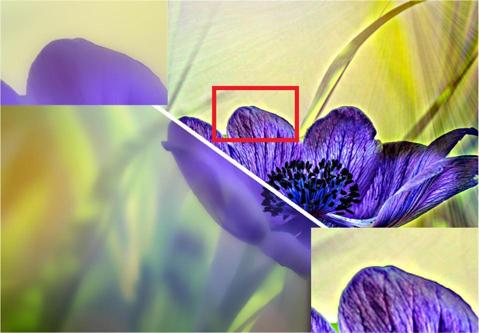}&
\includegraphics[width=0.4582\linewidth]{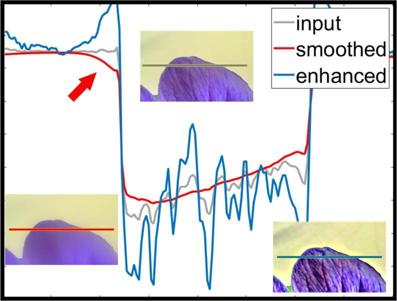}\\
(a) Smoothed/detail & (b) Intensity 1D plot\\
\ \ \ \ \ enhanced image &
\end{tabular}
\caption{Illustration of halo artifacts caused by edge blurring. (a) Smoothed image and the corresponding $3\times$ detail boosted image produced by the adaptive manifold filter \cite{gastal2012adaptive} ($\sigma_s=20,\sigma_r=0.3$). (b) 1D intensity plot of the region labeled with the line. The input image is shown in the first row of Fig.~\ref{FigIntensityShift}(a).}\label{FigLocalMethodLimits}
\end{figure}

\subsection{Parameter Discussion}
\label{SecParameters}

\subsubsection{$\lambda$ and iteration number $N$}

We first compare the difference between the smoothing behavior of our ILS under different iteration numbers of Eq.~(\ref{EqILSObj}). We then show that the smoothing strength of our ILS can be controlled by varying the value of $\lambda$ in Eq.~(\ref{EqILSObj}) with the iteration number fixed as $N=4$. Fig.~\ref{FigIterNumVSLambda}(b) and (c) show two smoothing results obtained with $N=4$ and $N=30$, the other parameters are fixed as $p=0.8,\lambda=0.5$. The results are compared in the following two aspects. First, for details of small amplitudes (labeled with the green arrows), both $N=4$ and $N=30$ can lead to similar smoothing results. Second, for large-amplitude details (labeled with the blue arrows), they are better smoothed by the ILS with $N=30$. This indicates that a larger iteration number can lead to stronger smoothing on large-amplitude details, but it is achieved at the expense of a much higher computational cost. In fact, a similar smoothing effect can also be achieved with a larger $\lambda$  and $N=4$. Fig.~\ref{FigIterNumVSLambda}(d) shows the smoothing result obtained with $\lambda=3.5$ and $N=4$. The large-amplitude details labeled with the blue arrow in Fig.~\ref{FigIterNumVSLambda}(d) achieves a similar smoothing effect as that in Fig.~\ref{FigIterNumVSLambda}(c). Based on the above facts, the smoothing strength of the ILS can be controlled by simply varying the value of $\lambda$ with the iteration number $N$ fixed. If not specified, the iteration number of the ILS in Eq.~(\ref{EqILSObj}) will be fixed as $N=4$ for all the experiments.

\begin{figure*}
\centering
 \setlength{\tabcolsep}{0.25mm}
\begin{tabular}{cccc}
\includegraphics[width=0.253\linewidth]{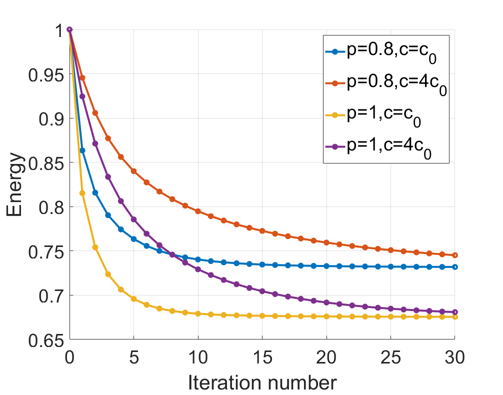}&
\includegraphics[width=0.253\linewidth]{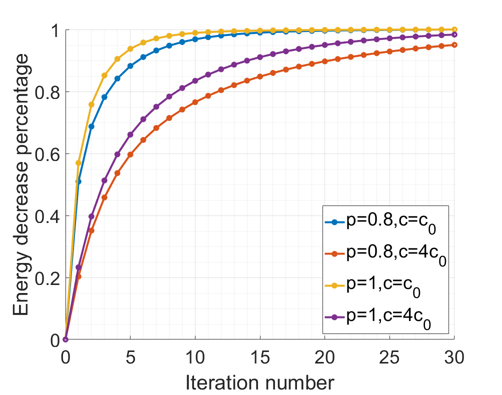}&
\includegraphics[width=0.253\linewidth]{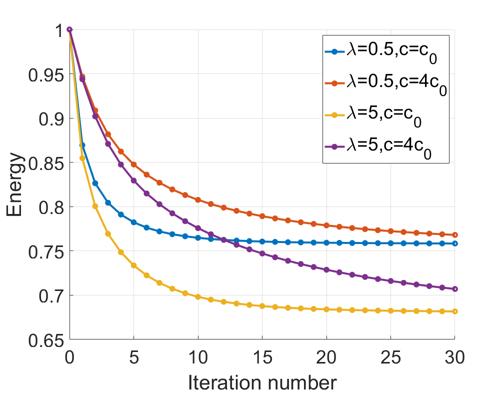}&
\includegraphics[width=0.253\linewidth]{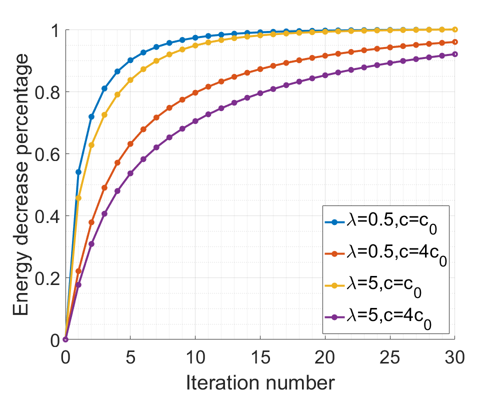}\\
(a) $\lambda=1$ & (b) $\lambda=1$  & (c) $p=0.8$ & (d) $p=0.8$
\end{tabular}
\caption{Convergency speed comparison of Eq.~(\ref{EqILSObj}) under different values of $c$. The plotted values in (a) and (c) are normalized in the same way as that in Fig.~\ref{FigEnergyVSIter}. The relative energy decrease percentage in (b) and (d) are computed as the ratio between the current energy decrease and the final energy decrease of  (b) the same $p$ with $c=c_0$ and (d) the same $\lambda$ with $c=c_0$, respectively. A smaller $c$ leads to a higher convergency speed.}\label{FigCvsEnergyDecrease}
\end{figure*}

\begin{figure*}
\centering
\setlength{\tabcolsep}{0.5mm}
\begin{tabular}{ccc}
\includegraphics[width=0.328\linewidth]{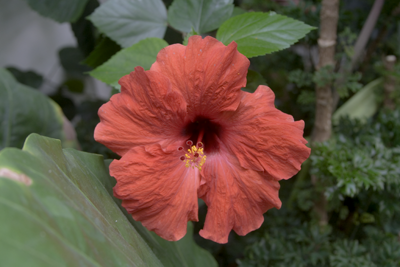} &
\includegraphics[width=0.328\linewidth]{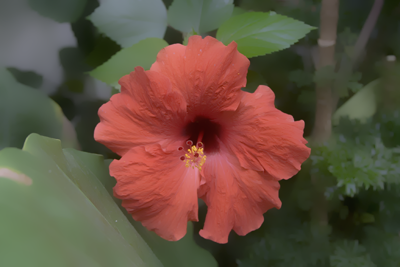} &
\includegraphics[width=0.328\linewidth]{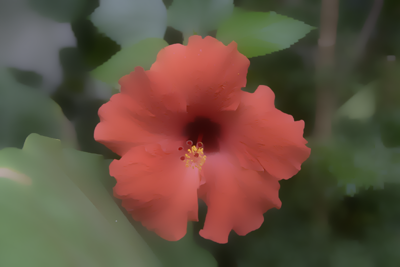} \\
(a) Input & (b) $c=4c_0$ & (c) $c=c_0$
\end{tabular}
\caption{Smoothing results of the ILS with different values of $c$ in Eq.~(\ref{EqILSObj}), other parameters are fixed as $p=0.8,\lambda=1$. A smaller $c$ leads to stronger smoothing on the input image.}
\label{FigCValue}
\end{figure*}

\begin{figure*}
\centering
 \setlength{\tabcolsep}{0.25mm}
\begin{tabular}{ccccc}
\includegraphics[width=0.203\linewidth]{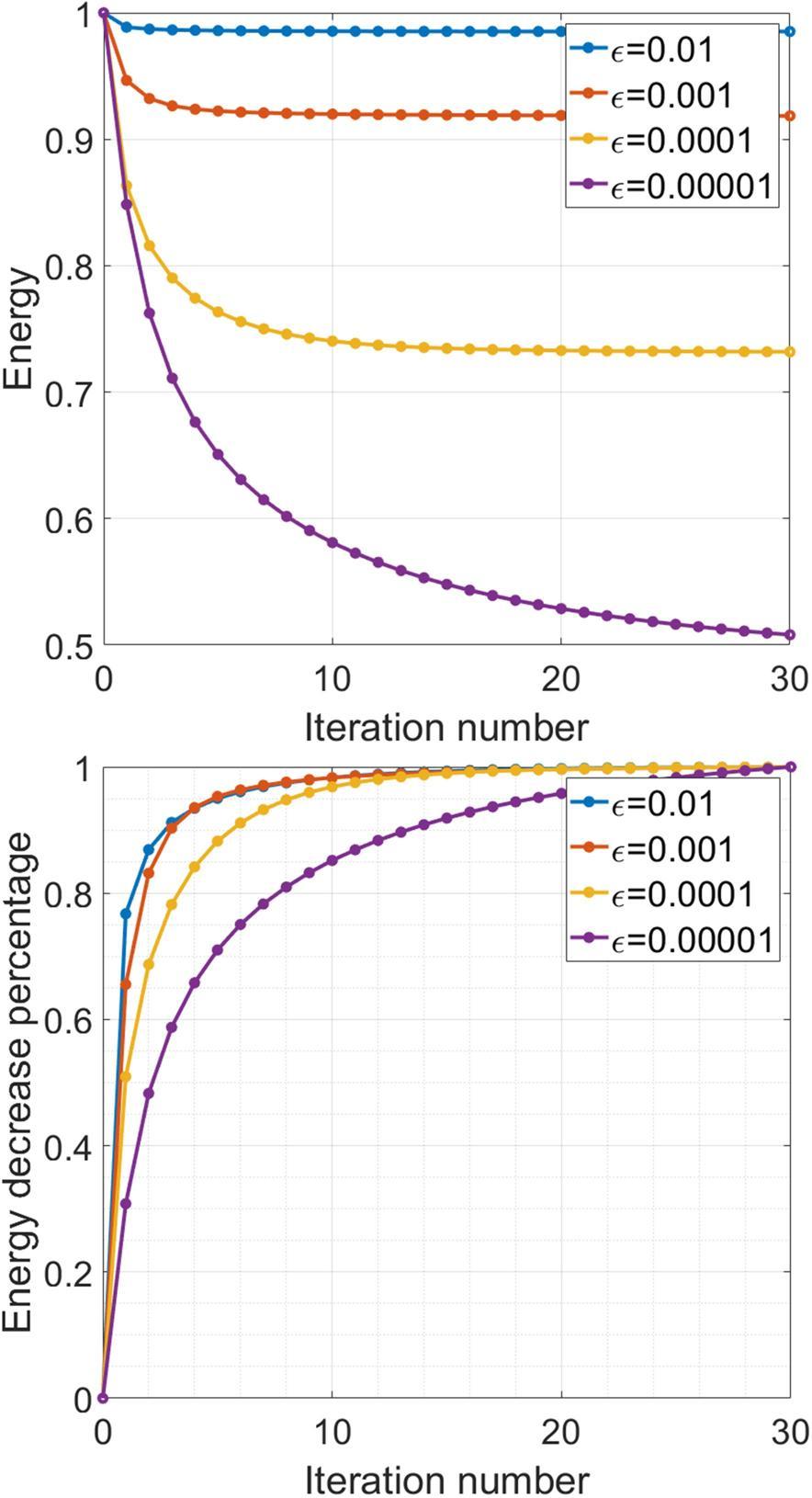}&
\includegraphics[width=0.195\linewidth]{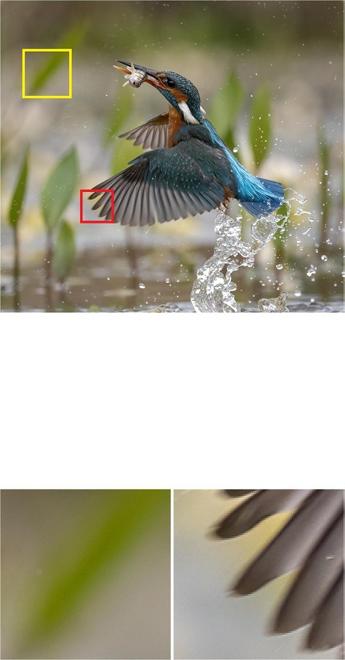}&
\includegraphics[width=0.195\linewidth]{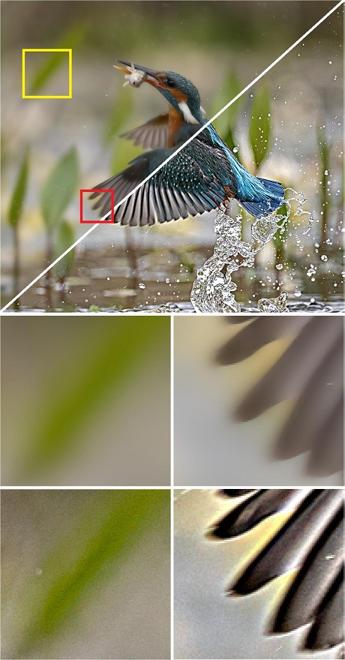}&
\includegraphics[width=0.195\linewidth]{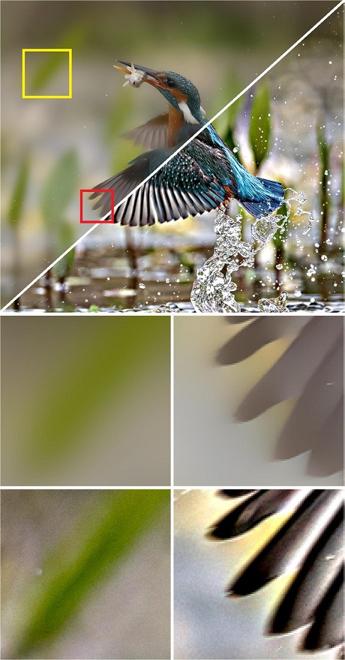}&
\includegraphics[width=0.195\linewidth]{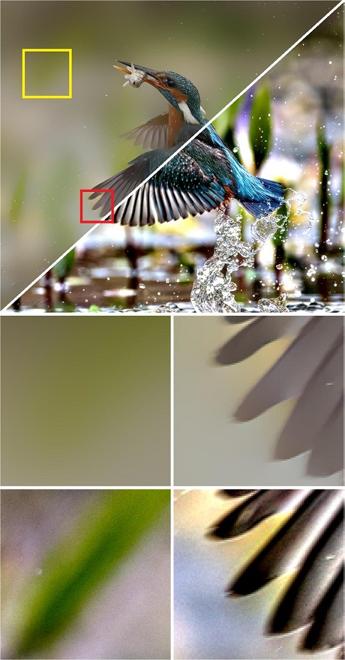}\\
(a) $p=0.8, \lambda=1$  & (b) Input & (c) $\epsilon=0.01, N=1$  & (d) $\epsilon=0.001, N=2$  & (e) $\epsilon=0.0001, N=4$
\end{tabular}
\caption{Comparison of different $\epsilon$ values in Eq.~(\ref{EqSmoothedLpNorm}). (a) Convergency speed comparison of Eq.~(\ref{EqILSObj}) under different values of $\epsilon$ . The plotted values are normalized in the same way as that in Fig.~\ref{FigEnergyVSIter}. The upper left parts of images in (c) $\sim$ (e) are smoothed images and the bottom right parts are detail enhanced images with details $3\times$ boosted. The other parameters are fixed as $p=0.8,\lambda=1$. The second row shows the highlighted regions of the smoothed images, the corresponding detail enhanced patches are illustrated in the third row.  Input image courtesy of the flickr user Andy Harris.}\label{FigEpsValue}
\end{figure*}

A larger $\lambda$ can lead to stronger smoothing on the input image, but it can also enhance the intensity shift effect as labeled with the red arrow in Fig.~\ref{FigIterNumVSLambda}(d). On the one side, this property can be used to enhance image contrast as shown in the first row of Fig.~\ref{FigIntensityShift}, and this is desired in the image enhancement task. On the other side, it can sometimes incorrectly cause artifacts as illustrated in the second row of Fig.~\ref{FigIntensityShift} (labeled with the red arrow). The artifacts are called compartmentalization in the work of Hessel et~al. \shortcite{hessel2018quantitative}. They are also known as the closing effect in the total variation based approaches \cite{meyer2001oscillating, strong1996exact, getreuer2012rudin}. The 1D plot in the second row of Fig.~\ref{FigIntensityShift}(d) shows an example. However, as we will show in Sec.~\ref{SecSmoothQualityComp}, the compartmentalization artifacts of our ILS is much milder than that of the WLS smoothing \cite{farbman2008edge} and the edge-avoiding wavelets \cite{fattal2009edge}.

Note that the intensity shift does not lead to halo artifacts which are caused by edge blurring. To demonstrate the difference between intensity shift and edge blurring, Fig.~\ref{FigLocalMethodLimits} shows a result containing halos produce by the adaptive manifold filter \footnote{Note that Fig.~\ref{FigLocalMethodLimits} is used to illustrate the difference between intensity shift and edge blurring. The halo artifacts in Fig.~\ref{FigLocalMethodLimits} can be avoided by setting $\sigma_r$ to a smaller value, e.g., $\sigma_r=0.2$. However, this will also weaken the smoothing strength.}\cite{gastal2012adaptive}. The 1D intensity plot in Fig.~\ref{FigLocalMethodLimits}(b) (labeled with the red arrow) shows an illustration of edge blurring which leads to the halos. The difference between intensity shift and edge blurring can be clearly observed from the 1D plots in Fig.~\ref{FigIntensityShift}(d) and Fig.~\ref{FigLocalMethodLimits}(b). Fig.~\ref{FigLocalMethodLimits} also demonstrates the fact that edges can be blurred under strong smoothing strength, which is known as the tradeoff between the smoothing abilities and the edge-preserving abilities of weighted average based approaches \cite{farbman2008edge, he2013guided}.

\begin{figure*}
\centering
\setlength{\tabcolsep}{1mm}
\begin{tabular}{ccccc}
\includegraphics[width=0.19\linewidth]{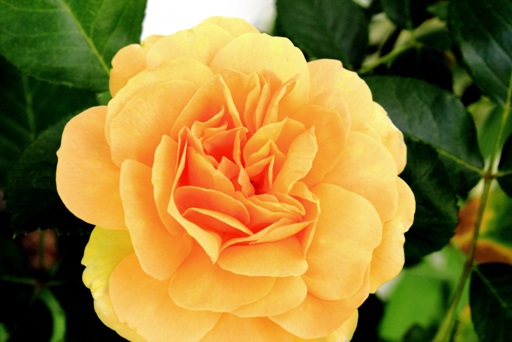}&
\includegraphics[width=0.19\linewidth]{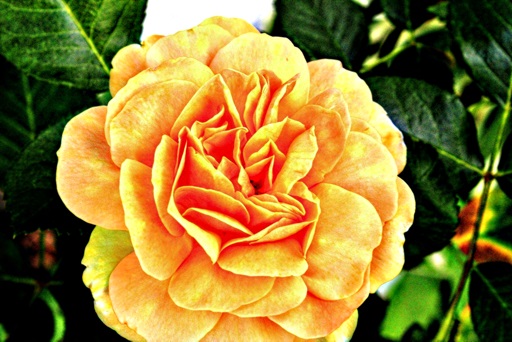}&
\includegraphics[width=0.19\linewidth]{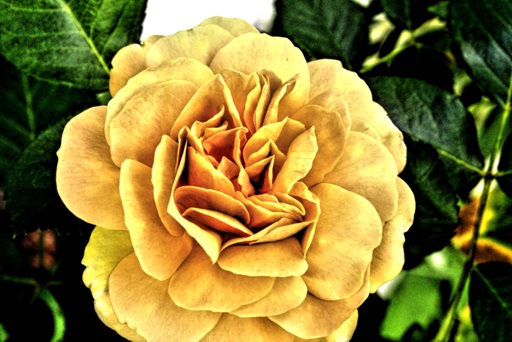}&
\includegraphics[width=0.19\linewidth]{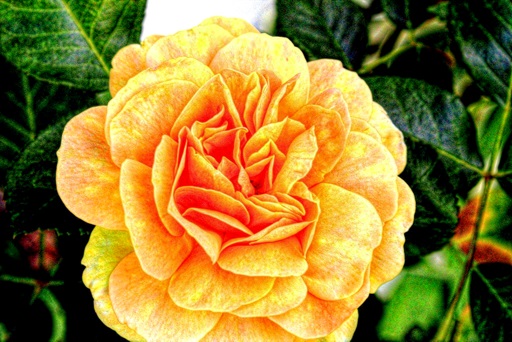}&
\includegraphics[width=0.19\linewidth]{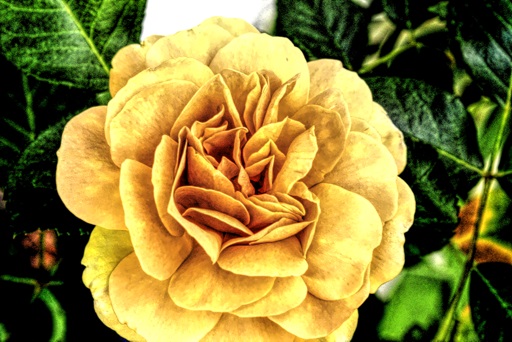}\\
(a) Input  & (b) ILS RGB & (c) ILS luminance only & (d) LLF RGB & (e) LLF luminance only
\end{tabular}
\caption{Comparison of different ways of processing color images. Image detail enhancement results obtained by enhancing each channel of the input image, shown in (b) and (d). Results obtained by only enhancing the luminance channel, shown in (c) and (e). The input image is shown in (a). Results in (b) and (c) are obtained with the proposed ILS ($p=0.8,\lambda=1$), and details are $3\times$ boosted. Results in (d) and (e) are obtained with local Laplacian filter \cite{paris2011local} ($\sigma_r=0.4,\alpha=0.25,\beta=1$). Input image courtesy of the flickr user DJ SINGH.}\label{FigLumVSSeperate}
\end{figure*}

\subsubsection{Power norm $p$}

The parameter $p$ controls the sensitivity to the edges in the input image. Fig.~\ref{FigPValue}(b) and (c) show the smoothing results of our ILS with different values of $p$. The ILS with a smaller $p$ tends to have a larger penalty on smooth regions (labeled with the red arrow) but a smaller penalty on salient edges (labeled with the blue arrow). This can be explained through the concept of edge stopping function \cite{black1998robust, huber2011robust}. The edge stopping function of a penalty function $\phi(x)$ is defined as $\varphi(x)=\frac{\phi'(x)}{2x}$ where $\phi'(x)$ is the derivative of $\phi(x)$ with respect to $x$. A larger value of $\varphi(x)$ refers to a larger penalty on the input $x$. The edge stopping function $\varphi_{p}(x)$ of $\phi_{p}(x)$ in Eq.~(\ref{EqSmoothedLpNorm}) is defined as:
\begin{equation}\label{EqLpEdgeStop}
  \varphi_{p}(x)=\frac{p}{2}(x^2+\epsilon)^{\frac{p}{2} - 1}.
\end{equation}
Eq.~(\ref{EqLpEdgeStop}) indicates that increasing $p$ will result in a larger value of $\varphi_{p}(x)$ for $x$ around zero and a smaller value of $\varphi_{p}(x)$ for $x$ far away from zero. An example of the comparison between $\phi_{p}(x)$ with $p=0.8$ and $p=0.2$ is shown in Fig.~\ref{FigPValue}(d). Their corresponding edge stopping functions are illustrated in Fig.~\ref{FigPValue}(e). Note that the gradients of smooth image regions are quite small (close to zero) while salient edges correspond to large image gradients (far away from zero), thus smaller $p$ will have larger penalties on smooth regions but smaller penalties on salient edges. Our experiments show that $p=0.8\sim1$ is suitable for tasks of tone and detail manipulation, which can produce results with little visible artifacts.

\subsubsection{Constant $c$}
\label{SecCAnalysis}
For the value of constant $c$, it depends on both $p$ and $\epsilon$. Due to the reason that $g(x)=\frac{c}{2}x^2-\phi_{p}(x)$ should be strictly convex, then based on the analysis in Appendix B, the constraint on the value of $c$ is formulated as:
\begin{equation}\label{EqCValue}
    c\geq c_0, \text{where}\ c_0=p\epsilon^{\frac{p}{2}-1}.
\end{equation}

Theoretically, any value of $c$ that meets Eq.~(\ref{EqCValue}) can be used in Eq.~(\ref{EqILSObj}). However, our experiments show that a smaller $c$ leads to higher convergency speed of Eq.~(\ref{EqILSObj}) as illustrated in Fig.~\ref{FigCvsEnergyDecrease}. This means that for a fixed iteration number $N=4$, smaller $c$ will lead to more energy decrease of Eq.~(\ref{EqLpSmoothObj}) and thus stronger smoothing on the input image, as demonstrated in Fig.~\ref{FigCValue}. Thus, $c$ should be set as small as possible. In this paper, we set $c=c_0$ in all the experiments.

\subsubsection{$\epsilon$}

In all the experiments above, the value of $\epsilon$ in Eq.~(\ref{EqSmoothedLpNorm}) is fixed as $\epsilon=0.0001$. This subsection further explores how the smoothing results and the processing speed are affected by the value of $\epsilon$. Fig.~\ref{FigEpsValue}(a) shows plots of the energy decrease of Eq.~(\ref{EqLpSmoothObj}) with respect to the iteration number $N$ of Eq.~(\ref{EqILSObj}). The plots indicate that a larger $\epsilon$ leads to higher convergency speed of Eq.~(\ref{EqILSObj}). For example, when $\epsilon=0.01$, one iteration ($N=1$) of Eq.~(\ref{EqILSObj}) is able to yield almost 80\% of the total energy decrease. In contrast, it needs seven iterations ($N=7$) to achieve a similar energy decrease with $\epsilon=0.00001$. The results are obtained with $p=0.8, \lambda=1$, but other parameter settings of $p$ and $\lambda$ also show similar observations. The above observation further indicates that we can adopt a smaller iteration number $N$ by increasing the value of $\epsilon$ to reduce the computational cost of our ILS. However, a larger $\epsilon$ has the risk of resulting in halo artifacts as shown in the highlighted regions labeled with the red boxes in Fig.~\ref{FigEpsValue}. The comparison between the highlighted regions labeled with the yellow boxes in Fig.~\ref{FigEpsValue} also indicates that larger values of $\epsilon$ result in smaller smoothing strength on the input image. This can be easily explained by considering the value change of $\varphi_{p}(x)$ in Eq.~(\ref{EqLpEdgeStop}) with respect to $\epsilon$: $\varphi_{p}(x)$ decreases as $\epsilon$ increases. Through comprehensive experiments, we find that setting $\epsilon=0.0001$ can seldom lead to visible halos. Thus, to maintain both the smoothing quality and the processing speed, we fix $\epsilon=0.0001$ in all the experiments.

After a detailed analysis of all the parameters in our ILS, we can adopt it to perform edge-preserving image smoothing. The following parameters are fixed throughout all the experiments: $N=4,c=c_0,\epsilon=0.0001$. $\lambda$ and $p$ are used to control the smoothing strength and the edge-preserving property, respectively. Then the smoothing procedure of ILS is an iterative process of the following two steps for $N=4$ times: (I) computing the intermediate variables $\mu^n_\ast$ ($\ast\in\{x,y\}$) using Eq.~(\ref{EqIntermidiateVar}) and (II) solving Eq.~(\ref{EqILSObj}) using Eq.~(\ref{EqILSSolution}). Algorithm~\ref{Alg} summarizes the smoothing procedure of our ILS.

\begin{algorithm}[t]
\caption {Iterative Least Squares for Image Smoothing}\label{Alg}
\begin{algorithmic}[1]
\REQUIRE
Input image $f$, parameter $p, \lambda, N=4$, $c=c_0$, $u^0=f$\\
pre-compute $\mathcal{F}(f)$, $\overline{\mathcal{F}(\nabla_\ast)}\cdot\mathcal{F}(\nabla_\ast)$ where $\ast\in\{x,y\}$

\FOR{$n=0:N-1$}
\STATE With $\nabla u^n_\ast$, solve for $\mu^n_\ast$ in Eq.~(\ref{EqIntermidiateVar})
\STATE With $\mu^n_\ast$, solve for $u^{n+1}$ in Eq.~(\ref{EqILSSolution})
\ENDFOR
\ENSURE
Smoothed image $u^{N}$
\end{algorithmic}
\end{algorithm}

\subsection{Processing Color Images}
\label{SecHandleColor}

\begin{figure*}
\centering
\setlength{\tabcolsep}{0.5mm}
\begin{tabular}{cc}
\includegraphics[width=0.45\linewidth]{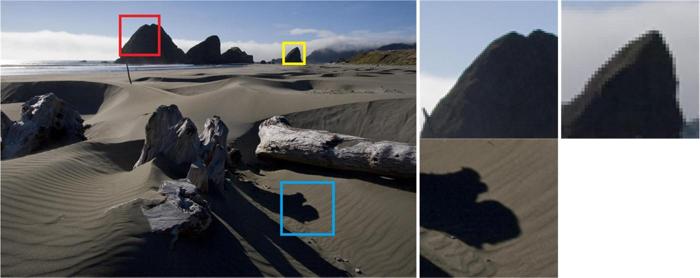} &
\includegraphics[width=0.541\linewidth]{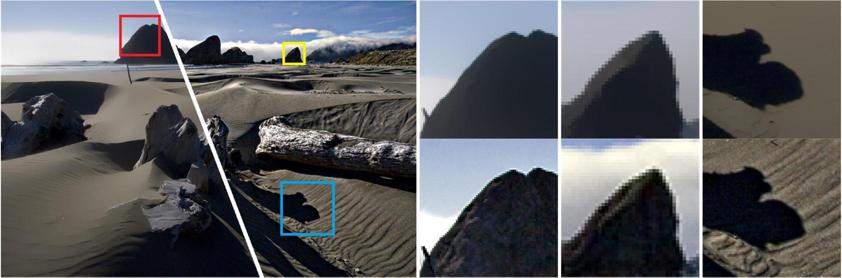} \\
(a) Input & (b) ILS\\
\end{tabular}
\begin{tabular}{cccc}
\includegraphics[width=0.245\linewidth]{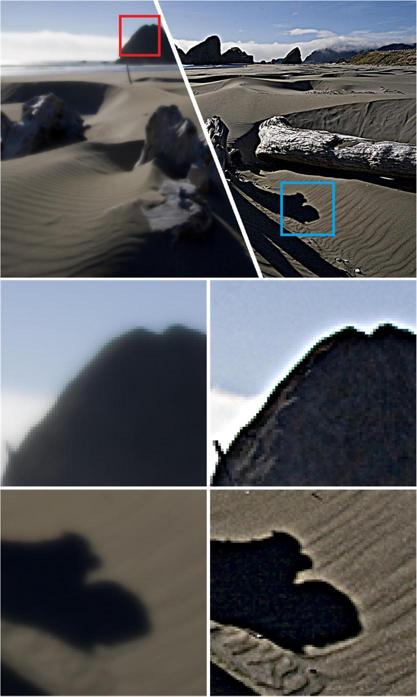} &
\includegraphics[width=0.245\linewidth]{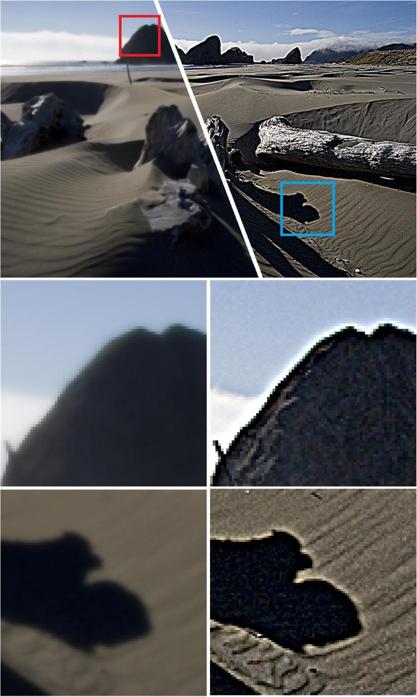} &
\includegraphics[width=0.245\linewidth]{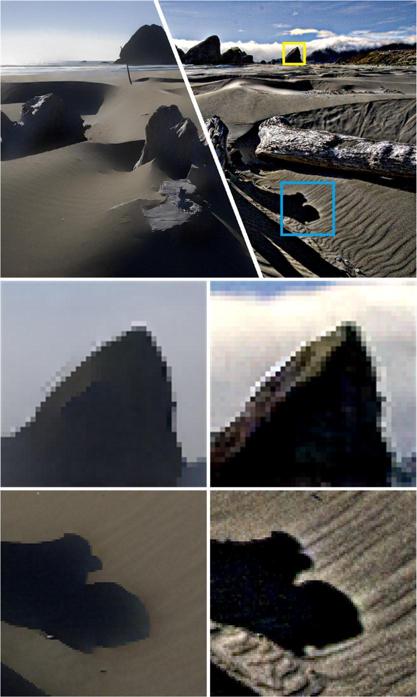} &
\includegraphics[width=0.245\linewidth]{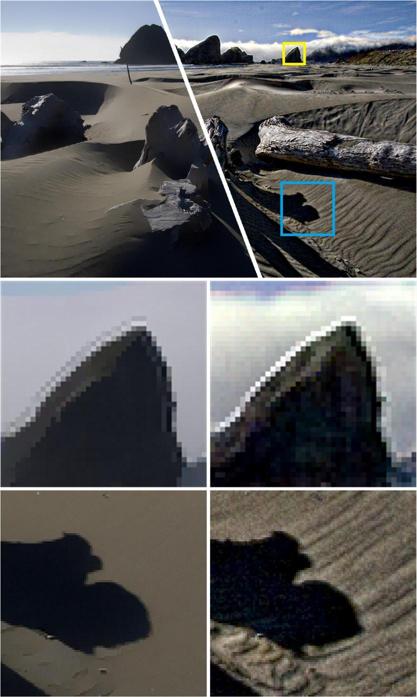} \\
(c) Eq.~(\ref{EqHalfQuadSplitIterate}) with $\phi(x)=|x|$ & (d) Eq.~(\ref{EqHalfQuadSplitIterate}) with $\phi(x)=|x|^{0.8}$ & (e) FOPA with $\gamma=\frac{15}{255}$ & (f)  FOPA with $\gamma=\frac{7.5}{255}$
\end{tabular}
\caption{Comparison of image detail enhancement results obtained with different approaches. (a) Input. (b) ILS with $p=0.8,\lambda=1$. (c) Eq.~(\ref{EqHalfQuadSplitIterate}) with $\phi(x)=|x|, \lambda=0.25, \beta_0=2\lambda$. (d) Eq.~(\ref{EqHalfQuadSplitIterate}) with $\phi(x)=|x|^{0.8}, \lambda=0.1, \beta_0=2\lambda$. The results in (c) and (d) are obtained with 4 iterations of Eq.~(\ref{EqHalfQuadSplitIterate}). The FOPA of Badri et~al. \shortcite{badri2015fast} with (e) $\phi(x)=\phi_W(x), \lambda=50,\gamma=\frac{15}{255}$ and (f) $\phi(x)=\phi_W(x), \lambda=50,\gamma=\frac{7.5}{255}$. The results in (e) and (f) are obtained with the iteration number set as 3 in the FOPA. Smoothed images are shown on the left side of each figure and detail enhanced images are shown on the right side. Details are $3\times$ boosted in the detail enhanced images. Input image courtesy of Norman Koren.}\label{FigILSVSL0L1}
\end{figure*}

The proposed ILS in Eq.~(\ref{EqILSObj}) is formulated to only handle single-channel images. When the input is  a color image, there are two ways to handle it: (I) only processing the luminance channel in the YUV color space, (II) separately processing each channel in the RGB color space. The advantage of the former one is its lower computational cost because only one channel needs to be processed. However, its drawback is that it can cause color shift in some cases as shown in Fig.~\ref{FigLumVSSeperate}(c). Note that this phenomenon is not unique for the proposed ILS, a reference produced by the local Laplacian filter \cite{paris2011local} is shown in Fig.~\ref{FigLumVSSeperate}(e). In contrast, the color shift artifacts can be properly eliminated if each channel in the RGB color space is filtered separately, as shown in Fig.~\ref{FigLumVSSeperate}(b) and (d). The disadvantage is the high computational cost which is almost three times of that of only processing the luminance channel. To maintain the smoothing quality, we process each channel of the input color images in all of our experiments, which is also widely adopted in other approaches \cite{xu2011image, paris2011local, aubry2014fast}.

\subsection{Comparison with the Related Methods}
\label{SecHalfSplitComp}

In this subsection, we first compare the additive half-quadratic minimization technique with another closely related minimization technique called half-quadratic splitting minimization. Then based on the comparison, we further compare our ILS with its highly related approaches which are derived from the half-quadratic splitting minimization.

Unlike the additive half-quadratic minimization, the half-quadratic splitting minimization is one of the most widely used minimization techniques \cite{wang2008new, xu2010two, xu2011image, xu2013unnatural} especially for the $L_p$ norm regularized optimization problems. We first define a general optimization problem as:
\begin{equation}\label{EqGeneralObj}
      \underset{u}{\min}(u - f)^2 +\lambda\phi(\nabla u).
\end{equation}
To simplify the notations in the following analysis, we drop all the subscripts and use $\nabla u$ to denote the gradients of $u$. $\phi(\cdot)$ defines a penalty function. Then the half-quadratic splitting minimization for Eq.~(\ref{EqGeneralObj}) is formulated as:
\begin{equation}\label{EqHalfQuadSplit}
     \underset{u,\mu}{\min}(u - f)^2 +\beta(\nabla u-\mu)^2 + \lambda\phi(\mu).
\end{equation}
Eq.~(\ref{EqHalfQuadSplit}) is iteratively optimized as:
\begin{equation}\label{EqHalfQuadSplitIterate}
\begin{split}
     & Q1: \ \ \ \mu^{n} = \underset{\mu}{\arg\min}\beta_n(\nabla u^n-\mu)^2 + \lambda\phi(\mu),\\
     & Q2: \ \ \ u^{n+1}= \underset{u}{\arg\min}(u - f)^2 +\beta_n(\nabla u-\mu^n)^2.
\end{split}
\end{equation}
The parameter $\beta_n$ also increases as $\beta_n=\kappa\cdot\beta_{n-1}$ with a constant factor $\kappa>1$. The initial value of $\beta$ is usually set as $\beta_0=1$ \cite{wang2008new, krishnan2009fast} or $\beta_0=2\lambda$ \cite{xu2011image}.

The optimization of Eq.~(\ref{EqHalfQuadSplitIterate}) shares a similar procedure with that of the additive half-quadratic minimization: The solution to Q1 is usually a point-wise operator on the image gradients $\nabla u$, which is similar to Eq.(\ref{EqIntermidiateVar}).  The LS optimization problem in Q2 can also be solved in the same way as that of Eq.~(\ref{EqILSObj}).

In the view of mathematical formulation, there are two differences between the half-quadratic splitting minimization and the additive half-quadratic minimization: (I) In Eq.~(\ref{EqAugmentedAdd}), the coefficient $\lambda$ in the second term keeps constant throughout the iterations. In contrast, the coefficient $\beta$ in Eq.~(\ref{EqHalfQuadSplit}) needs to increase in each iteration as detailed in Eq.~(\ref{EqHalfQuadSplitIterate}). (II) $\psi(\cdot)$ in the third term of Eq.~(\ref{EqAugmentedAdd}) is not equal to $\phi_{p}(\cdot)$ in Eq.~(\ref{EqLpSmoothObj}) (see Appendix A for more details), while $\phi(\cdot)$ in the third term of Eq.~(\ref{EqHalfQuadSplit}) is exactly the same as the $\phi(\cdot)$ in Eq.~(\ref{EqGeneralObj}).

The differences mentioned above also indicate the different mathematical mechanisms used in the two different minimization techniques: (I) The additive half-quadratic minimization iteratively minimizes the upper bound Eq.~(\ref{EqAugmentedAdd}) of the original problem in Eq.~(\ref{EqLpSmoothObj}), which can be easily observed from  Eq.~(\ref{EqEnergyInequalityAdd}). (II)  The half-quadratic splitting minimization iteratively minimizes the upper bound Eq.~(\ref{EqHalfQuadSplit}) of a series of other objective functions, which we denote as $E_k (k=0,\cdots,n)$, instead of the original objective function in Eq.~(\ref{EqGeneralObj}). However, after a number of iterations, $E_n$ will be sufficiently close to the one in Eq.~(\ref{EqGeneralObj}).

\begin{figure*}
\centering
\setlength{\tabcolsep}{0.75mm}
\begin{tabular}{ccc}
\includegraphics[width=0.33\linewidth]{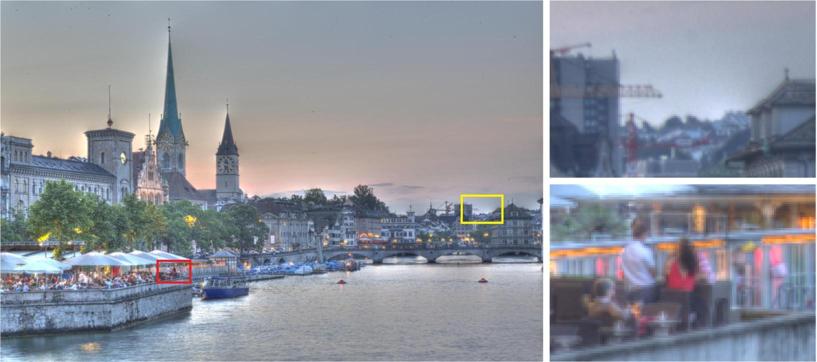} &
\includegraphics[width=0.33\linewidth]{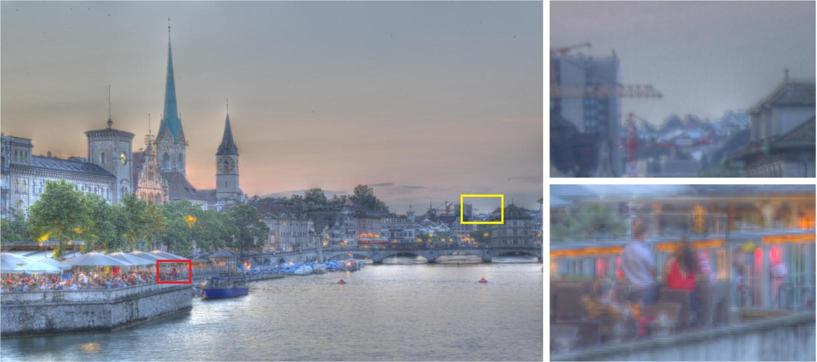} &
\includegraphics[width=0.33\linewidth]{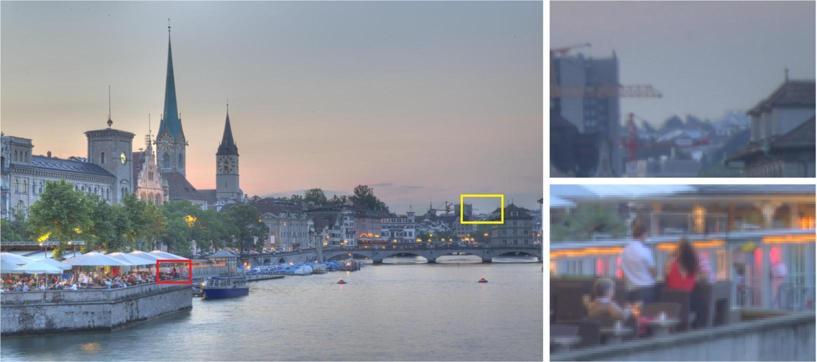} \\
(a) FOPA with $\gamma=0.35$  & (b) FOPA with $\gamma=0.2$ & (c) ILS
\end{tabular}
\caption{HDR tone mapping comparison between the proposed ILS and the FOPA of Badri et~al. \shortcite{badri2015fast}.  Results of FOPA with (a) $\phi(x)=\phi_C(x),\lambda=500, \gamma=0.35$, (b) $\phi(x)=\phi_C(x),\lambda=500, \gamma=0.2$. The iteration number of FOPA is set as 3. Smaller $\gamma$ can effectively alleviate halo artifacts (labeled with the yellow boxes), but it also limits the smoothing ability (labeled with the red boxes). (c) Result of the proposed ILS with $p=1, \lambda=10$. HDR image \textcopyright Mark Fairchild.}
\label{FigILSvsFOPA_HDR}
\end{figure*}

\begin{figure}
\centering
\setlength{\tabcolsep}{0.25mm}
\begin{tabular}{cc}
\includegraphics[width=0.492\linewidth]{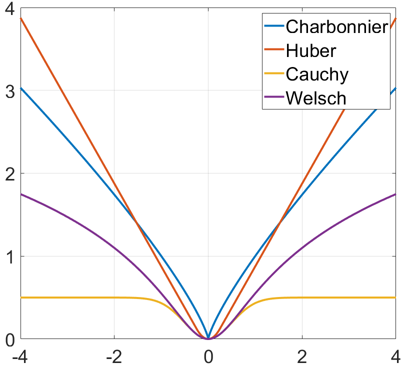} &
\includegraphics[width=0.5\linewidth]{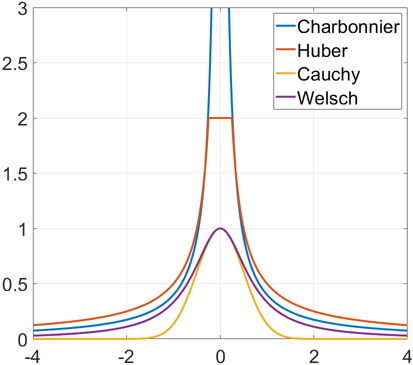}\\
(a) Penalty functions & (b) Edge stopping functions
\end{tabular}
\caption{Comparison of (a) different penalty functions and (b) their corresponding edge stopping functions.}\label{FigNormComp}
\end{figure}

We take an example of anisotropic total variation to further explain the above statement (II). In this case, we have $\phi(x)=|x|$ in Eq.~(\ref{EqGeneralObj}). Note that the generalized Charbonnier penalty with $p=1$ in Eq.~(\ref{EqSmoothedLpNorm}) is a close approximation to this penalty function. First, we re-write the second term and the third term in Eq.~(\ref{EqHalfQuadSplit}) as:
\begin{equation}\label{EqHuberUpperBound}
  g_H(\nabla u, \mu)=\frac{1}{2\alpha}(\nabla u - \mu)^2 + |\mu|,
\end{equation}
where $\frac{1}{2\alpha}=\frac{\beta}{\lambda}$. Based on the analysis in Appendix C, Eq.~(\ref{EqHuberUpperBound}) is the upper bound of the following Huber penalty function:
\begin{equation}\label{EqHuber}
\phi_H(\nabla u)=\left\{
\begin{split}
     &  \frac{1}{2\alpha}(\nabla u)^2, \ \ |\nabla u| \leq \alpha\\
     &  |\nabla u| - \frac{\alpha}{2}, \ \  |\nabla u| > \alpha
\end{split}
\right.,
\end{equation}
which means:
\begin{equation}\label{EqHuberOptimum}
\phi_H(\nabla u)=\underset{\mu}{\min}g_H(\nabla u, \mu).
\end{equation}
The optimum condition of Eq.~(\ref{EqHuberOptimum}) is a soft threshold on the image gradients $\nabla u$, which is also a point-wise operation (detailed in Appendix C) similar to that of Eq.~(\ref{EqIntermidiateVar}).

Clearly, on the one side, Eq.~(\ref{EqHuberOptimum}) indicates that the objective function in Eq.~(\ref{EqHalfQuadSplit}) is an upper bound of Eq.~(\ref{EqGeneralObj}) with $\phi(\nabla u)=\phi_H(\nabla u)$, however, this is different from the original objective function where the penalty function is $\phi(\nabla u)=|\nabla u|$. On the other side, since $\beta$ increases in each iteration as stated in Eq.~(\ref{EqHalfQuadSplitIterate}), then based on $\frac{1}{2\alpha}=\frac{\beta}{\lambda}$, $\alpha$ will decrease in each iteration. It will be quite small after a number of iterations, and $\phi_H(\nabla u)$ in Eq.~(\ref{EqHuber}) will thus be sufficiently close to $|\nabla u|$. These are right what the above statement (II) summarizes.

Due to the mathematical mechanism above, applying a similar strategy (stopping the iteration at $N=4$) to the half-quadratic splitting is not suitable for edge-preserving smoothing, which can result in noticeable halo artifacts. This is because $\phi_H(\cdot)$ in Eq.~(\ref{EqHuber}) can be dominated by the $L_2$ norm part at the beginning of the iteration procedure. For example, when we set $\beta_0=2\lambda$, then based on  $\frac{1}{2\alpha}=\frac{\beta}{\lambda}$, we have $\alpha=0.25$ in Eq.~(\ref{EqHuber}). This means that image gradients smaller than $0.25$  will be penalized with the $L_2$ norm\footnote{All the image intensities are assumed to be normalized into $[0,1]$ in the $L_p$ norm regularized optimization problem.} which is not an edge-preserving penalty function. As a result, visible halo artifacts may occur in the results. Fig.~\ref{FigILSVSL0L1}(c) shows one example obtained with 4 iterations of Eq.~(\ref{EqHalfQuadSplitIterate}). Fig.~\ref{FigILSVSL0L1}(d) further shows an example obtained by adopting the same strategy to Eq.~(\ref{EqGeneralObj}) regularized by the $L_{0.8}$ norm penalty $\phi(x)=|x|^{0.8}$ (Eq.~(\ref{EqSmoothedLpNorm}) with $p=0.8$ is a close approximation to this penalty function). There are also significant halos in the result. In contrast, there are seldom visible halos in the results produced by our ILS, as shown in Fig.~\ref{FigILSVSL0L1}(b).

The method proposed by Badri et~al. \shortcite{badri2013fast, badri2015fast} is also highly related to the proposed ILS. Their approach is based on the half-quadratic splitting minimization technique, but they adopt the Welsch penalty function and the Cauchy penalty function instead of the $L_p$ norm penalty function. These two penalty functions are defined as:
\begin{equation}\label{EqWelschCauchy}
\begin{split}
     & \text{Welsch:} \  \phi_W(x)=2\gamma^2\left(1-\exp\left(\frac{x^2}{2\gamma^2}\right)\right)\\
     & \text{Cauchy:} \  \phi_C(x)=2\gamma^2\log\left(1+\frac{x^2}{2\gamma^2}\right)
\end{split}
\end{equation}

Unlike $L_{0.8}$ norm and $L_1$ norm penalty functions where closed-form solutions to Q1 exist \cite{wang2008new, krishnan2009fast}, there are no close-form solutions when $\phi(x)=\phi_W(x)$ and $\phi(x)=\phi_C(x)$. Badri et~al. \shortcite{badri2013fast, badri2015fast} thus approximate $\phi_W(x)$ and $\phi_C(x)$ with their first order Taylor expansions, which leads to a closed-form solution to Q1. They denote it as the first order proximal approximation to  Q1. They further fix $\beta=1$ but the iteration number is not fixed \footnote{In their paper, some of their results are produced with 1 iteration while some other results are obtained with 3 iterations. In this paper, we fix the iteration number of their method to 3 due to the reason that a larger iteration number can lead to better results. The warm start proposed by Badri et~al. \shortcite{badri2015fast} is adopted to initialize the result.}. In this case, a theoretical analysis on which upper bound the half-quadratic splitting exactly minimizes is hard to obtain. However, we need to point out that the Welsch and Cauchy penalty functions themselves are not suitable for tasks of tone and detail manipulation. Fig.~\ref{FigNormComp} shows a comparison of different penalty functions. The Welsch and Cauchy penalty functions are similar to $L_2$ norm when the input is small, which has the risk of resulting in halos. To simplify the notation in the following sections, we abbreviate the method proposed by  Badri et~al. \shortcite{badri2013fast, badri2013fast} as FOPA in short of the first order proximal approximation used in their method. We show examples produce by FOPA in Fig.~\ref{FigILSVSL0L1}(e) and Fig.~\ref{FigILSvsFOPA_HDR}(a). Please also refer to their paper \cite{badri2015fast} where visible halos exist in their results in Fig.~13 (around the book edges) and Fig.~14 (around the picture frames and the light fixture). The halo artifacts can be alleviated by setting $\gamma$ in Eq.~(\ref{EqWelschCauchy}) to a smaller value. However, this will also limit the smoothing ability, please refer to the comparison between the highlighted regions in Fig.~\ref{FigILSVSL0L1}(e) $\&$ (f) and Fig.~\ref{FigILSvsFOPA_HDR}(a) $\&$ (b). For the Welsch penalty function, reducing $\gamma$ can also lead to visible gradient reversals as shown in the highlighted regions (labeled with the yellow boxes) in Fig.~\ref{FigILSVSL0L1}(e) and (f). This is because the Welsch penalty function seldom penalizes salient edges and can sharpen them instead. This can also be observed from its edge stopping function shown in Fig.~\ref{FigNormComp}(b) where its value rapidly reduces to the one close to zero for the large input value. Its decreasing speed can be even faster for a smaller $\gamma$.

\section{Experimental Results and Comparison with the State-of-the-art Approaches}
\label{SecComp}

We apply the proposed ILS to tasks of tone and detail manipulations and compare it against many state-of-the-art approaches: fast bilateral filter (fast BLF) \cite{paris2006fast}, adaptive manifold filter (AMF) \cite{gastal2012adaptive}, domain transform filter (DTF) \cite{gastal2011domain}, guided filter (GF) \cite{he2013guided}, local Laplacian filter (LLF) \cite{paris2011local}, fast local Laplacian filter (fast LLF) \cite{aubry2014fast}, edge-avoiding wavelets (EAW) \cite{fattal2009edge}, fast global smoother (FGS) \cite{min2014fast}, semi-global weighted least squares (SG-WLS) \cite{liu2017semi}, fast bilateral solver (FBS) \cite{barron2016fast}, weighted least squares filter (WLS) \cite{farbman2008edge}, gradient $L_0$ norm smoothing ($L_0$ norm) \cite{xu2011image}, the FOPA of Badri et~al. \shortcite{badri2013fast, badri2013fast}, and the deep learning based approaches proposed by Xu et~al. \shortcite{xu2011image}, Liu et~al. \shortcite{liu2016learning} and Chen et~al. \shortcite{chen2017fast}. We use DTF-NC, DTF-IC and DTF-RF to denote the normalized convolution (NC), interpolated convolution (IC) and recursive filtering (RF) versions of DTF, respectively. The comparison is performed in two aspects: processing speed and smoothing quality. In terms of smoothing quality, we omit the comparison with the deep learning based approaches because the filters approximated by them are included in the above mentioned ones.

The implementation details of all the methods are as follows: We adopt the C++ implementation of Paris et~al. \shortcite{paris2006fast} \footnote{http://people.csail.mit.edu/sparis/bf/} for fast BLF. AMF, DTF, GF and FBS are implemented with the build-in C++ functions in OpenCV \footnote{AMF: \texttt{amFilter()}, DTF: \texttt{dtFilter()}, GF: \texttt{guidedFilter()}, FBS: \texttt{fastBilateralSolverFilter()}}. LLF is implemented with the author-provided MATLAB code \footnote{http://people.csail.mit.edu/sparis/publi/2011/siggraph/}. The fast LLF is implemented with the MATLAB build-in $\texttt{MEX}$  function \footnote{MATLAB build-in function \texttt{locallapfilt()}}. EAW, FGS and SG-WLS are implemented with the author-provided MATLAB \texttt{MEX} code \footnote{\url{http://www.cs.huji.ac.il/~raananf/projects/eaw/},  \url{https://sites.google.com/site/globalsmoothing/}}. WLS is implemented with the author-provided MATLAB code \footnote{https://www.cs.huji.ac.il/~danix/epd/wlsFilter.m}, but we adopt the MATLAB build-in PCG solver and incomplete Cholesky factorization preconditioner \footnote{MATLAB build-in functions \texttt{pcg()} and \texttt{ichol()}.} to solve the inverse problem, which is much faster than the MATLAB ``$\setminus$'' operator in the original implementation. $L_0$ norm, FOPA and our ILS are implemented by us in MATLAB (and \texttt{gpuArray} if running on a GPU). These three methods share a similar two-step smoothing procedure and only slightly differ in the point-wise operation on image gradients. Deep learning based approaches proposed by Xu et~al. \shortcite{xu2011image}, Liu et~al. \shortcite{liu2016learning} and Chen et~al. \shortcite{chen2017fast} are implemented with the author-provided code \footnote{\url{https://github.com/jimmy-ren/vcnn_double-bladed/tree/master/applications/deep_edge_aware_filters}, \url{https://github.com/Liusifei/caffe-lowlevel}, \url{https://github.com/CQFIO/FastImageProcessing}}.

\subsection{Speed Comparison}

\begin{table*}
\centering
\caption{Running time (in seconds) of different methods for different image sizes. The value on the left in each cell is the running time for gray images and the right one is for color images.}\label{TabTimeComp}
\resizebox{1\linewidth}{!}
{
\begin{tabular}{l|cccccc}
  \Xhline{1.5pt}
                 & QVGA($320\times240$ ) & VGA($640\times480$)  & 720p($1280\times720$) & 1080p($1920\times1080$) & 2k($2048 \times 1080$) & 4k($3840 \times 2160$)\\

  \hline
  Xu et~al. \shortcite{xu2015deep} (GPU, MATLAB gpuArray) & $\ast$ | 0.33 & $\ast$ | 1.32 & $\ast$ | 3.88 & $\ast$ | 8.81 & $\ast$ | 9.83 & $\ast$ | 32.19 \\

  Liu et~al. \shortcite{liu2016learning} (GPU, caffe \cite{jia2014caffe}) & $\ast$ | 0.036 & $\ast$ | 0.43 & $\ast$ | 1.06 & $\ast$ | 1.96 & $\ast$ | 2.51 & $\ast$ | 5.26 \\

  Chen et~al. \shortcite{chen2017fast} (GPU, tenserflow \cite{tensorflow2015-whitepaper}) & $\ast$ | {0.016} & $\ast$ | {0.19} & $\ast$ | {0.47} & $\ast$ | {0.87} & $\ast$ | {1.12} & $\ast$ | {2.19}\\

  \hline
  AMF \cite{gastal2012adaptive} (CPU, single-thread, C++) & 0.011 | 0.028 & 0.043 | 0.11 & 0.12 | 0.29 & 0.28 | 0.68 & 0.30 | 0.71 & 1.25 | 3.16 \\

  DTF-NC \cite{gastal2011domain} (CPU, single-thread, C++)  & 0.0034 | {0.0044} & 0.012 | {0.018} & 0.038 | {0.051} & 0.090 | {0.14} & 0.11 | {0.17} & 0.38 | {0.65} \\

  DTF-IC \cite{gastal2011domain} (CPU, single-thread, C++)  & 0.038 | 0.0059 & 0.015 | 0.028 & 0.051 | 0.073 & 0.11 | 0.18 & 0.12 | 0.22 & 0.45 | 0.77 \\

  DTF-RF \cite{gastal2011domain} (CPU, single-thread, C++)  & 0.084 | 0.015 & 0.028 | 0.054 & 0.079 | 0.17 & 0.18 | 0.42 & 0.19 | 0.46 & 0.86 | 2.06 \\

  Fast BLF \cite{paris2006fast}  (CPU, single-thread, C++)        & 0.0047 | 0.014 & 0.019 | 0.054 & 0.059 | 0.17 & 0.13 | 0.38 & 0.14 | 0.41 & 0.47 | 1.42 \\

  GF \cite{he2013guided}  (CPU, single-thread, C++)          & {0.0028} | 0.013 & {0.0079} | 0.058 & {0.026} | 0.16 & {0.064} | 0.35 & {0.066} | 0.37 & {0.24} | 1.43 \\

  \hline
  Fast LLF \cite{aubry2014fast} \\(automatic intensity levels, CPU, single-thread, MATLAB MEX)   & 0.078 | 0.22 & 0.34 | 0.94 & 1.14 | 3.36 & 2.53 | 7.61 & 4.88 | 15.05 & 11.83 | 36.08 \\

  Fast LLF \cite{aubry2014fast} \\(256 intensity levels, CPU, single-thread, MATLAB MEX)  & 0.37 | 1.09 & 1.69 | 5.24 & 5.58 | 17.95 & 13.58 |  40.74 & 25.79 | 79.86 & 62.93 | 189.06 \\

  EAW \cite{fattal2009edge} (CPU, single-thread, MATLAB MEX) & 0.0048 | 0.014 & 0.022 | 0.066 & 0.071 | 0.21 & 0.17 | 0.51 & 0.18 | 0.54 & 0.70 | 2.12 \\

  \hline
  FGS \cite{min2014fast} (CPU, single-thread, MATLAB MEX)          & {0.0089} | {0.019} & {0.026} | {0.059} & {0.091} | {0.23} & {0.21} | {0.34} & {0.24} | {0.38} & {0.71} | {1.59}\\

  SG-WLS \cite{liu2017semi}  (CPU, single-thread, MATLAB MEX)      & 0.016 | 0.055 & 0.061 | 0.21 & 0.19 | 0.63 & 0.42 | 1.31 & 0.47 | 1.51 & 1.76 | 6.05 \\

  FBS \cite{barron2016fast}  (CPU, single-thread, C++)      & 0.085 | 0.25 & 0.13 | 0.37 & 0.38 | 1.13 & 0.96 | 2.87 & 1.04 | 3.11 & 4.47 | 13.37\\

  \hline
  WLS \cite{farbman2008edge}   (CPU, single-thread, MATLAB)    & 0.19 | 0.51 & 0.74 | 2.03 & 2.36 | 6.48 & 5.64 | 15.51 & 6.07 | 16.68 & 24.81 | 68.21 \\

  $L_0$ norm \cite{xu2011image}  (CPU, single-thread, MATLAB) & 0.12 | 0.33 & 0.47 | 1.37 & 1.53 | 4.40 & 3.59 | 10.75 & 4.09 | 11.72 & 17.53 | 51.65\\

  $L_0$ norm \cite{xu2011image}  (CPU, multi-thread, MATLAB)  & 0.069 | 0.18 & 0.24 | 0.70 & 0.75 | 2.21 & 1.82 | 5.46 & 2.09 | 6.09 & 8.87 | 26.59\\

  $L_0$ norm \cite{xu2011image}  (GPU, MATLAB gpuArray) & 0.036 | 0.043 & 0.042 | 0.067 & 0.074 | 0.15 & 0.128 | 0.31 & 0.133 | 0.33 & 0.51 | 1.41\\

  FOPA \cite{badri2015fast} (CPU, single-thread, 2 iterations, MATLAB)  & 0.019 | 0.051 & 0.73 | 0.20 & 0.22  | 0.62 & 0.41  | 1.21 & 0.52 | 1.49 & 2.21 | 6.53\\

  FOPA \cite{badri2015fast} (CPU, multi-thread, 2 iterations, MATLAB)  & 0.011 | 0.027 & 0.036 | 0.10 & 0.11 | 0.31 & 0.21  | 0.61 & 0.27 | 0.76 & 1.12 | 3.36\\

  FOPA \cite{badri2015fast} (GPU, 2 iterations, MATLAB gpuArray)  & 0.0056 | 0.0074 & 0.0064  | 0.013 & 0.010 | 0.020 & 0.016 | 0.039 & 0.017 | 0.040 & 0.056 | 0.16\\

   Proposed ILS (CPU, single-thread, $N=4$, MATLAB)  & 0.028 | 0.076 & 0.11  | 0.30 & 0.32  | 0.92 & 0.61  | 1.81 & 0.78 | 2.23 & 3.32 | 9.79\\

  Proposed ILS (CPU, multi-thread, $N=4$, MATLAB)  & 0.016 | 0.041 & 0.054 | 0.15 & 0.16 | 0.46 & 0.31  | 0.92 & 0.40 | 1.13 & 1.68 | 5.04\\

  Proposed ILS (GPU, $N=4$, MATLAB gpuArray)  & 0.0073 | 0.0096 & 0.0083  | 0.013 & 0.013 | 0.026 & 0.021 | 0.050 & 0.022 | 0.052 & 0.073 | 0.20\\

  \Xhline{1.5pt}
\end{tabular}
}
\end{table*}

The running time of some methods can be highly related to the parameter setting. To have a fair speed comparison, for each image resolution, we test the running time of these methods under different parameter settings, and the average time is used as the final running time for the current image resolution. To make the following statements clear, the original parameter notations used in their papers are adopted for each method. For AMF and fast BLF, smaller $\sigma_s$ and $\sigma_r$ can lead to longer execution time. We set $\sigma_s\in\{2, 4, 8, 16, 32\}$ , $\sigma_r\in\{0.01,0.05,0.1,0.5\}$ and use their different combinations for time measurement. The running time of FBS is sensitive to $\sigma_{s}$, and setting it to a smaller value can greatly slow down the running speed of FBS. We set $\sigma_s\in\{2, 4, 8, 16, 32\}$ with other parameters kept as default values. A larger $\lambda$ in WLS makes it more time-consuming, we set $\lambda\in\{0.1,0.5,1,5,10\}$ for time measurement. The speed of fast LLF can be affected by the number of intensity levels used to sample the intensity range. A smaller intensity level number can speed up the execution but result in lower smoothing quality. The interface of the  MATLAB build-in function offers two kinds of choices for the intensity levels \footnote{The ``NumIntensityLevels'' parameter option in \texttt{locallapfilt()}.}: predefined intensity levels and automatically computed intensity levels. We test the speed of fast LLF with predefined 256 intensity levels and automatically computed intensity levels, and report their running time separately as shown in Table~\ref{TabTimeComp}. Except for the $L_0$ norm smoothing, FOPA and the proposed ILS, all the other compared methods are tested in single-thread computing because the adopted implementations do not support multi-thread processing.

Table~\ref{TabTimeComp} shows the running time of different approaches for processing images of different sizes. Six widely used image resolutions are tested: QVGA ($320\times240$), VGA ($640\times480$), 720p ($1280\times720$), 1080p ($1920\times1080$), 2k ($2048\times1080$) and 4k ($3840\times2160$). Both color images and gray images are used for the  speed measurement. We adopt 10 different images for each image size when measuring the running time, and the average time is used as the final running time for each parameter setting. For approaches formulated to only handle single-channel images, they are applied to each channel of the input when processing color images. All the compared approaches are evaluated on an i5-7600 CPU with 8GB memory and a GeForce GTX1080 GPU.  For a clear comparison, the compared methods are divided into five groups in Table~\ref{TabTimeComp}: deep learning based approaches in the first group, weighted average based methods in the second group, fast LLF and EAW of pyramid based methods in the third group, approaches that accelerate global optimization based methods in the fourth group and global optimization based methods in the fifth group. We do not include the running time of LLF in Table~\ref{TabTimeComp} because there is no C/C++ implementation of LLF, its MATLAB implementation is extremely slow, which is not fair for comparison. However, we have included the running time of fast LLF which should be faster than the original LLF.

Among all the global optimization based approaches, the proposed ILS is slightly slower than FOPA. Note that due to the reason that Badri et~al. \shortcite{badri2015fast} do not fix the iteration number in their FOPA, we fix the iteration number to 2 when measuring the running time. The processing time of their proposed warm start is also included in the measured time. The iteration number is chosen as the medium number between the minimum iteration number (1 iteration) and the maximum iteration number (3 iterations) used in their paper. However, all of their results in this paper are produced with the iteration number fixed to 3 for better results. The proposed ILS is generally around $6\times$ faster than $L_0$ norm smoothing. When compared with WLS, it is $6\times\sim8\times$ faster when running in single-thread, and $13\times\sim17\times$ times faster when running in multi-thread.

All the weighted average based methods in the second group are faster than our ILS. However, the ILS also shows a comparable speed with some approaches. For example, the AMF is generally $3\times$ faster than the single-thread ILS and less than $2\times$ faster than the multi-thread ILS. The DTF-RF and fast BLF are generally around $6\times$ faster than the single-thread ILS and around $3\times$ faster than the multi-thread ILS.

The EAW in the third group and the FGS in the fourth group share a comparable execution speed. They are generally $3\times\sim5\times$ faster than the single-thread ILS and less than $3\times$ faster than the multi-thread ILS. Our ILS can run at a similar speed as the SG-WLS does. It is faster than FBS and fast LLF.

\begin{figure*}
  \centering
  \setlength{\tabcolsep}{0.25mm}
  \begin{tabular}{cccc}
  \includegraphics[width=0.247\linewidth]{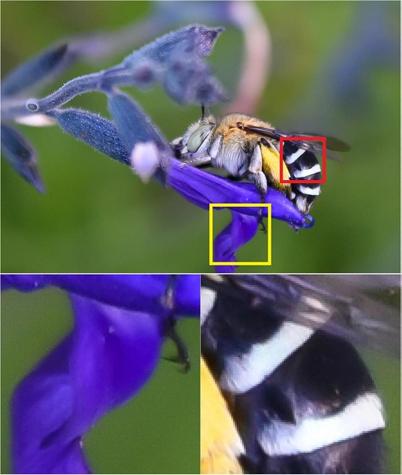}&
  \includegraphics[width=0.247\linewidth]{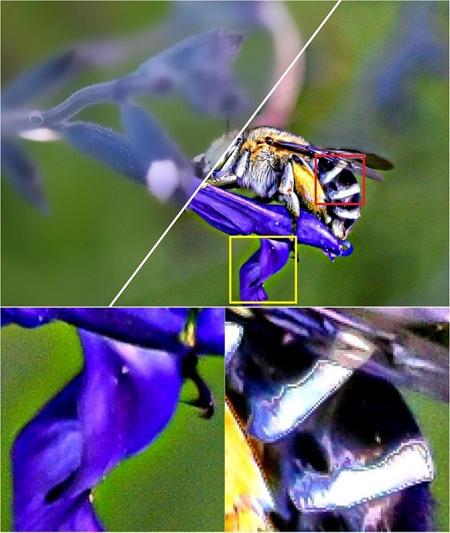}&
  \includegraphics[width=0.247\linewidth]{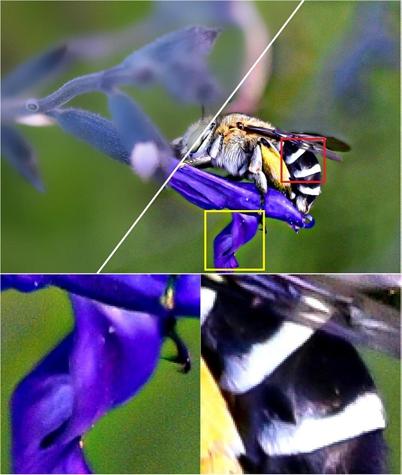}&
  \includegraphics[width=0.247\linewidth]{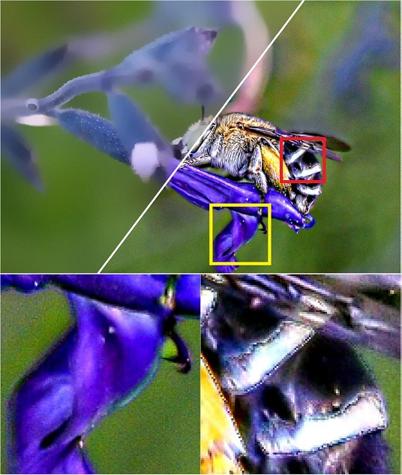}\\
  (a) Input & (b) AMF & (c) GF & (d) DTF-NC \\

  \includegraphics[width=0.247\linewidth]{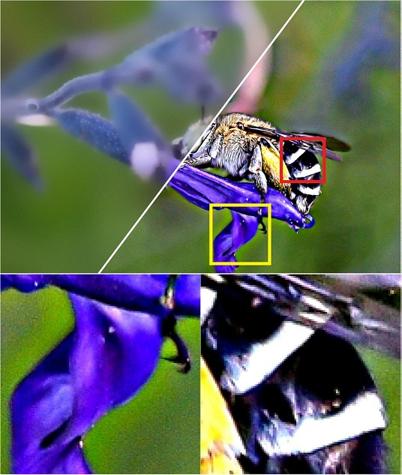}&
  \includegraphics[width=0.247\linewidth]{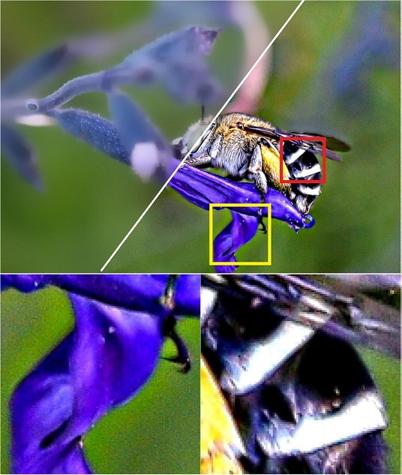}&
  \includegraphics[width=0.247\linewidth]{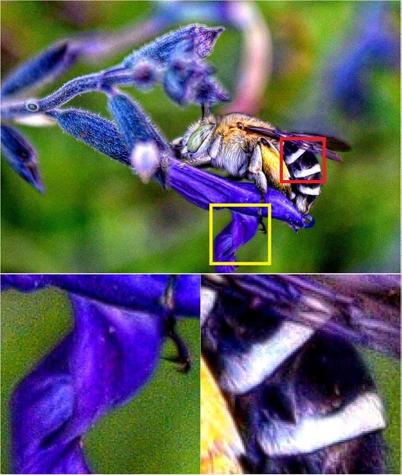}&
  \includegraphics[width=0.247\linewidth]{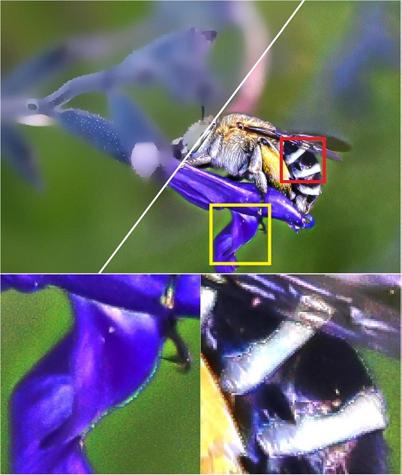}\\
  (e) DTF-IC & (f) DTF-RF & (g) LLF & (h) FGS \\

  \includegraphics[width=0.247\linewidth]{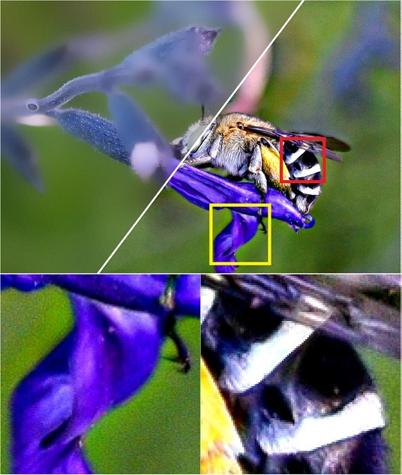}&
  \includegraphics[width=0.247\linewidth]{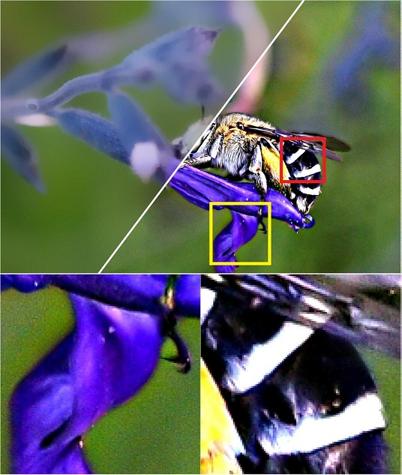}&
  \includegraphics[width=0.247\linewidth]{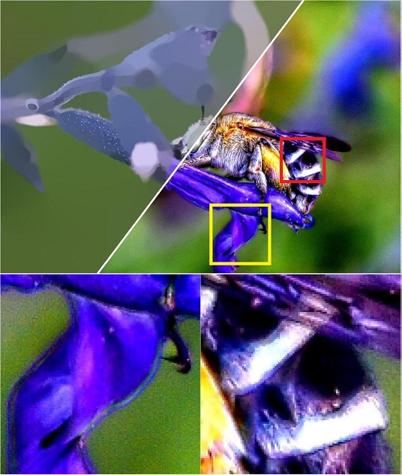}&
  \includegraphics[width=0.247\linewidth]{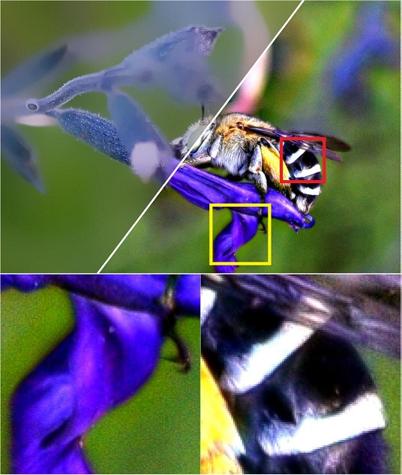}\\
  (i) FOPA & (j) WLS & (k) $L_0$ norm & (l) ILS
  \end{tabular}
  \caption{Visual comparison of image detail enhancement in terms of  gradient reversals (highlighted regions in the red boxes) and halos (highlighted regions in the yellow boxes). (a) Input. Result of (b) AMF ($\sigma_s=20,\sigma_r=0.2$), (c) GF ($r=16,\epsilon=0.08^2$), (d) DTF-NC ($\sigma_s=60,\sigma_r=0.4$), (e) DTF-IC ($\sigma_s=40,\sigma_r=0.25$), (f) DTF-RF ($\sigma_s=60,\sigma_r=0.4$), (g) LLF ($\alpha=0.25, \beta=1, \sigma_r=0.4$), (h) FGS ($\sigma_c=0.03,\lambda=400$), (i) FOPA (Cauchy penalty function, $\lambda=100,\gamma=\frac{3.5}{255}$), (j) WLS ($\lambda=1,\alpha=1.2$), (k) $L_0$ norm ($\lambda=0.02$) and (l) the proposed ILS ($\lambda=1,p=0.8$). The upper left part in each image is the smoothed image and the bottom right part is the corresponding $3\times$ detail enhanced image. Input image courtesy of the flickr user Conall.}\label{FigDetailEnhancement}
\end{figure*}

It is worthwhile to point out that the computational complexity of most compared approaches in Table~\ref{TabTimeComp} is $\mathcal{O}(N)$ given $N$ as the image size while it is $\mathcal{O}(Nlog N)$ for our ILS. However, one merit of the proposed ILS is that its two steps, i.e., the point wise computation in Eq.~(\ref{EqIntermidiateVar}) and the FFT/IFFT in Eq.~(\ref{EqILSSolution}), are highly parallel and can be easily accelerated. For example, the FFT and IFFT in MATLAB are computed in multi-thread under the default configuration, and they can be further accelerated through the GPU hardware. The GPU counterpart of our ILS can yield a up to $49\times$  speedup over its single-thread CPU implementation, and a $25\times$ speedup compared with its multi-thread CPU implementation. This means that the ILS is able to process images of 1080p resolution ($1920\times1080$) at the rate of 20fps for color images and 47fps for gray images, as illustrated in Table~\ref{TabTimeComp}. When compared with the GPU implementation of the $L_0$ norm smoothing, our ILS is around $6\times$ faster. It is also generally $15\times\sim20\times$ faster than the deep learning based method proposed by Chen et~al. \shortcite{chen2017fast} in the first group of Table~\ref{TabTimeComp}. Their method is the state-of-the-art deep learning based approach in both running speed and approximation accuracy.

\begin{figure*}
  \centering
  \setlength{\tabcolsep}{0.75mm}
  \begin{tabular}{cccc}
  \includegraphics[width=0.23\linewidth]{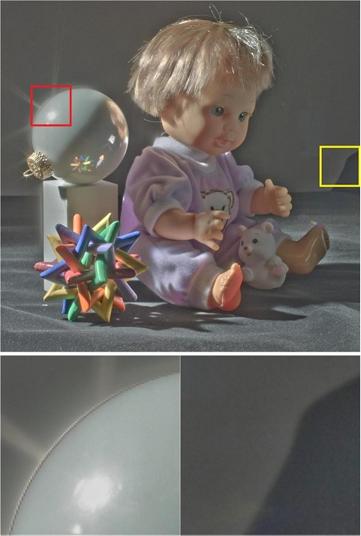}&
  \includegraphics[width=0.23\linewidth]{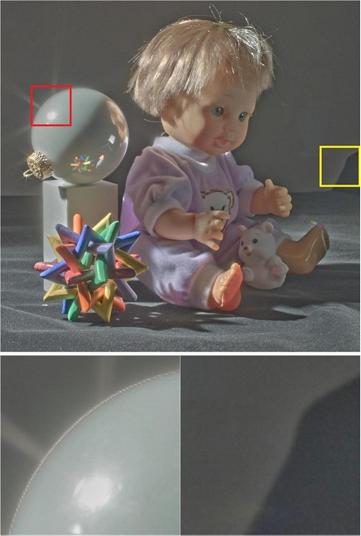}&
  \includegraphics[width=0.23\linewidth]{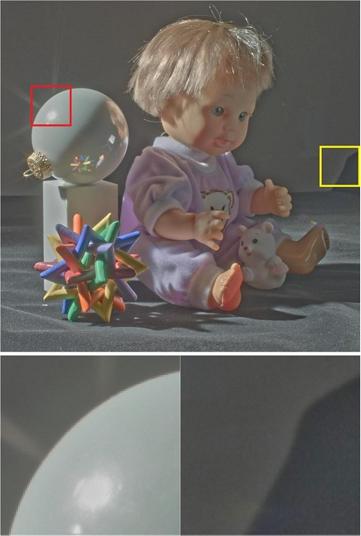}&
  \includegraphics[width=0.23\linewidth]{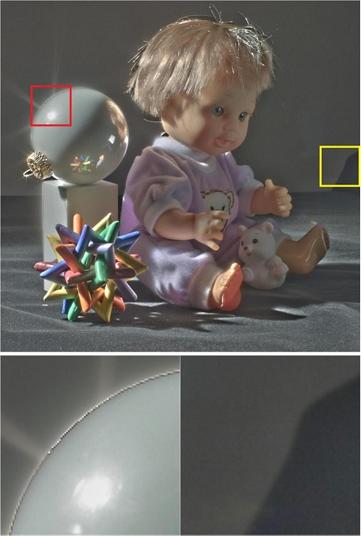}\\
  (a) Fast BLF & (b) AMF & (c) GF & (d) DTF-NC \\

  \includegraphics[width=0.23\linewidth]{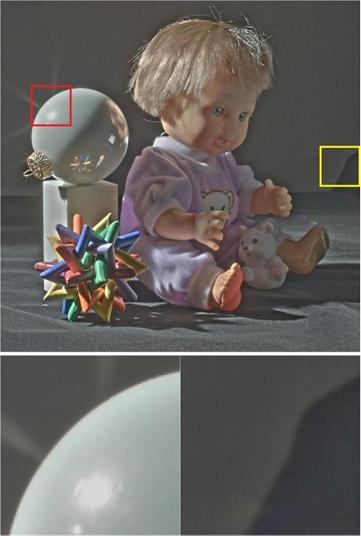}&
  \includegraphics[width=0.23\linewidth]{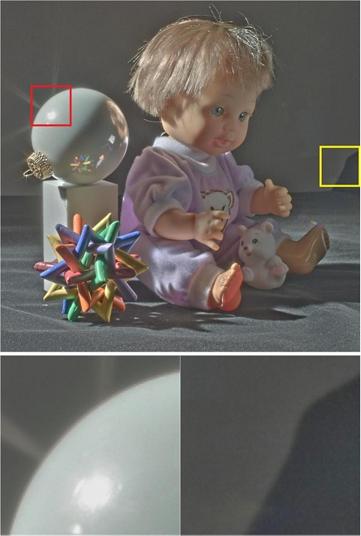}&
  \includegraphics[width=0.23\linewidth]{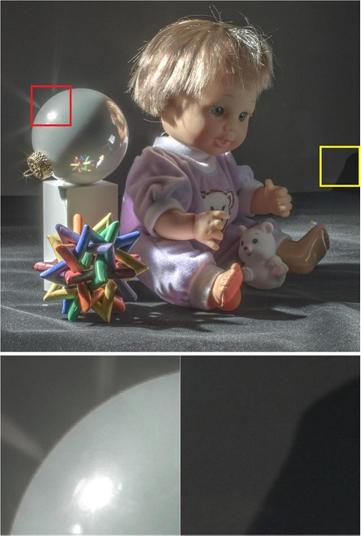}&
  \includegraphics[width=0.23\linewidth]{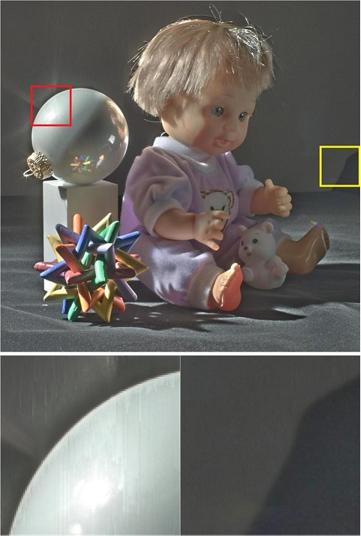}\\
  (e) DTF-IC & (f) DTF-RF & (g) LLF & (h) FGS \\

  \includegraphics[width=0.23\linewidth]{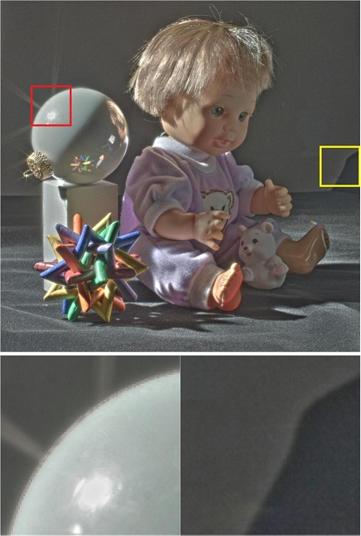}&
  \includegraphics[width=0.23\linewidth]{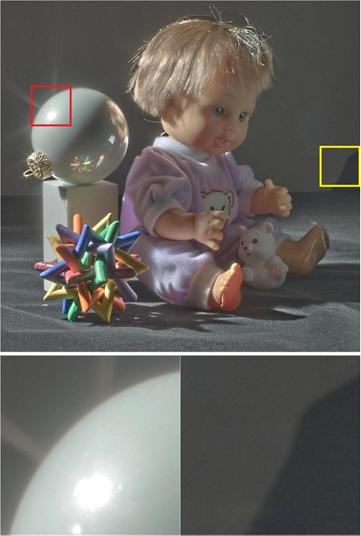}&
  \includegraphics[width=0.23\linewidth]{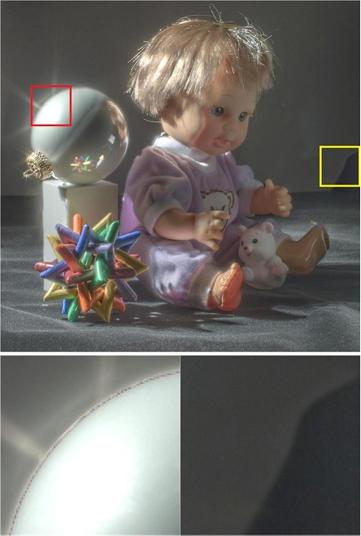}&
  \includegraphics[width=0.23\linewidth]{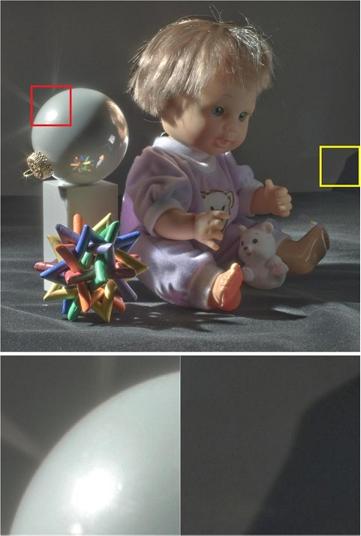}\\
  (i) FOPA & (j) WLS & (k) $L_0$ norm & (l) ILS
  \end{tabular}
  \caption{Visual comparison of single-scale HDR tone mapping in terms of  gradient reversals (highlighted regions in the red boxes) and halos (highlighted regions in the yellow boxes). Result of (a) fast BLF ($\sigma_s=20,\sigma_r=0.2$), (b) AMF ($\sigma_s=20,\sigma_r=0.15$), (c) GF ($r=16,\epsilon=0.08^2$), (d) DTF-NC ($\sigma_s=30,\sigma_r=0.35$), (e) DTF-IC ($\sigma_s=20,\sigma_r=0.2$), (f) DTF-RF ($\sigma_s=30,\sigma_r=0.2$), (g) LLF ($\alpha=1, \beta=0, \sigma_r=log(2.5)$), (h) FGS ($\sigma_c=0.3,\lambda=200$), (i) FOPA (Welsch penalty function, $\lambda=100,\gamma=0.5$), (j) WLS ($\lambda=10,\alpha=1.2$), (k) $L_0$ norm ($\lambda=0.01$) and (l) the proposed ILS ($\lambda=10,p=1$).}\label{FigHDR}
\end{figure*}

\begin{figure*}
  \centering
  \setlength{\tabcolsep}{0.25mm}
  \begin{tabular}{ccc}
  \includegraphics[width=0.315\linewidth]{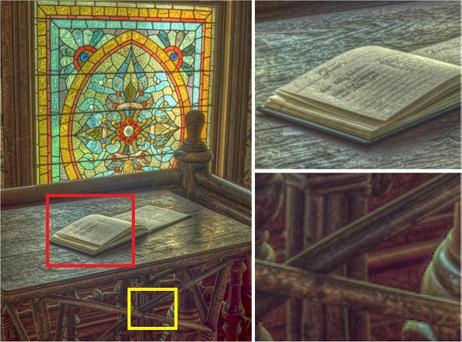}&
  \includegraphics[width=0.315\linewidth]{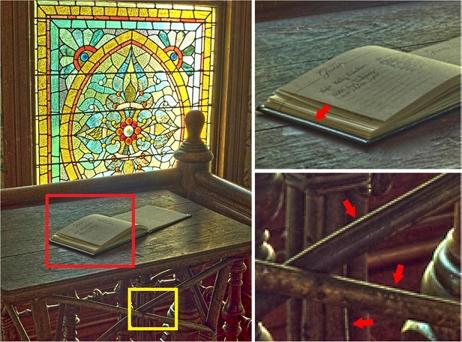}&
  \includegraphics[width=0.315\linewidth]{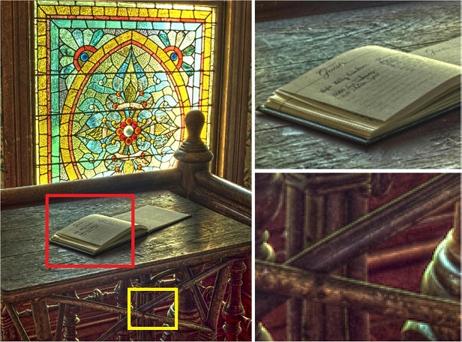}\\

  \includegraphics[width=0.315\linewidth]{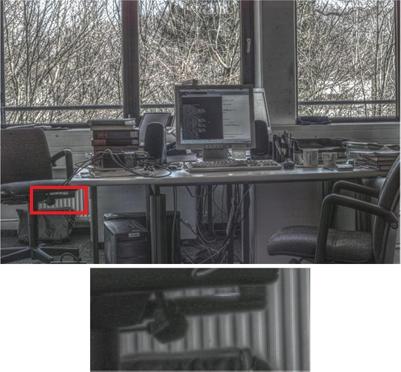}&
  \includegraphics[width=0.315\linewidth]{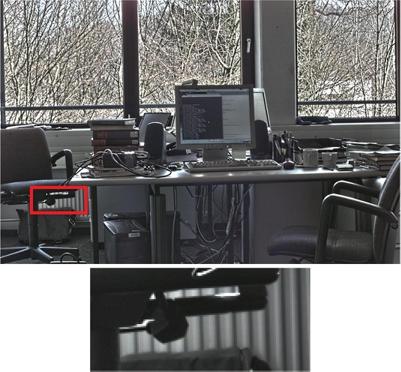}&
  \includegraphics[width=0.315\linewidth]{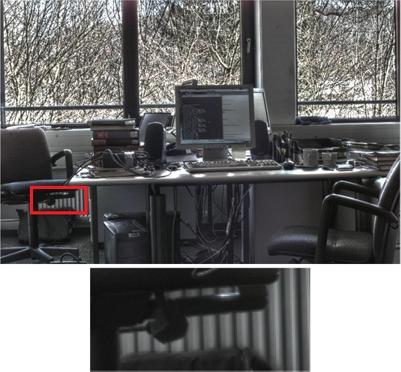}\\

  \includegraphics[width=0.315\linewidth]{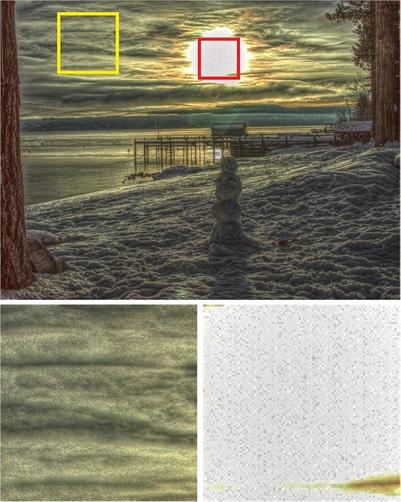}&
  \includegraphics[width=0.315\linewidth]{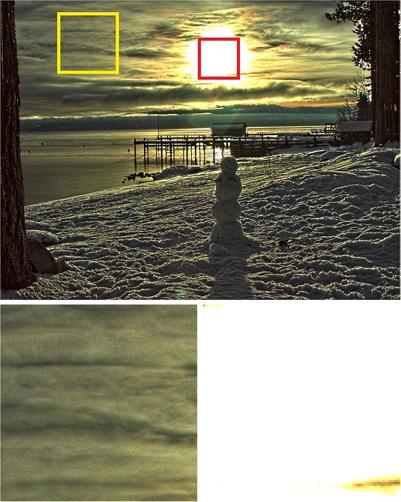}&
  \includegraphics[width=0.315\linewidth]{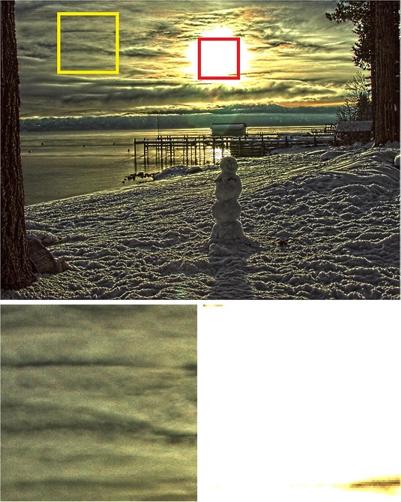}\\
   (a) LLF & (b) WLS & (c) ILS
  \end{tabular}
  \caption{HDR tone mapping results of (a) LLF ($\alpha=0.25,\beta=0,\sigma_r=log(0.25)$) with details boosted. Multi-scale HDR tone mapping results of (b) WLS and (c) ILS. The parameters of WLS for producing the results in the first to the third row are $\alpha=1.2,\lambda=\frac{1}{16}/\frac{1}{2}/4$, $\lambda=\frac{1}{4}/2/16$ and $\lambda=\frac{1}{8}/1/8$, respectively. The corresponding parameters of the proposed ILS are $p=1, \lambda=\frac{1}{2}/4/32$, $\lambda=\frac{1}{16}/\frac{1}{2}/4$ and $\lambda=\frac{1}{4}/2/16$. }\label{FigHDRMultiScale}
\end{figure*}

\subsection{Smoothing Quality Comparison}
\label{SecSmoothQualityComp}

\subsubsection{Handling Gradient Reversals and Halos}
In this subsection, we compare all the approaches mentioned above in terms of their smoothing quality.  Gradient reversals and halos have been widely considered as two kinds of artifacts that affect the smoothing quality. When the edges are sharpened in the smoothed image, they will be boosted in a reverse direction, which results in gradient reversals. The halo artifacts will occur when the edges are blurred in the smoothed image. We test the smoothing quality of all the methods through image detail enhancement and HDR tone mapping. The tone mapping results of LLF and fast LLF are produced based on the framework proposed by Paris et~al. \shortcite{paris2011local}. Results of the other compared methods are obtained with the framework proposed by Farbman et~al. \shortcite{farbman2008edge} \footnote{http://www.cs.huji.ac.il/~danix/epd/}. For HDR tone mapping, only the luminance channel in YUV color space of the input HDR image is compressed and the smoothing is performed on the logarithm of the luminance channel.

Fig.~\ref{FigDetailEnhancement} and Fig.~\ref{FigHDR} show the image detail enhancement and single-scale HDR tone mapping results produced by different approaches. Parameters of all the compared methods are carefully tuned to ensure that they have similar smoothing on the input image. Results produced by all the weighted average based approaches in the second group of Table~\ref{TabTimeComp} contain either gradient reversals or halos. Some results even contain both these two kinds of artifacts, e.g., the results of fast BLF and AMF shown in Fig.~\ref{FigDetailEnhancement}(b), Fig.~\ref{FigHDR}(a) and (b). One the one side, the gradient reversals in the results of fast BLF, AMF and DTF-NC are due to the small $\sigma_r$ used in the smoothing, and a larger $\sigma_r$  helps to mitigate the artifacts. On the other side, a smaller $\sigma_r$ in fast BLF, AMF and DTF can properly eliminate halos. Thus, for the results shown in Fig.~\ref{FigDetailEnhancement}(b), Fig.~\ref{FigHDR}(a) and (b), either increasing or decreasing the $\sigma_r$ in fast BLF and AMF will lead to stronger halos or gradient reversals. For the results of GF, DTF-IC and DTF-RF, although decreasing the $\epsilon$ in GF or the $\sigma_r$ in DTF-IC and DTF-RF  is an effective way to avoid halos. However, it will also weaken their smoothing strength, which will lead to insufficient smoothing on the input. This is known as the drawback of the weighted average based approaches: there is a tradeoff between their smoothing abilities and edge-preserving abilities \cite{farbman2008edge, he2013guided}.

\begin{figure*}
  \centering
  \setlength{\tabcolsep}{0.25mm}
  \begin{tabular}{ccccc}
  \includegraphics[width=0.1975\linewidth]{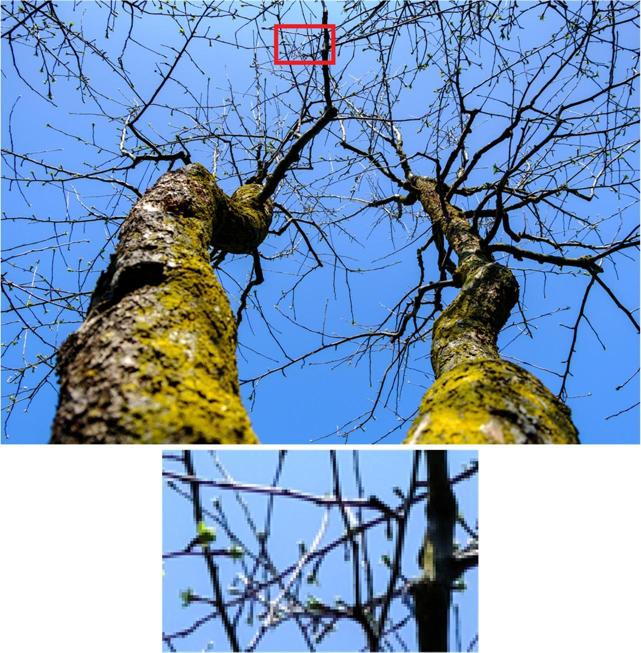}&
  \includegraphics[width=0.1975\linewidth]{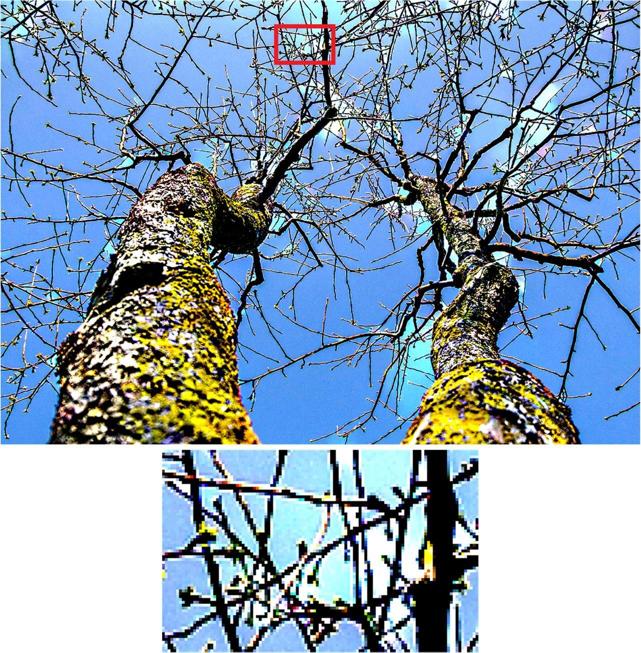}&
  \includegraphics[width=0.1975\linewidth]{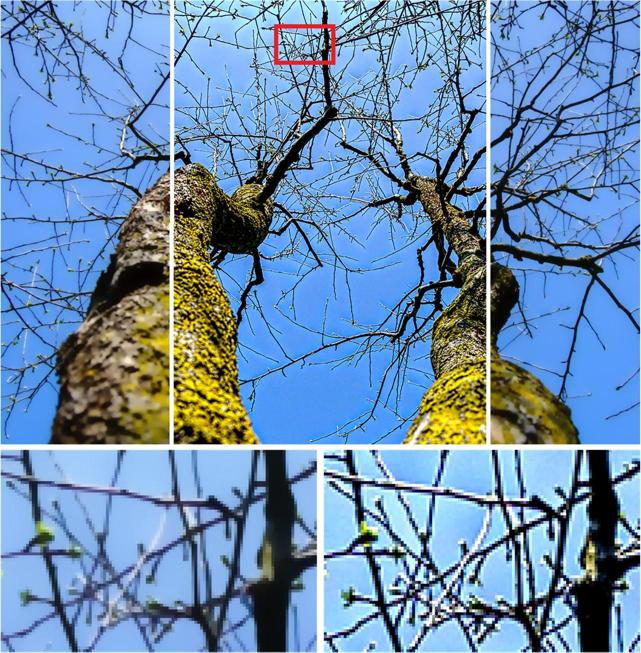}&
  \includegraphics[width=0.1975\linewidth]{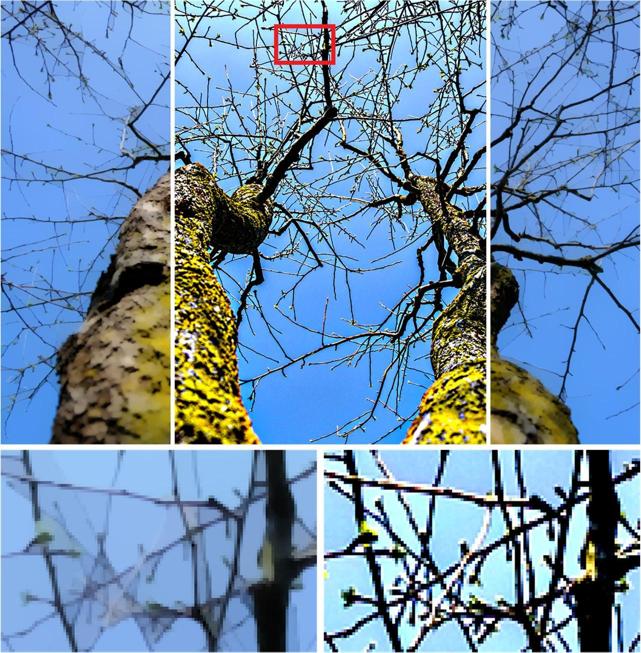}&
  \includegraphics[width=0.1975\linewidth]{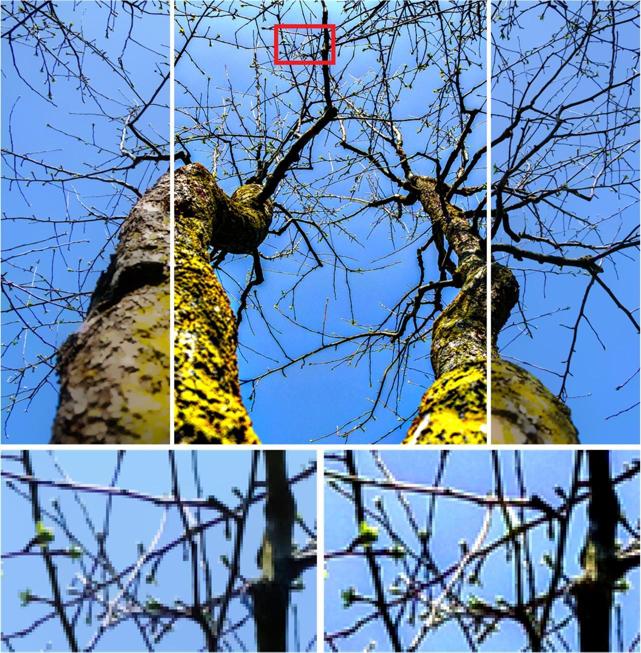}\\
  (a) Input & (b) EAW & (c) DTF-IC & (d) WLS & (e) ILS
  \end{tabular}
  \caption{Comparison of compartmentalization artifacts in image detail enhancement. Result of (b) EAW (WRB wavelets, $\alpha=1,\beta=1,\gamma=1.4$), (c) DTF-IC ($\sigma_s=20, \sigma_r=0.2$), (d) WLS ($\lambda=1,\alpha=1.2$) and (e) ILS ($p=0.8,\lambda=1$). The input image is shown in (a). All the detail enhanced images in (c)$\sim$(e) are produced with details $3\times$ boosted.}\label{FigILSvsWLS_DetailEnhancement}
\end{figure*}

\begin{figure*}
  \centering
  \setlength{\tabcolsep}{0.5mm}
  \begin{tabular}{ccc}
  \includegraphics[width=0.328\linewidth]{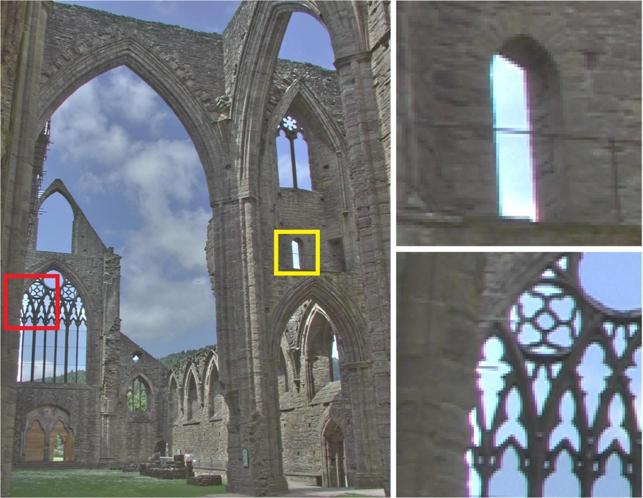}&
  \includegraphics[width=0.328\linewidth]{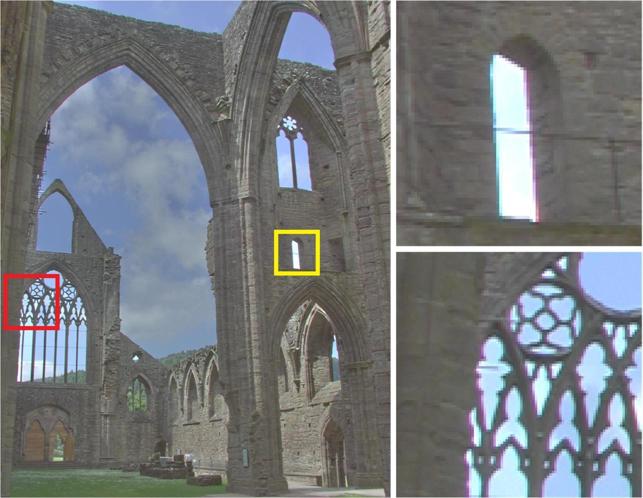}&
  \includegraphics[width=0.328\linewidth]{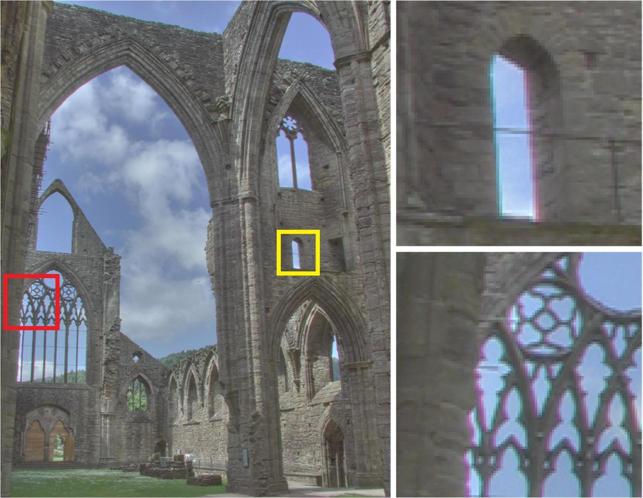}\\
  (a) WLS & (b) EAW & (c) ILS
  \end{tabular}
  \caption{Comparison of compartmentalization artifacts in single-scale HDR tone mapping. Result of (a) WLS ($\lambda=20,\alpha=1.2$), (b) EAW (WRB wavelets, $\alpha=0.8, \beta=0.125,\gamma=0.8$) and (c) the proposed ILS ($p=1,\lambda=20$). Results of WLS and EAW contain visible compartmentalization artifacts.}\label{FigILSvsWLS_HDR}
\end{figure*}

The results of FGS in Fig.~\ref{FigDetailEnhancement}(h) and Fig.~\ref{FigHDR}(h) also contain clear artifacts. Fig.~\ref{FigDetailEnhancement}(i) shows the detail enhancement result produced by FOPA (Cauchy penalty function) where halos can be observed. Fig.~\ref{FigHDR}(i) further shows the HDR tone mapping result of FOPA (Welsch penalty function) where both gradient reversals and halos exist. There are gradient reversals in the results of $L_0$ norm smoothing as shown in Fig.~\ref{FigDetailEnhancement}(k) and Fig.~\ref{FigHDR}(k). Generally, there are seldom any visible gradient reversals and halos in the results of the proposed ILS as shown in Fig.~\ref{FigDetailEnhancement}(l) and Fig.~\ref{FigHDR}(l). In terms of handling gradient reversals and halos, the ILS, WLS and LLF show comparable results as demonstrated in Fig.~\ref{FigDetailEnhancement} and Fig.~\ref{FigHDR}.

\subsubsection{Handling Other Artifacts}

Besides the widely considered gradient reversals and halos, there are also other kinds of artifacts that can affect the smoothing quality. We demonstrate them through different examples.

Fig.~\ref{FigHDRMultiScale} further shows the multi-scale HDR tone mapping results of WLS, ILS and the detail boosted results of LLF. The LLF can produce visually ``noisy'' results in some cases when the parameters are set to boost the details. Examples are shown in the highlighted regions (labeled with the red boxes) in the first row and the third row of Fig.~\ref{FigHDRMultiScale}(a). The WLS is prone to over enhance small structures. The highlighted regions in the first row (labeled with the red arrows) and the second row of Fig.~\ref{FigHDRMultiScale}(b) illustrate examples. As we will demonstrate in the next paragraph, this kind of artifacts are known as compartmentalization \cite{hessel2018quantitative}. Compared with their results, the results of our ILS in Fig.~\ref{FigHDRMultiScale}(c) are less prone to contain the artifacts mentioned above.

\begin{figure*}
  \centering
  \setlength{\tabcolsep}{0.5mm}
  \begin{tabular}{ccccc}
  \includegraphics[width=0.195\linewidth]{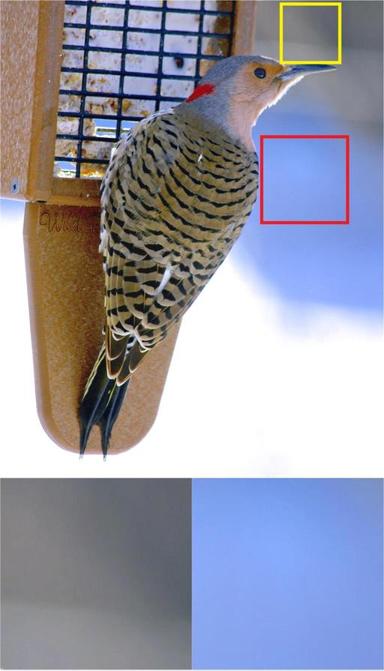}&
  \includegraphics[width=0.195\linewidth]{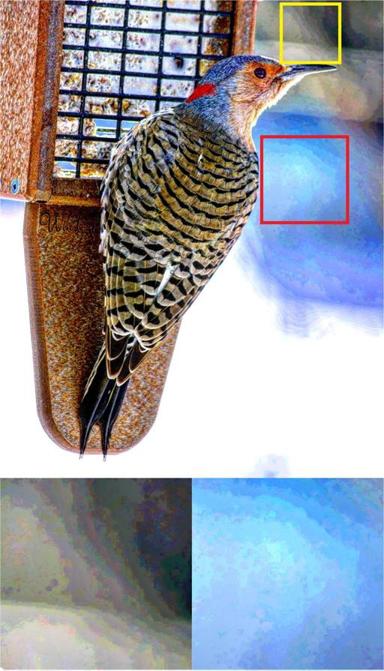}&
  \includegraphics[width=0.195\linewidth]{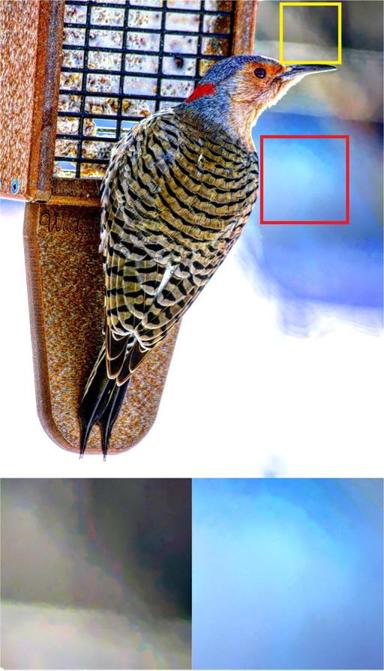}&
  \includegraphics[width=0.195\linewidth]{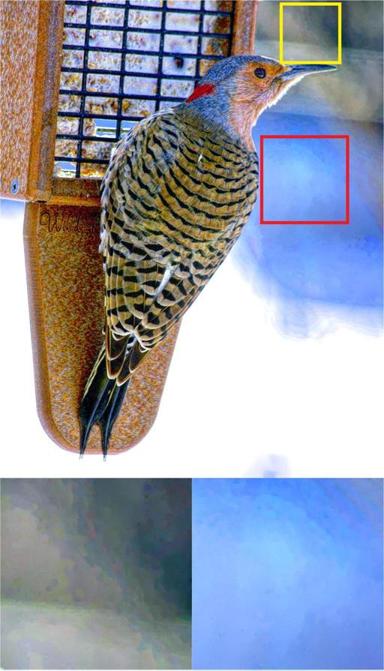}&
  \includegraphics[width=0.195\linewidth]{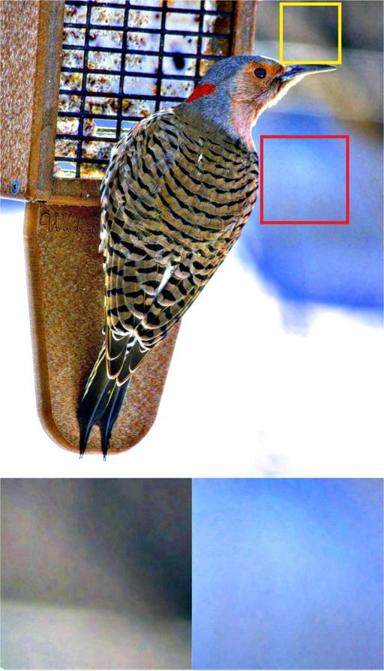}\\
  (a) Input & (b) Fast LLF & (c) Fast LLF & (d) LLF & (e) ILS\\
  & (auto intensity levels) & (256 intensity levels) & &\\
  \end{tabular}
  \caption{Image detail enhancement comparison between the result of (b) fast LLF with automatically computed intensity levels ($\alpha=0.25,\beta=1,\sigma_r=0.4$), (c) fast LLF with predefined 256 intensity levels ($\alpha=0.25,\beta=1,\sigma_r=0.4$), (d) LLF ($\alpha=0.25,\beta=1,\sigma_r=0.4$) and (e) the proposed ILS ($p=0.8,\lambda=1,3\times$ detail boosting). The input image is shown in (a). Fast LLF with automatically computed intensity levels can lead to visible quantization artifacts. Input image courtesy of the flickr user Missy Mandel.}\label{FigILSvsFastLaplacian_DetailEnhancement}
\end{figure*}

As we have mentioned in Sec.~\ref{SecParameters}, the intensity shift produced by our ILS can lead to compartmentalization artifacts in some cases. We show that there are also other approaches which share the same drawback. Fig.~\ref{FigILSvsWLS_DetailEnhancement} and Fig.~\ref{FigILSvsWLS_HDR} show comparisons of image detail enhancement and single-scale HDR tone mapping results produced by different methods. The EAW, DTF-IC and WLS produce clearly visible compartmentalization artifacts as shown in the highlighted regions. Compared with their results, the compartmentalization artifacts in the results produced by our ILS are much milder. This can also be validated through the comparison between the highlighted regions in the second row of Fig.~\ref{FigHDRMultiScale}(b) and (c).

\begin{figure*}
  \centering
  \setlength{\tabcolsep}{0.5mm}
  \begin{tabular}{cccc}
  \includegraphics[width=0.245\linewidth]{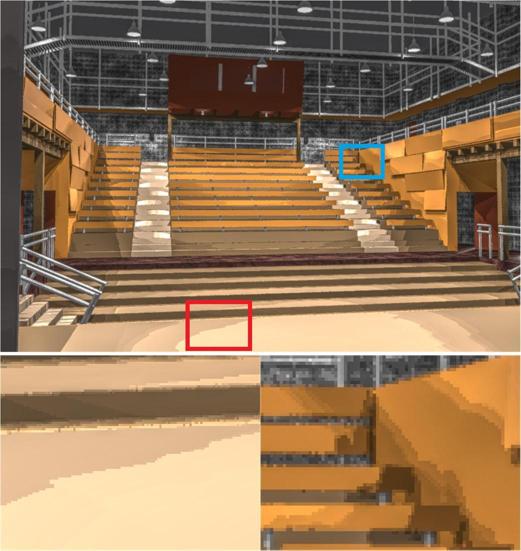}&
  \includegraphics[width=0.245\linewidth]{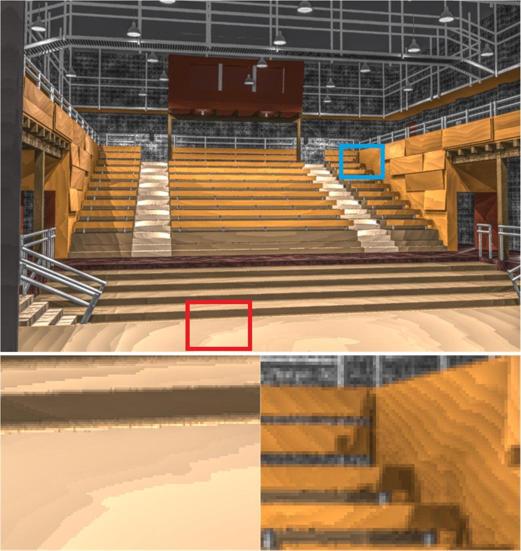}&
  \includegraphics[width=0.245\linewidth]{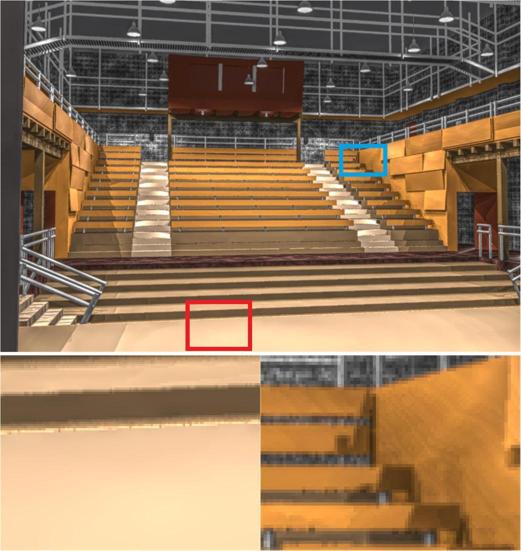}&
  \includegraphics[width=0.245\linewidth]{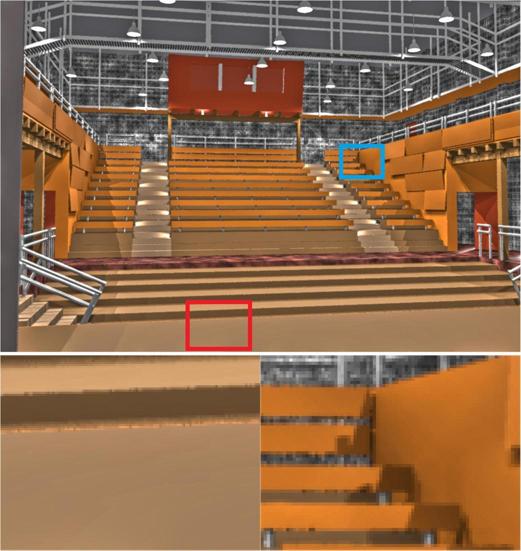}\\
  (a) Fast LLF  & (b) Fast LLF  & (c) LLF & (d) ILS\\
  (auto intensity levels) & (256 intensity levels) & &\\
  \end{tabular}
  \caption{HDR tone mapping comparison between the result of (a) fast LLF with automatically computed intensity levels ($\alpha=0.25,\beta=0,\sigma_r=log(2.5)$), (b) fast LLF with predefined 256 intensity levels ($\alpha=0.25,\beta=0,\sigma_r=log(2.5)$), (c) LLF ($\alpha=0.25,\beta=0,\sigma_r=log(2.5)$) and (d) multi-scale ILS tone mapping ($p=1,\lambda=\frac{1}{8}/1/8)$. The parameters of LLF and fast LLF are set to boost details. Quantization artifacts exist in the results of fast LLF.}\label{FigILSvsFastLaplacian_HDR}
\end{figure*}
\begin{figure*}
  \centering
  \setlength{\tabcolsep}{0.5mm}
  \begin{tabular}{ccc}
  \includegraphics[width=0.33\linewidth]{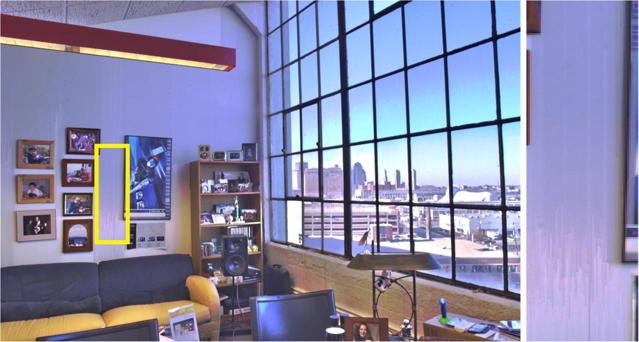}&
  \includegraphics[width=0.33\linewidth]{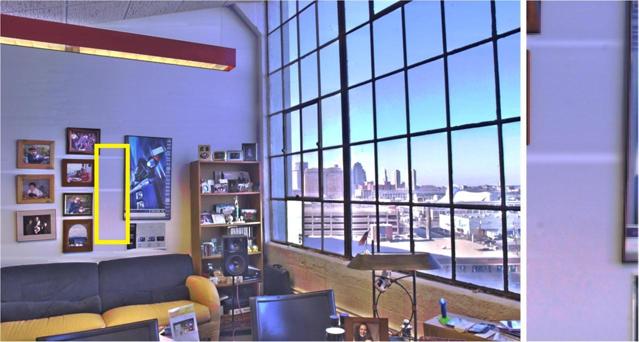}&
  \includegraphics[width=0.33\linewidth]{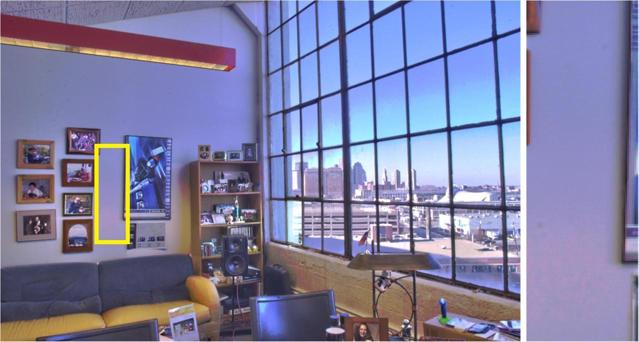}\\
  (a) FGS & (b) SG-WLS & (c) ILS
  \end{tabular}
  \caption{HDR tone mapping comparison between the result of (a) FGS ($\sigma_c=0.5,\lambda=500$), (b) SG-WLS ($r=1,\alpha_s=\alpha_r=0.5,\lambda=500$) and (c) the proposed ILS ($p=1, \lambda=10$). Results of FGS and SG-WLS contain blocky artifacts.}\label{FigILSvsFGSandSGWLS}
\end{figure*}
\begin{figure*}
  \centering
  \setlength{\tabcolsep}{0.5mm}
  \begin{tabular}{ccc}
  \includegraphics[width=0.33\linewidth]{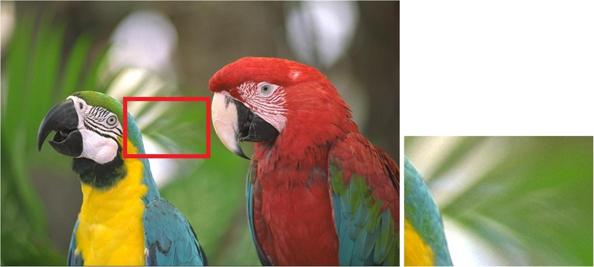}&
  \includegraphics[width=0.33\linewidth]{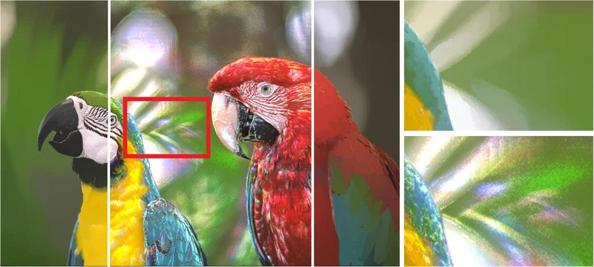}&
  \includegraphics[width=0.33\linewidth]{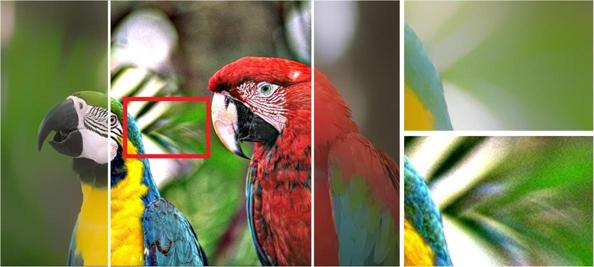}\\
  (a) Input & (b) FBS & (c) ILS
  \end{tabular}
  \caption{Comparison of image detail enhancement results produced by (b) FBS ($\sigma_{xy}=8,\sigma_l=8,\sigma_{uv}=4,\lambda=1$) and (c) ILS ($p=1,\lambda=0.75$). The input image is shown in (a). Details are $3\times$ boosted. The result of FBS contains gradient reversals and unnatural artifacts.}\label{FigILSvsFBS}
\end{figure*}

\begin{figure*}
  \includegraphics[width=1\linewidth]{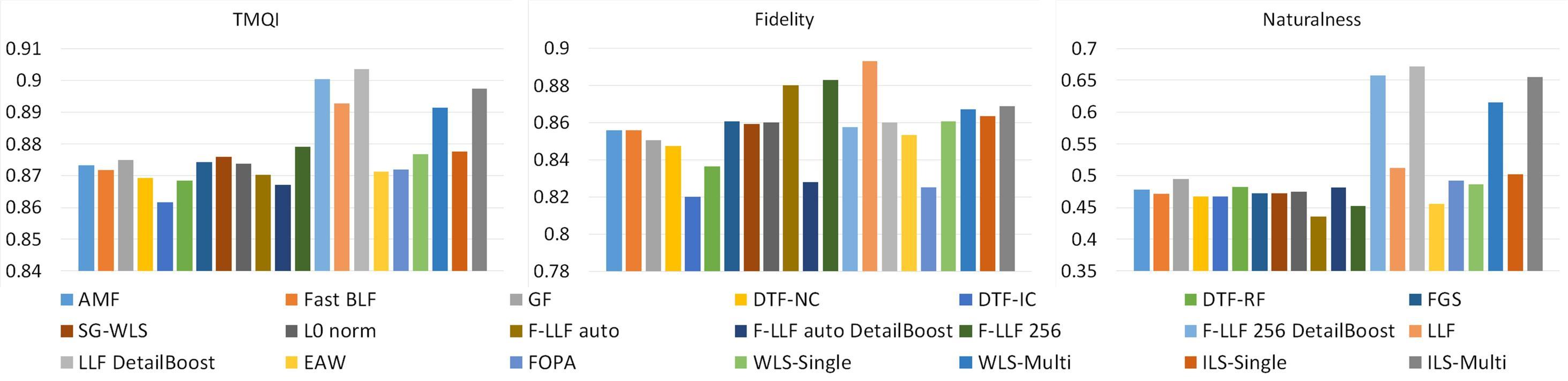}
  \caption{Quantitative comparison of HDR tone mapping results produced by different approaches. \emph{F-LLF auto} represents the fast LLF with automatically computed intensity levels, \emph{F-LLF 256} denote the fast LLF with predefined 256 intensity levels. \emph{F-LLF auto DetailBoost, F-LLF 256 DetailBoost, LLF DetailBoost} denote the corresponding approaches with the parameters set to boost details. \emph{WLS-Multi} and \emph{ILS-Multi} refer to the multi-scale tone mapping of WLS and ILS, respectively. \emph{WLS-Single} and \emph{ILS-Single} denote the corresponding single-scale HDR tone mapping. The other methods perform single-scale HDR tone mapping.}\label{FigHDRQuantitative}
\end{figure*}

We also compare our ILS against the fast LLF with two different parameter settings: automatically computed intensity levels and predefined 256 intensity levels. Fig.~\ref{FigILSvsFastLaplacian_DetailEnhancement} and Fig.~\ref{FigILSvsFastLaplacian_HDR} show comparison of image detail enhancement and HDR tone mapping results produced by fast LLF, LLF and our ILS. As shown in the highlighted regions in Fig.~\ref{FigILSvsFastLaplacian_DetailEnhancement}(b) and Fig.~\ref{FigILSvsFastLaplacian_HDR}(a), the fast LLF with automatically computed intensity levels can result in visible quantization artifacts. These artifacts can be mitigated by using a predefined 256 intensity levels as demonstrated in Fig.~\ref{FigILSvsFastLaplacian_DetailEnhancement}(c). However, predefined 256 intensity levels are usually larger than the automatically computed ones. This means that the computational cost will be increased, which can also be observed from Table.~\ref{TabTimeComp}. The fast LLF with predefined 256 intensity levels is able to have close performance to the original LLF, but there are still quantization artifacts that cannot be completely removed in some cases, as shown in Fig.~\ref{FigILSvsFastLaplacian_HDR}(c). The results of our ILS do not suffer from quantization artifacts. Note that the color difference between our result and those of LLF and fast LLF in Fig.~\ref{FigILSvsFastLaplacian_HDR} is due to the different HDR tone mapping frameworks adopted for producing the results, however, this does not affect the above analysis.

Fig.~\ref{FigILSvsFGSandSGWLS} shows a comparison of single-scale HDR tone mapping results produced by FGS, SG-WLS and our ILS. There are blocky artifacts in the results of FGS and SG-WLS as demonstrated in the highlighted regions in Fig.~\ref{FigILSvsFGSandSGWLS}(a) and (b). This is due to the reason that FGS and SG-WLS smooth images in each row/column patch separately. The proposed ILS does not smooth images in this way and thus no blocky artifacts exist in its results, as shown in  Fig.~\ref{FigILSvsFGSandSGWLS}(c).

A comparison of image detail enhancement results produced by FBS and our ILS is shown in Fig.~\ref{FigILSvsFBS}. The FBS is prone to produce images with piece-wise constant regions separated by sharpened edges, which can lead to gradient reversals and visually unnatural artifacts in the result as shown in Fig.~\ref{FigILSvsFBS}(b). The smoothed image of our ILS is piece-wise smooth and the corresponding detail enhanced image in Fig.~\ref{FigILSvsFBS}(c) is more visually pleasant than the one in Fig.~\ref{FigILSvsFBS}(b).

Finally, a quantitative evaluation of all the compared methods is performed. Since there are no measurements for the results of image detail enhancement, we only perform a quantitative evaluation on the HDR tone mapping results produced by different approaches. Due to the reason that there are no public datasets for HDR tone mapping, we collect 25 different HDR images containing different scenes. All the tone mapping results are evaluated with the tone mapped image quality index (TMQI) proposed by Yeganeh et~al. \shortcite{yeganeh2012objective}. TMQI first evaluates the structural fidelity and naturalness of the tone mapped images. The two measurements are then adjusted by a power function and averaged to give a final score ranging from 0 to 1. Larger values of TMQI indicate better quality of the tone mapped images, and vice versa. Fig.~\ref{FigHDRQuantitative} illustrates the mean TMQI score of each method. Generally, except for the LLF and fast LLF with predefined 256 intensity levels, our ILS surpasses all the other compared methods. Note that our ILS is much faster than LLF and fast LLF with predefined 256 intensity levels, as shown in Table~\ref{TabTimeComp}.

\section{Model Extension}
\label{SecExtension}

\begin{figure*}
\centering
\setlength{\tabcolsep}{0.75mm}
\begin{tabular}{ccc}
\includegraphics[width=0.32\linewidth]{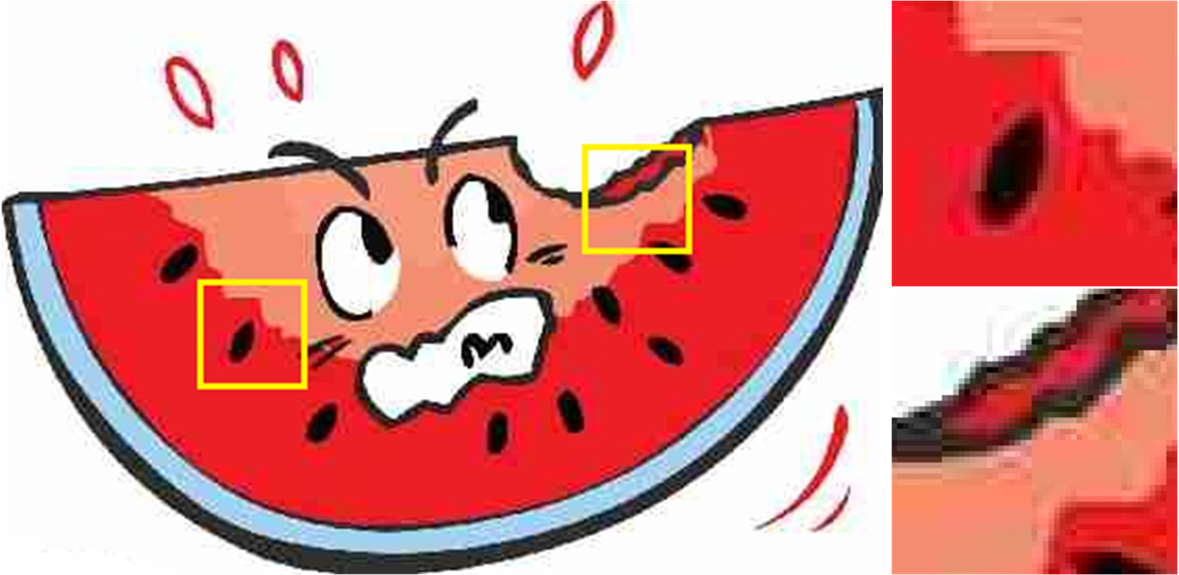} &
\includegraphics[width=0.32\linewidth]{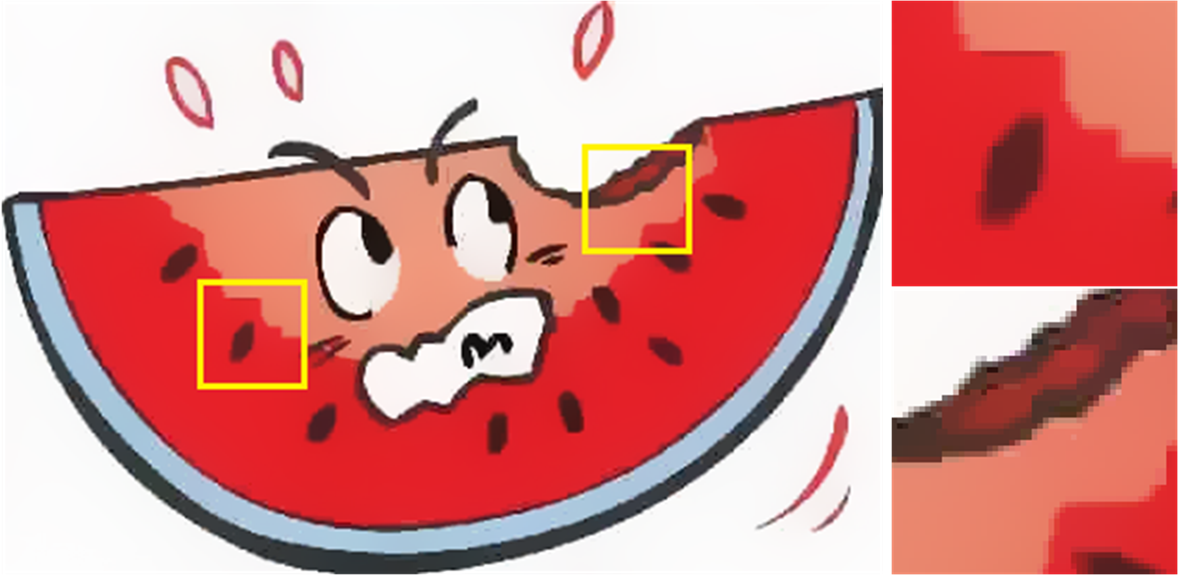} &
\includegraphics[width=0.32\linewidth]{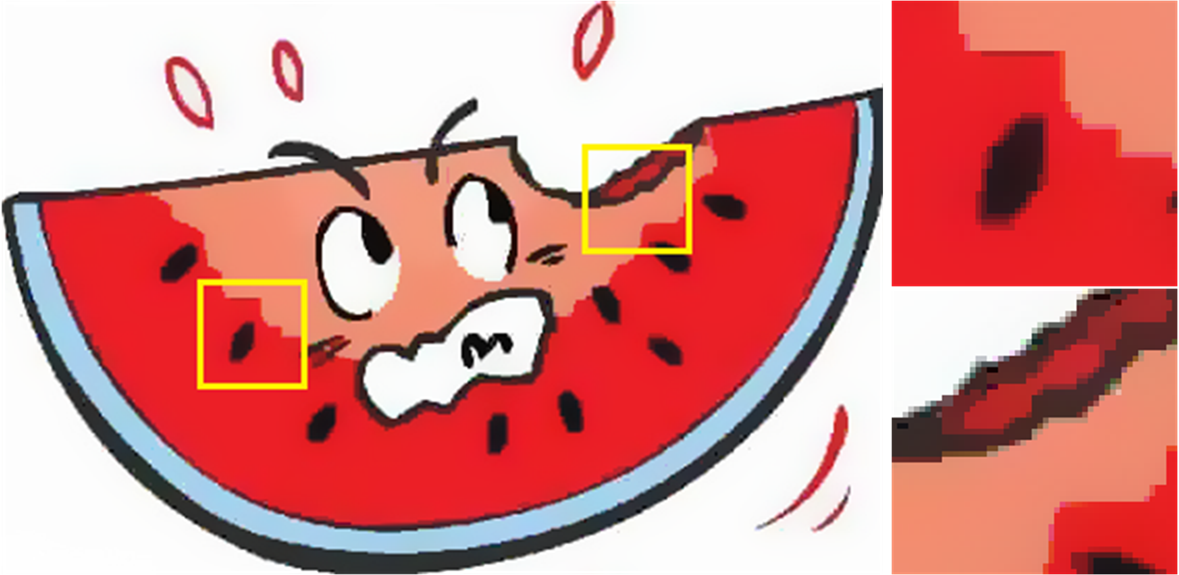} \\

(a) Input & (b) Generalized Charbonnier penalty & (c) Welsch penalty
\end{tabular}
\caption{Comparison of clip-art compression artifacts removal results obtained with the ILS using different penalty functions.  (a) Input compressed image. Smoothing result of (b) ILS using the generalized Charbonnier penalty function ($p=0.2, \lambda=0.5, N=10$) and (c) ILS using the Welsch penalty function ($\gamma=\frac{5}{255}, \lambda = 30, N = 10$).}
\label{FigWelschAdvantage}
\end{figure*}

The ILS described in Sec.~\ref{SecILS} is suitable for applications of tone and detail manipulation. When we need to handle other tasks such as clip-art compression artifacts removal which require to sharpen edges, the ILS is not suitable because the adopted generalized Charbonnier penalty function $\phi_{p}(\cdot)$ in Eq.~(\ref{EqSmoothedLpNorm}) is not able to sharpen edges. Fig.~\ref{FigWelschAdvantage}(b) shows the corresponding clip-art compression artifacts removal result where the edges are not sharpened. In this example, we adopt a larger iteration number $N=10$ because the compression artifacts need to be completely removed.

In contrast, the Welsch penalty function $\phi_W(\cdot)$ in Eq.~(\ref{EqWelschCauchy}) is able to sharpen edges as explained in Sec.~\ref{SecHalfSplitComp}. It also meets the requirements of the additive half-quadratic minimization in Sec.~\ref{SecHalfQuadratic}. Thus, we can simply replace the generalized Charbonnier penalty function with the Welsch penalty function and apply the same procedure in Eqs.~(\ref{EqLpUpperBound})$\sim$(\ref{EqILSSolution}) to obtain another ILS with different smoothing properties. There are two differences when adopting the Welsch penalty function. First, the point-wise operation on the image gradients in Eq.~(\ref{EqIntermidiateVar}) should be updated as:
\begin{equation}\label{EqGradientOperate_Welsch}
{
\begin{split}
    & \mu^n_{\ast,s}=c\nabla u^n_{\ast,s} - \phi_{W}'\left(\nabla u^n_{\ast,s}\right)\\
    & \ \ \ \ \ \ =c\nabla u^n_{\ast,s}-2\nabla u^n_{\ast,s}\exp\left(-\frac{(\nabla u^n_{\ast,s})^2}{2\gamma^2} \right), \ast\in\{x,y\}.
\end{split}
}
\end{equation}
Second, the constraint on $c$ should be:
\begin{equation}\label{EqCValueWelsch}
  c\geq c_0 \ \text{where}\ c_0=2.
\end{equation}
A detailed proof of Eq.~(\ref{EqCValueWelsch}) can be found in Appendix B. Similar to the analysis in Sec.~\ref{SecCAnalysis}, we set $c=c_0$ in all the experiments. Fig.~\ref{FigWelschAdvantage}(c) shows the result of our ILS using the Welsch penalty function, and the edges are much sharper than that in Fig.~\ref{FigWelschAdvantage}(b).

\subsection{Applications and Results}
\label{SecExtExperiments}

\begin{figure*}
\centering
\setlength{\tabcolsep}{0.25mm}
\begin{tabular}{cccccc}
\includegraphics[width=0.165\linewidth]{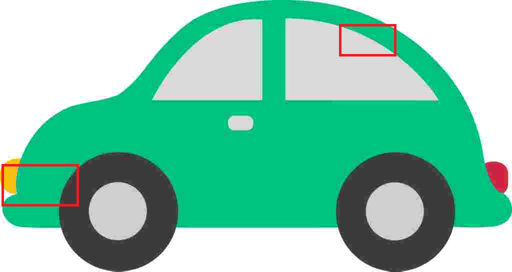} &
\includegraphics[width=0.165\linewidth]{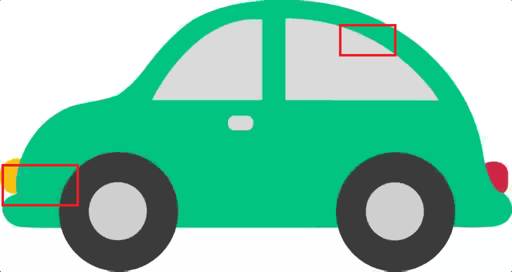} &
\includegraphics[width=0.165\linewidth]{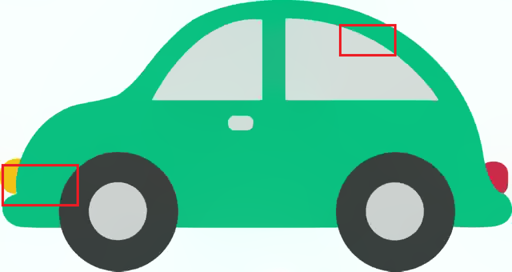} &
\includegraphics[width=0.165\linewidth]{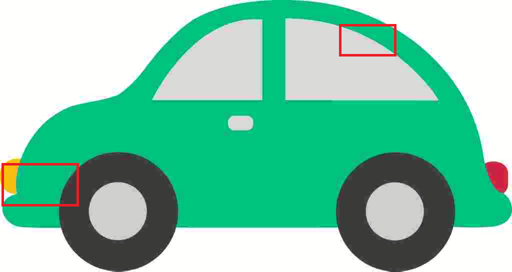} &
\includegraphics[width=0.165\linewidth]{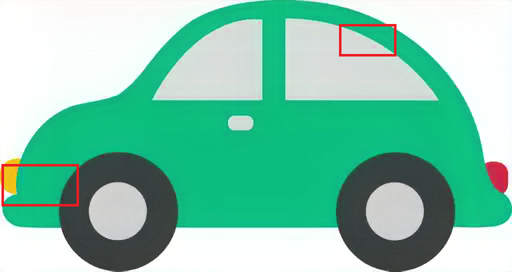} &
\includegraphics[width=0.165\linewidth]{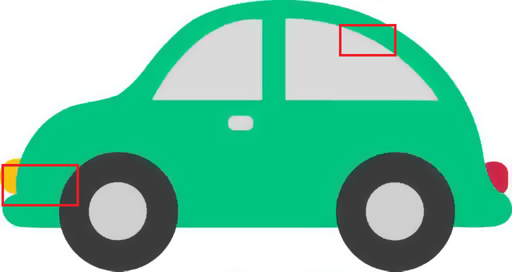}\\

\includegraphics[width=0.165\linewidth]{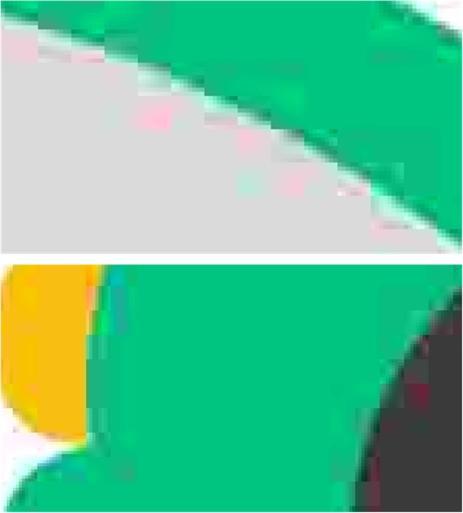} &
\includegraphics[width=0.165\linewidth]{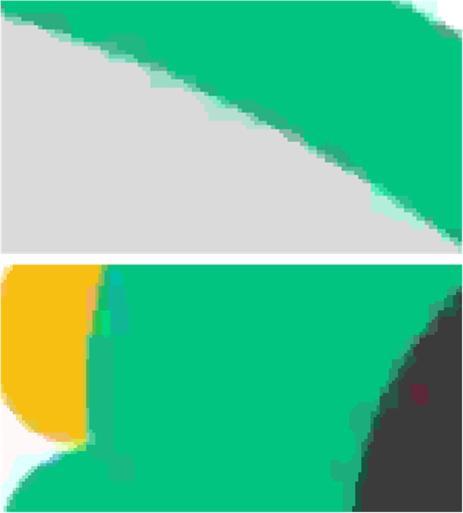} &
\includegraphics[width=0.165\linewidth]{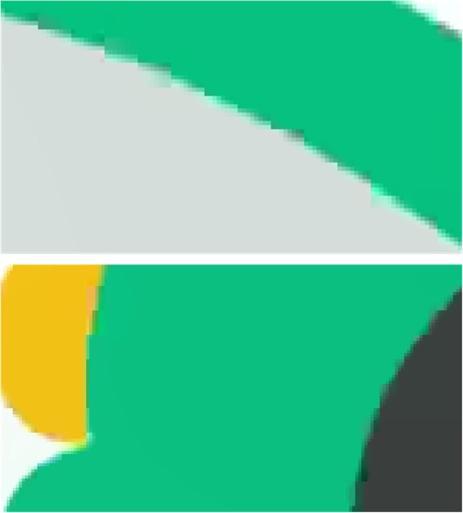} &
\includegraphics[width=0.165\linewidth]{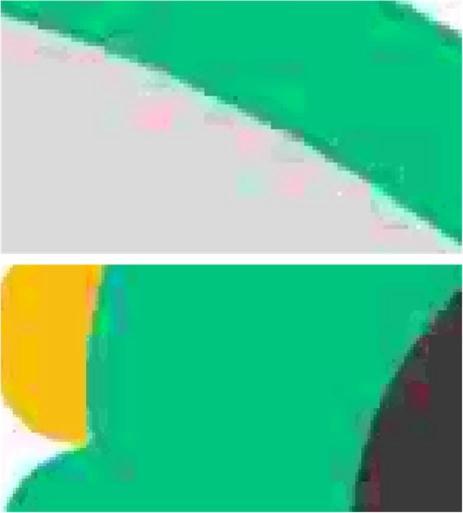} &
\includegraphics[width=0.165\linewidth]{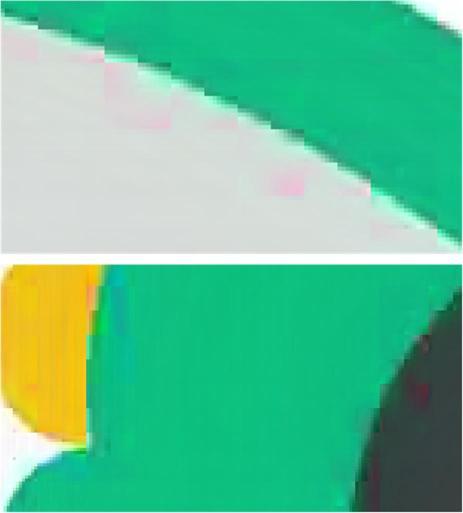} &
\includegraphics[width=0.165\linewidth]{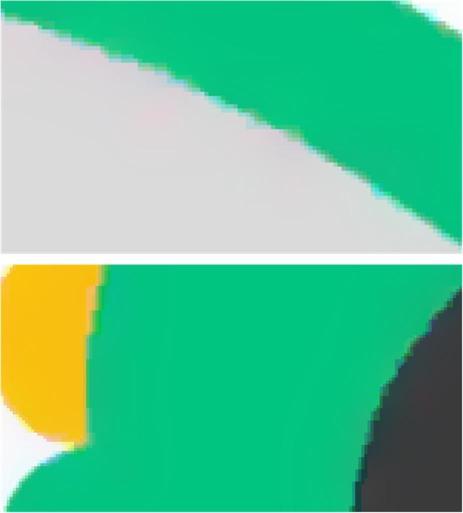} \\
(a) Input & (b) Region fusion & (c) $L_0$ norm smoothing & (d)  Wang et~al. & (e) FCN & (f) ILS\\
\end{tabular}
\caption{Clip-art compression artifacts removal results of different approaches. (a) Input image of size $543\times1024$. (b) Region fusion ($\lambda=0.02$) \cite{nguyen2015fast}, time cost is 0.51 seconds on CPU. (c) $L_0$ norm smoothing ($\lambda=0.02$) \cite{xu2011image}, time cost is 2.21 seconds (multi-thread)/ 4.32 seconds (single-thread) on CPU  and 0.15 seconds on GPU. (d) Wang et~al. \cite{wang2006deringing}. (e) The FCN based approach proposed by Chen et~al. \shortcite{chen2017fast}, time cost is 0.095 seconds on GPU. (f) The proposed ILS using the Welsch penalty function ($\lambda=20, \gamma=\frac{10}{255}, N=10$), time cost is 1.06 seconds  (multi-thread)/2.01 seconds (single-thread) on CPU and 0.032 seconds on GPU.}
\label{FigClipArt}
\end{figure*}

\begin{figure*}
 \centering
 \setlength{\tabcolsep}{0.75mm}
 \begin{tabular}{cc}
 \includegraphics[width=0.45\linewidth]{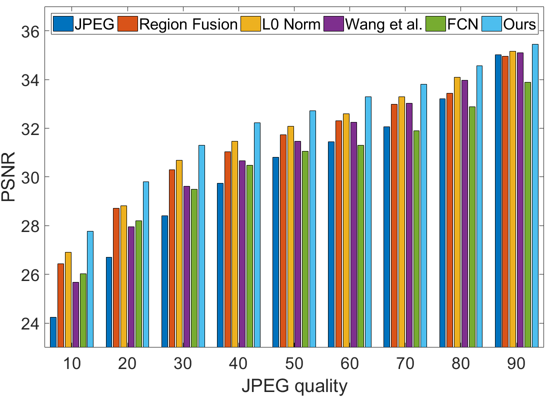}&
  \includegraphics[width=0.46\linewidth]{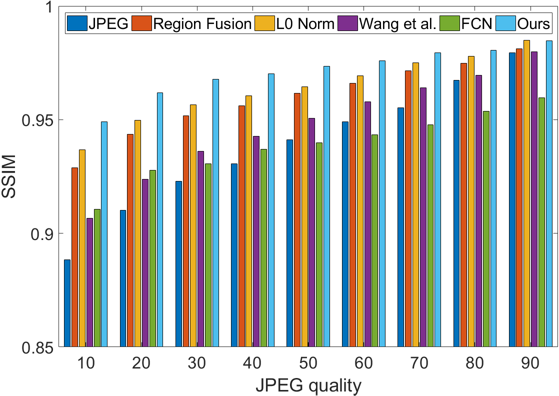}\\
 \end{tabular}
  \caption{Quantitative evaluation of the clip-art compression artifacts removal results produce by different approaches.}\label{FigClipArtEvaluation}
\end{figure*}


\begin{figure*}
\centering
\setlength{\tabcolsep}{0.25mm}
\begin{tabular}{cccccc}
\includegraphics[width=0.165\linewidth]{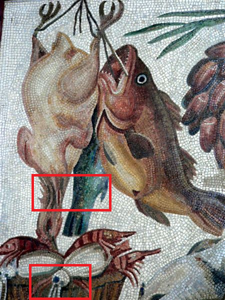} &
\includegraphics[width=0.165\linewidth]{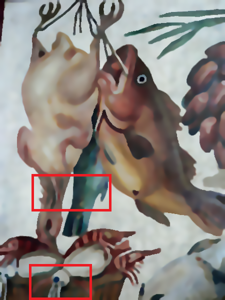} &
\includegraphics[width=0.165\linewidth]{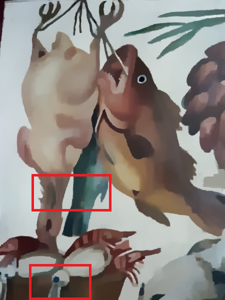} &
\includegraphics[width=0.165\linewidth]{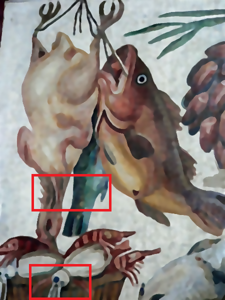} &
\includegraphics[width=0.165\linewidth]{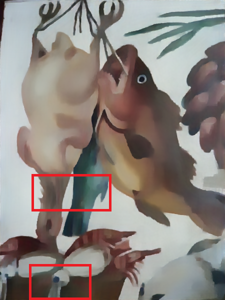} &
\includegraphics[width=0.165\linewidth]{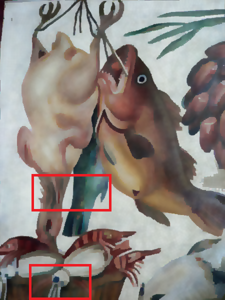} \\

\includegraphics[width=0.165\linewidth]{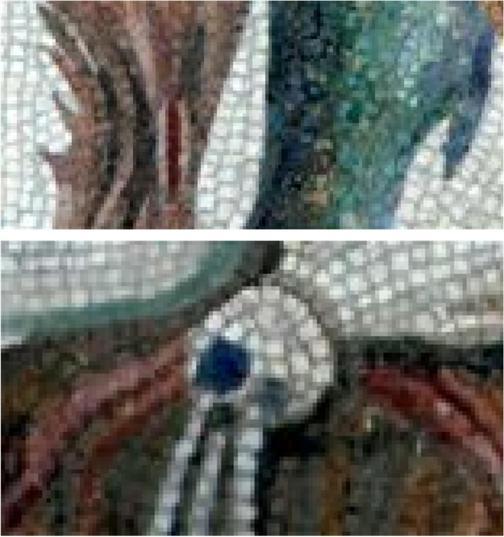} &
\includegraphics[width=0.165\linewidth]{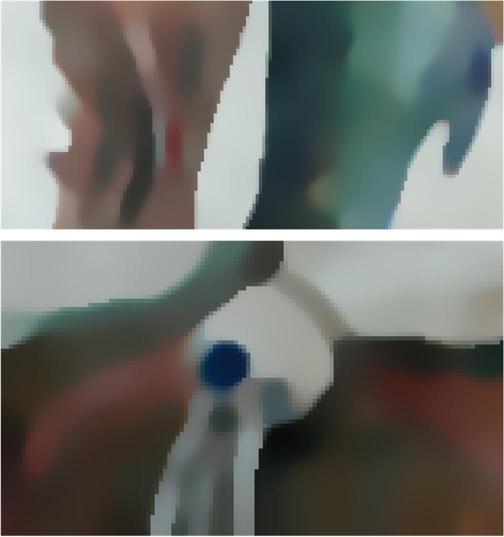} &
\includegraphics[width=0.165\linewidth]{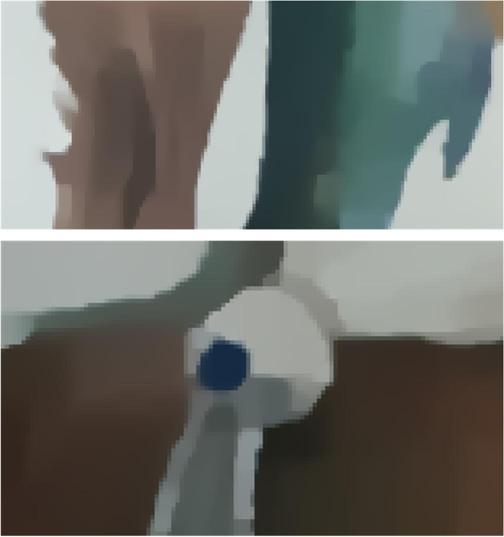} &
\includegraphics[width=0.165\linewidth]{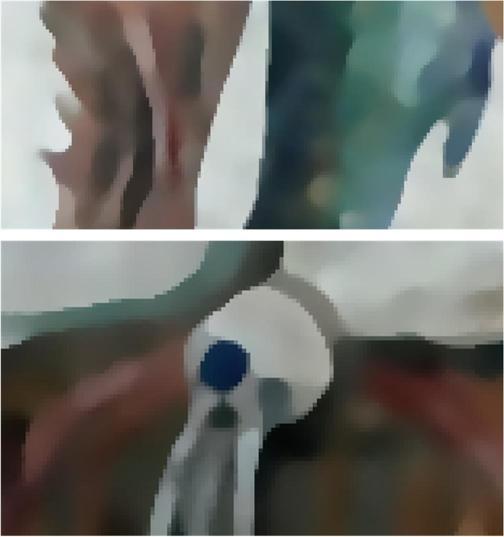} &
\includegraphics[width=0.165\linewidth]{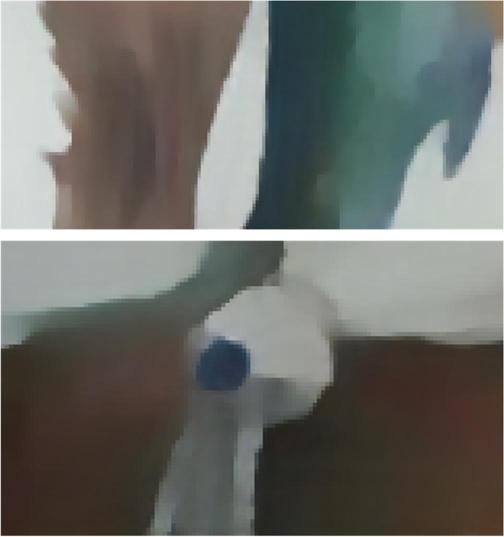} &
\includegraphics[width=0.165\linewidth]{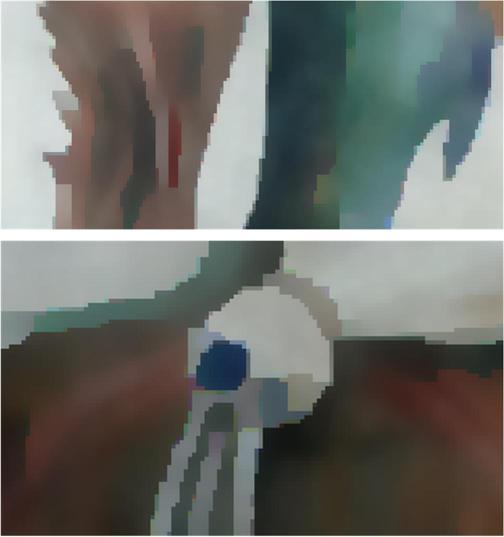} \\
(a) Input & (b) BTF & (c) RTV & (d) RGF & (e) FCN & (f) ILS\\
\end{tabular}
\caption{Texture smoothing results of different approaches. (a) Input image of size $600\times450$. (b) BTF ($k=5, n_{itr}=3$) \cite{cho2014bilateral}, time cost is 3.64 seconds on CPU. (c) RTV ($\lambda=0.025,\sigma=3$) \cite{xu2012structure}, time cost is 5.1 seconds on CPU. (d) RGF \cite{zhang2014rolling} ($\sigma_s=5,\sigma_r=0.07,N^{iter}=5$), time cost is 0.23 seconds on CPU. (e) The FCN based approach proposed by Chen et~al. \shortcite{chen2017fast}, time cost is 0.053 seconds on GPU. (f) The proposed ILS using the Welsch penalty function ($\lambda=30, \gamma=\frac{10}{255}, N=15$), time cost is 0.61 seconds (multi-thread)/1.19 seconds (single-thread) on CPU  and 0.027 seconds on GPU. Input image courtesy of Wikimedia Commons.}
\label{FigTextureSmooth}
\end{figure*}

Based on the analysis above, we apply the modified ILS to clip-art compression artifacts removal and texture smoothing. For these tasks, a larger iteration number $N$ is adopted. Specifically, we set $N=10$ for clip-art compression artifacts removal and $N=15$ for texture smoothing.

Fig.~\ref{FigClipArt} shows clip-art compression artifacts removal results of different approaches. The parameters of all the compared methods are carefully tuned to best suppress the compression artifacts. The region fusion method \cite{nguyen2015fast} cannot completely remove the compression artifacts, it even produces noticeable artifacts along edges as shown in the highlighted regions. The $L_0$ norm smoothing \cite{xu2011image} can eliminate most compression artifacts, but some compression artifacts along edges still exist. In addition, the color of the car window is also shifted as shown in the highlighted region. The approach of Wang et~al. \shortcite{wang2006deringing} can seldom handle heavy compression artifacts. Fig.~\ref{FigClipArt}(e) shows the result of the fully convolutional networks (FCN) based approach proposed by Chen et~al. \shortcite{chen2017fast}. The result is produced by the model that approximates the $L_0$ norm smoothing \cite{xu2011image}. When compared with the result of $L_0$ norm smoothing in Fig.~\ref{FigClipArt}(c), the compression artifacts are less removed in the result of the FCN based approach in Fig.~\ref{FigClipArt}(e). The compression artifacts are properly eliminated in our result. The running time of different approaches is detailed in the caption of Fig.~\ref{FigClipArt}. Our approach is $\sim2\times$ faster than the $L_0$ norm smoothing and $\sim3\times$ faster than the FCN based approach. The region fusion approach is $\sim2\times$ faster than ours. In addition, Fig.~\ref{FigClipArtEvaluation} shows the quantitative evaluation in terms of  PSNR and SSIM metrics for the results produced by different methods. Our approach consistently shows better performance for different JPEG compression quality.

Fig.~\ref{FigTextureSmooth} shows the texture smoothing results. Since the Welsch penalty function seldom penalizes very large gradients and textures can sometimes have very high contrast, we pre-smooth the input image with a Gaussian filter of a small variation to properly reduce the contrast of small structures. Our result is shown in Fig.~\ref{FigTextureSmooth}(f). Compared with the results of bilateral texture filtering (BTF) \cite{cho2014bilateral} and rolling guidance filter (RGF) \cite{zhang2014rolling}, our ILS has better performance in terms of texture smoothing and structure preserving as shown in the highlighted regions. The relative total variation (RTV) \cite{xu2012structure} tends to over smooth the shading on the object surface while our approach can effectively restore it. The model that approximates the RTV is adopted in the FCN based approach \cite{chen2017fast}. The corresponding texture smoothing result is shown in Fig.~\ref{FigTextureSmooth}(e), which is similar to the result of RTV in Fig.~\ref{FigTextureSmooth}(c). In comparison of running time, our approach is $\sim6\times$ faster than BTF, $\sim10\times$ faster than RTV, and $\sim2\times$ faster than the FCN based approach. The RGF is $\sim3\times$ faster than ours.

\section{Conclusions and Limitations}
\label{SecConclusion}

We proposed a new global optimization based approach, named iterative least squares (ILS), for efficient edge-preserving image smoothing. The proposed method is derived from the additive half-quadratic minimization technique. It is able to produce results that are on par with that of the state-of-the-art approaches but at a much lower computational cost. The computation of our method is also highly parallel. It can be easily accelerated through either multi-thread computing or the GPU hardware. When running on a GTX 1080 GPU, our ILS is able to process images of 1080p resolution ($1920\times1080$) at the rate of 20fps for color images and 47fps for gray images, which is suitable for real-time image processing. In addition, we show its flexibility with the extension of handling more applications that require different smoothing properties.

\noindent\textbf{Limitations\ \ }One limitation of our approach is that it cannot use additional information of a guidance image to perform guided image filtering. Another limitation of our method is that it cannot be used for edge-preserving sparse propagation. These limitations, however, are also shared by the $L_0$ norm smoothing \cite{xu2011image}. The compartmentalization artifacts produced by the ILS in some cases should also be counted as its limitation.

\section*{Acknowledgement}
This work was partially supported by Ministry of Science and Technology, China (No. 2019YFB1311605), National Natural Science Foundation of China (No. 61977046), National Key Research Development Project (No. 2018AAA0100702), Committee of Science and Technology, Shanghai, China (No. 19510711200), Australian Research Council through the Centre of Excellence for Robotic Vision CE140100016 and the ARC  Laureate Fellowship FL130100102 to IR. \emph{Xiaolin Huang} and \emph{Jie Yang} are the corresponding authors of this paper.

{
\bibliographystyle{ACM-Reference-Format}
\bibliography{ref}
}

\section*{Appendix A}
\label{SecAppendixA}
This appendix briefly presents the derivation of the additive half-quadratic minimization used in this paper. This technique is based on the following theory: \emph{If $f(x)$ is strictly convex, then $f(u)\geq f(v) + f'(v)(u - v) = f'(v)u + f(v) - f'(v)v$. The equality holds with $u=v$.}

Given $g(x) = \frac{c}{2}x^2 - \phi(x)$ is strictly convex, we have:
\begin{equation*}
\phi(x)\leq  \frac{c}{2}x^2 - g'(y)x - g(y) + g'(y)y.
\end{equation*}
Define $\mu=g'(y)$, then we have $y=(g')^{-1}(\mu)$ which is based on the fact that $g(\cdot)$ is strictly convex. The above inequality can now be re-written as:
\begin{equation*}
\small
    \phi(x)\leq \frac{c}{2}x^2 - \mu x + \frac{1}{2c}\mu^2 - \frac{1}{2c}\mu^2 - g((g')^{-1}(\mu)) + \mu (g')^{-1}(\mu).
\end{equation*}

By defining $\psi(\mu) = - \frac{1}{2c}\mu^2 - g((g')^{-1}(\mu)) + \mu (g')^{-1}(\mu)$, we have the following inequality:
\begin{equation}
    \phi(x)\leq \frac{1}{2}(\sqrt{c}x - \frac{1}{\sqrt{c}}\mu)^2 + \psi(\mu).
\end{equation}

The equality holds if and only if $y=x$, which means $\mu = g'(y) = g'(x) = cx - \phi'(x)$

The above analysis can be formulated as:
\begin{equation}\label{EqAddFormInequality}
    \phi(x)= \underset{\mu}{\min} \left\{\frac{1}{2}(\sqrt{c}x - \frac{1}{\sqrt{c}}\mu)^2 + \psi(\mu)\right\},
\end{equation}
with $\mu = cx - \phi'(x)$ as the optimum condition.

\section*{Appendix B}
\label{AppendixB}

This section presents how to get the constraint on the constant value $c$ in the proposed ILS. The ILS is based on the fact that $g(x)=\frac{c}{2}x^2-\phi(x)$ is strictly convex. When $\phi(x)=\phi_{p}(x)$ in Eq.~(\ref{EqSmoothedLpNorm}), we have:
\begin{equation*}
    g''(x)= c - p(x^2+\epsilon)^{\frac{p}{2}-2}\left[(p-1)x^2+\epsilon\right].
\end{equation*}
By defining $h(x)=p(x^2+\epsilon)^{\frac{p}{2}-2}\left[(p-1)x^2+\epsilon\right]$, for $g''(x)>0$, we should have $c>h(x_m)$ where $x_m$ is the point that reaches the maximum of $h(x)$. To get $x_m$, we set $h'(x)=0$ where $h'(x)$ is the first derivative of $h(x)$. We have:
\begin{equation*}
    h'(x)= p(p-2)x(x^2+\epsilon)^{\frac{p}{2} - 3}\left[(p-1)x^2+3\epsilon\right].
\end{equation*}
One also needs to keep in mind that $0<p\leq1$. Thus, for $p=1$, equation $h'(x)=0$ only has one real root $x=0$. Further calculation shows that $x_m=0$ reaches the maximum of $h(x)$. For $0<p<1$, equation $h'(x)=0$ has three real roots $x=0,x=\pm\sqrt{\frac{3\epsilon}{1-p}}$. Further calculation shows that $x_m=0$ reaches the maximum of $h(x)$. Thus, the constant value $c$ in the ILS needs to meet $c>h(0)$. However, if we set $c=h(0)=p\epsilon^{\frac{p}{2}-1}$, then we have $g''(x)=0$ only for $x=0$ and $g''(x)>0$ for any $x\neq0$ because of the monotonicity of $h(x)$. Thus, if we set $c=h(0)=p\epsilon^{\frac{p}{2}-1}$, the strict convexity of $g(x)$ is still guaranteed. Based on the above statements, the final constraint on the value of $c$ is:
\begin{equation}\label{EqCValueCondition}
    c\geq h(0)\rightarrow c\geq p\epsilon^{\frac{p}{2}-1}.
\end{equation}

Similarly, when $\phi(x)=\phi_W(x)$ in Eq.~(\ref{EqWelschCauchy}), we have:
\begin{equation*}
    g''(x)= c - 2\exp\left({-\frac{x^2}{2\kappa^2}}\right)\left(1-\frac{x^2}{\kappa^2}\right).
\end{equation*}
By defining $h(x)=2\exp\left({-\frac{x^2}{2\kappa^2}}\right)\left(1-\frac{x^2}{\kappa^2}\right)$, the maximum of $h(x)$ is obtained by solving $h'(x)=0$ where we have:
\begin{equation*}
    h'(x)= 2\frac{x}{\kappa^2}(\frac{x^2}{\kappa^2}-2)\exp\left(-\frac{x^2}{2\kappa^2}\right).
\end{equation*}
The roots of $h'(x)=0$ are $x=0,\pm\sqrt{2}\kappa$. Simple calculation shows that only $x=0$ reaches the maximum of $h(x)$. By setting $c=h(0)$, we have $g''(x)=0$ only when $x=0$ and $g''(x)>0$ for any $x\neq0$. Thus, $c=h(0)$ can also guarantee the strict convexity of $g(x)$. In this way, the final constraint on the value of $c$ is:
\begin{equation}\label{EqCValueConditionWelsch}
    c\geq h(0)\rightarrow c\geq 2.
\end{equation}

\section*{Appendix C}
\label{AppendixC}

Given $\phi_{H}(x)$ and $g_H(x,\mu)$ defined as Eq.~(\ref{EqHuber}) and Eq.~(\ref{EqHuberUpperBound}), respectively, we have $\phi_{H}(x) = \min\limits_\mu g_H(x,\mu)$, which can be obtained through the following proof.

Assuming the minimum of $g_H(x,\mu)$ with respect to $\mu$ is obtained at $\hat\mu$, and we use $ g'_H(x, \mu)$ to denote the gradient of $ g_H(x, \mu)$  with respect to $\mu$, then we have the following cases:

\noindent\textbf{Case 1:} When $\hat\mu=0$, we have $g_H(x,\mu=0)=\frac{1}{2\alpha}x^2$. In this case, the subgradient of $g_H(x,\mu)$ with respect to $\mu$ should meet the following condition:
\begin{equation*}
\begin{aligned}
  & \ \ \ \ \ \ \ g'_H(x, 0^-)\leq 0\leq g'_H(x, 0^+)\\
  & \Rightarrow -\frac{x}{\alpha} - 1\leq0\leq  -\frac{x}{\alpha} + 1\\
  & \Rightarrow |x|\leq\alpha.
\end{aligned}
\end{equation*}

\noindent\textbf{Case 2:} When $\hat\mu\neq0$, we have $g'_H(x,\hat{\mu})=0$, i.e.,
\begin{equation*}
\begin{aligned}
  & \ \ \ \ \ \ \ \frac{\hat\mu - x}{\alpha} + \text{sign}(\hat\mu)=0\\
  & \Rightarrow x = \hat\mu + \alpha \cdot \text{sign}(\hat\mu)\\
  & \Rightarrow x = \hat\mu - \alpha <-\alpha, \ \text{if} \ \hat\mu<0\\
  & \ \ \ \ \ \ x = \hat\mu + \alpha > \alpha, \ \text{if} \ \hat\mu>0\\
  & \Rightarrow |x|>\alpha.
\end{aligned}
\end{equation*}
The above proof also indicates that $x$ has the same sign as $\hat\mu$, i.e., $\text{sign}(x)=\text{sign}(\hat\mu)$.  We thus have $\hat\mu=x-\alpha\cdot\text{sign}(\hat\mu)=x-\alpha\cdot\text{sign}(x)$. In this case, the minimum of $g_H(x, \mu)$ with respect to $\mu$ is obtained as:
\begin{equation*}
\begin{aligned}
  & g_H(x, \hat\mu=x-\alpha\cdot\text{sign}(x))=\frac{1}{2\alpha}\cdot \alpha^2 + |x-\alpha\cdot\text{sign}(x)|\\
  & \ \ \ \ \ \ \ \ \ \ \ \ \ \ \ \ \ \ \ \ \ \ \ \ \ \ \ \ \ \ \ \ \ \ \ \ \ \ \ \ \ = \frac{\alpha}{2} + |x|  - \alpha\\
  & \ \ \ \ \ \ \ \ \ \ \ \ \ \ \ \ \ \ \ \ \ \ \ \ \ \ \ \ \ \ \ \ \ \ \ \ \ \ \ \ \  = |x|  - \frac{\alpha}{2}.
\end{aligned}
\end{equation*}
The equality $|x-\alpha\cdot\text{sign}(x)|=|x|  - \alpha$ is based on the condition that $|x|>\alpha$ ($\hat\mu=x-\alpha\cdot\text{sign}(\hat\mu)$  and considering the cases where $\hat\mu>0$ and $\hat\mu<0$).

Based on the above proof, we have $\phi_{H}(x) =  g_H(x,\hat\mu) = \min\limits_\mu g_H(x,\mu)$  where the minimum condition is $\hat\mu=0$ if $|x|\leq\alpha$ and $\hat\mu=x-\alpha\cdot\text{sign}(x)$ if $|x|>\alpha$. The above optimization procedure is also called the \emph{soft threshold} of $x$ with respect to the parameter $\alpha$.

\end{document}